\documentclass[acmsmall,authorversion,nonacm]{acmart}%,anonymous

\AtBeginDocument{%
	}

\setcopyright{acmcopyright}
\copyrightyear{2023}
\acmYear{2023}
\acmDOI{XXXXXXX.XXXXXXX}

\acmJournal{JACM}
\usepackage{float}

\usepackage{subcaption}

\usepackage{enumitem}

\usepackage{booktabs} % For formal tables

\usepackage{csquotes}
\usepackage{amsmath}
\usepackage{outlines}

	\usepackage{siunitx}
\def\sym#1{\ifmmode^{#1}\else\(^{#1}\)\fi}
\DeclareSIUnit\eur{\officialeuro}
\DeclareSIUnit\M{M}
\DeclareSIUnit\k{k}

\def\capfirstletteraux#1#2\relax{\uppercase{#1}\lowercase{#2}}

\usepackage{marvosym}

\usepackage{graphicx}
\usepackage{graphbox} %loads graphicx package
\usepackage{array}
\newcolumntype{P}[1]{>{\centering\arraybackslash}m{#1}}
\newcolumntype{L}[1]{>{\arraybackslash}m{#1}}
\newcolumntype{R}{>{\raggedleft\arraybackslash}X}

\newcommand*\mc[1]{\multicolumn{1}{c}{#1}}

\usepackage{xspace}
\newcommand\ie{i.\,e.\xspace}
\newcommand\eg{e.\,g.\xspace}

\newcommand\cf{cf.\xspace}

\newcommand\US{U.\,S.\xspace}

\newcommand{\mathup}[1]{\mathrm{#1}}

\DeclareMathOperator{\logit}{logit}

\usepackage{url}
\usepackage[normalem]{ulem}
\usepackage{balance} 

\makeatletter
\renewcommand{\fps@figure}{htb}         % default {tbp}
\renewcommand{\fps@table}{htb}         % default {tbp}
\makeatother 

\usepackage{tabularx}

\usepackage{cleveref}

\allowdisplaybreaks

\definecolor[named]{REVCOLOR}{cmyk}{1,1,0,0}

\begin{document}
	
\title{Mechanisms of True and False Rumor Sharing in Social Media: Collective Intelligence or Herd Behavior?}

\author{Nicolas Pröllochs}
\affiliation{%
  \institution{JLU Giessen}
  \country{Germany}}
\email{nicolas.proellochs@wi.jlug.de}

\author{Stefan Feuerriegel}
\affiliation{%
  \institution{LMU Munich \& Munich Center for Machine Learning}
  \country{Germany}}
\email{feuerriegel@lmu.de}

%\author{Author names blinded for review}

%\renewcommand{\shortauthors}{Author names blinded for review}

\begin{abstract}
Social media platforms disseminate extensive volumes of online content, including true and, in particular, false rumors. Previous literature has studied the diffusion of offline rumors, yet more research is needed to understand the diffusion of online rumors. In this paper, we examine the role of lifetime and crowd effects in social media sharing behavior for true vs. false rumors. Based on \num[group-separator={,},group-minimum-digits=1]{126301} Twitter cascades, we find that the sharing behavior is characterized by lifetime and crowd effects that explain differences in the spread of true as opposed to false rumors. All else equal, we find that a longer lifetime is associated with less sharing activities, yet the reduction in sharing is larger for false than for true rumors. Hence, lifetime is an important determinant explaining why false rumors die out. Furthermore, we find that the spread of false rumors is characterized by herding tendencies (rather than collective intelligence), whereby the spread of false rumors becomes proliferated at a larger cascade depth. These findings explain differences in the diffusion dynamics of true and false rumors and further offer practical implications for social media platforms. 
\end{abstract}

%%%
%%% The code below is generated by the tool at http://dl.acm.org/ccs.cfm.
%%% Please copy and paste the code instead of the example below.
%%%
%


% the work being presented. Separate the keywords with commas.

\maketitle

\sloppy
\raggedbottom

%%%%%%%%%%%%%%%%%%%%%%%%%%%%%%%%%%%%%%%%%%%%%%%%%%%%%%%%%%%%%%%%%%%%%%

%\interfootnotelinepenalty=10000

%\TODO{consistent: tweet vs. post / resharing cascade etc.}

\section{Introduction}
\label{sec:introduction}

% context: online spreading

Social media has emerged as a prevalent information source for large parts of society, representing an unprecedented shift in how information is consumed \cite{Bakshy.2015}. It is currently estimated that almost \SI{62}{\percent} of the adult population consume news via social media platforms, and this proportion is expected to increase further \cite{Pew.2016}. Social media platforms, such as Facebook, Sina Weibo, and Twitter, also change how information is diffused, as they allow virtually all users to disseminate information simply by sharing (\eg, retweeting) content \cite[\eg,][]{Garg.2011,Han.2020,Stieglitz.2013}.

% problem: fake news

With the possibility for everyone to share information, social media essentially transfers quality control of the information from trained journalists to platform users. This also renders it possible for deceptive content in the form of false rumors{\footnotemark} to be disseminated, sometimes with severe consequences. False information on social media threatens not only the reputation of individuals and organizations \cite{Forbes.2017}, but also society at large \cite{Bar.2023}, especially if electoral discussions are manipulated. According to a survey on the 2016 \US election, social media users shared false rumors more often than news stories from mainstream media outlets \cite{TheEconomist.2017}. Moreover, the average \US adult was estimated to have seen and remembered \num{1.14} false news stories prior to the same election, which, at least for some, affected their voting \cite{Allcott.2017}. Due to their deceptive nature, false rumors on social media have become an immediate concern \cite{Lazer.2018,Webb.2016}.

% prior literature: false rumor spreading

The diffusion of rumors through social networks has been the subject of prior research in the context of offline settings. For instance, social psychology has examined the underlying dynamics of what makes people share rumors in offline settings \cite{Knapp.1944,Allport.1947}. Here rumor theory stipulates that people largely pass information along without validating the underlying veracity. This has been recognized as a potential risk, as false information might go unnoticed. A different stream of research has studied online rumors. The underlying propagation path within a social network is referred to as a cascade. For instance, some works model the diffusion of rumors on online platforms \cite[\eg,][]{Lee.2015,Tambuscio.2015}. In this regard, the structure of resharing cascades has been compared across true vs. false rumors \cite{Friggeri.2014,Vosoughi.2018}, suggesting that cascades associated with false rumors are deeper and wider in size. This has raised the question among researchers as to which determinants could explain differences in the resharing behavior with regard to true vs. false rumors. 
\footnotetext{\footnotesize A variety of terms are used to refer to online content that is deceptive. Examples include \textquote{misinformation,} \textquote{fake news,} \textquote{propaganda,} or \textquote{rumors,} each of which places a different emphasis on the underlying characteristics. For instance, the term \textquote{fake news} was conceived to indicate content that mimics traditional media, yet where the content is fabricated in order to serve a different intent \cite[\eg,][]{Allcott.2017,Lazer.2018}. We refer to the aforementioned sources for a taxonomy. In the case of fake news, the source derives a direct benefit by undermining the credibility of standard media outlets \cite{Lazer.2018}. In contrast, \cite{Vosoughi.2018} refer to information that can be true or false as \textquote{rumors.} This term highlights the possibility that the underlying veracity can be determined. Furthermore, it does not impose assumptions on the motif or the format of the content and, for this reason, we follow \cite{Vosoughi.2018} by adopting the same terminology throughout this work.}

In this work, we study how the veracity of a rumor is associated with resharing behavior on social media. Specifically, we contribute to rumor theory by studying whether there are lifetime and crowd effects acting as a form of quality control and thereby eventually diminishing the spread of false rumors on social media. There is compelling evidence from offline settings that time works against rumors, as new facts become available or, more generally, as the interest in their content declines. For offline rumors, a rumor's lifetime is a key barrier to its spread \cite{Shibutani.1966}, yet evidence from an online setting is lacking. Apart from lifetime effects, the collective sharing behavior of users itself might act as an implicit quality control measure, preventing the spread of rumors in online settings. Such crowd effects have been repeatedly observed \cite{Hong.2004,Hong.2014,Li.2018,Love.2017}. In the context of social media, sharing rumors can be seen as a problem-solving process in which individuals can pool their knowledge to falsify rumors. Following this reasoning, collective intelligence may take over with a larger cascade depth (\ie, after the rumor has spread over multiple generations and reached many follower bases) and thus diminish the spread of false rumors. Alternatively, sharing behavior might be characterized by conformity. Then users would transmit false rumors if the false rumor is of a larger cascade depth (\ie, as it has already spread over multiple generations). Different from offline rumors, users share online rumors through a simple click on the reshare button, which would imply that online rumors (as opposed to offline rumors) are especially prone to conformity. To the best of our knowledge, we are not aware of any research studying the role of lifetime and crowd behavior in the diffusion of true vs. false rumors. This represents the contribution of this work.

%% what we do

We examine differences in users' sharing behavior with regard to true vs. false online rumors based on two common characteristics of information diffusion \cite{Stieglitz.2013,Zaman.2014}, namely the volume and the speed with which users share online rumors. Specifically, we advance prior literature by providing an empirical regression model based on which we study resharing behavior across online rumors of different veracity. Our model further controls for both between-cascade heterogeneity and between-user heterogeneity (\eg, the number of followers, user engagement) as sources of different sharing behavior. We perform hypothesis testing concerning how resharing behavior is associated with lifetime and cascade depth. This yields insights into whether lifetime and crowd behavior act as a form of quality control and, eventually, diminish the spread of false rumors. 

% findings

Our empirical findings are based on \num[group-separator={,},group-minimum-digits=1]{126301} rumor cascades from Twitter. This yields $N=\num[group-separator={,},group-minimum-digits=1]{3724197}$ reshares (\ie, retweets). The sharing behavior differs for true vs. false rumors, which can be explained based on lifetime and crowd effects as follows. The number of reshares declines with a larger lifetime, yet the decline is more pronounced for false than for true rumors. The number of reshares also declines with a larger cascade depth (\ie, after the crowd has spread the rumor over multiple generations). However, the decline is stronger for true than for false rumors. Hence, lifetime is an important determinant explaining why false rumors die out. Moreover, the sharing behavior is not characterized by ``intelligent'' behavior of the crowd but rather by herding tendencies. This is in opposition to common assumptions according to which crowd behavior acts as a form of quality control (\ie, collective intelligence). On the contrary, higher cascade depths are associated with a greater virality of false rumors. Users thus do not curtail false rumors after multiple generations of a resharing cascade, but rather conform to the resharing behavior of their peers. The results remain robust when controlling for various sources of heterogeneity at the cascade and user levels, and also when considering rumors that are of mixed veracity. 

% implications

Our work has important implications for both theory and practice. First, we shed light on the viral effects of false rumors. By drawing upon rumor theory, we establish that both lifetime and crowd effects play a key role in online rumor diffusion and, in particular, explain differences in the diffusion of true vs. false rumors. Thereby, we expand upon existing research and offer a theoretical understanding of the underlying mechanism by which users share true vs. false rumors. Second, we demonstrate that, over time, the interest of users in false rumors declines and their cascade dies out. Third, a widespread -- but incorrect -- assumption is that the user base recognizes false rumors, which then die out as a result. Based on our findings, we must reject such premises according to which crowd behavior on social media platforms functions as an implicit form of quality control (\ie, collective intelligence) that eventually quashes false rumors. On the contrary, our work implies that user sharing is subject to conformity and, thus, is -- at least partially -- driven by herd behavior. This is an alarming discovery for social media platforms as it shows that they cannot rely upon collective intelligence but must rather mitigate false rumors through active and effective countermeasures.  

% outline

The rest of this paper is organized as follows. \Cref{sec:background} reviews rumor theory and information diffusion in social networks, revealing a dearth of works addressing online rumor diffusion. As a remedy, \Cref{sec:hypotheses} develops hypotheses in order to study social media sharing behavior. We then draw upon data from Twitter (\Cref{sec:dataset}) and report our empirical results (\Cref{sec:results}). Finally, \Cref{sec:discussion} discusses the implications of our work for theory and social media platforms, while \Cref{sec:conclusion} concludes. 

\section{Background}
\label{sec:background}

\subsection{Rumor Theory}

% rumor / definition

Social psychology literature commonly designates the spreading of information as \textquote{rumor,} when the content refers to a proposition that is passed from person to person without confirmation of its truthfulness \cite{Knapp.1944}. This definition implies that rumors can vary in their underlying veracity: rumors can disseminate statements that are either truthful, a distorted version of the truth, or bluntly false. When rumors convey false information, this is seen as a \textquote{failure of formal news channels} \cite{Shibutani.1966}. Knapp (1944) identified three basic principles that characterize rumors \cite{Knapp.1944}: first, rumors spread by word of mouth; second, their information addresses a \textquote{person, happening, or condition;} and third, rumors fulfill an emotional need. Rumors commonly pertain to information of public interest (rather than personal matters or trivia) and are thus viewed as a social phenomenon. This was later extended by the principle that rumors are characterized by rapid diffusion \cite{Rosnow.1991}. Rapid diffusion is especially fostered in online environments \cite{Vosoughi.2018}, thus making speed, \ie, the time users take to pass on a social media post, a distinctive characteristic of rumor diffusion.  

% offline diffusion

Research has theorized how rumors travel in offline settings. Rumors compete with other information for attention and therefore can only spread if their message is of sufficient importance \cite{Allport.1947}. Hence, false rumors are often composed in such a way that they are perceived as important, so that they attract attention and have a high chance of being disseminated. The dissemination of rumors is typically characterized by both the volume and speed of circulation \cite[\cf][]{Stieglitz.2013}. As rumors spread via word of mouth, their content can be altered throughout the process of diffusion in offline settings \cite{Shibutani.1966}. However, such a distortion is not possible in the case of social media, where resharing allows users only to forward the original message without changing its content. In addition, rumor theory stipulates differences in how true vs. false rumors are designed. Authors of false rumors frequently have the intention to deceive others and, in order to compete successfully for attention, false rumors tend to be designed in a way that makes them more likely to be propagated by others.

% time + crowd effects

In offline settings, rumor dissemination has been found to be a dynamic rather than a static process. The elapsed time since a rumor's initial appearance (\ie, lifetime) and the sharing behavior of the crowd itself are important drivers. (1)~Lifetime plays a key role in the survival of rumors. The reason is that rumormongers have the attention of their audience for only a short time, since there is only a limited window in which verification from official channels is absent \cite{Shibutani.1966}. Hence, lifetime can be regarded as a natural barrier to rumor dissemination. (2)~Crowd behavior is based on sociological aspects of rumor theory, according to which rumors are a \textquote{collective enterprise} \citep[p.\,17]{Shibutani.1966}. Following this notion, rumor dissemination is a dynamic problem-solving process in which individuals can refer to group activities. On the one hand, individuals can engage in debunking existing rumors or, on the other hand, they can conform to the collective understanding (\ie, exhibit herd tendencies) \cite{Rosnow.1991}. When sharing rumors, individuals pass them from one person to another by means of a serial transmission chain \cite{Allport.1947}, whereas exposing the falsehood of a given rumor will break the chain. In this transmission process, the role of users varies depending on their position in the chain (\ie, whether they are close to the root of the chain or further along in the dissemination process and thus located at a higher depth in the chain) \cite{Bordia.2004}.

A key difference between offline and online rumors is how they are shared. Offline rumors require extensive cognitive reasoning when transmitting a rumor as one has to retell the underlying story. This requires cognitive reasoning and thus increases the propensity toward verification behavior. In contrast to that, online rumors can be shared through a simple click on the reshare button, which makes it easier for users to act in conformity. Conformity in resharing behavior should be especially common in online settings given that users share online content without extensive cognitive reasoning and thus without assessing its believability. Hence, while the role of conformity was discussed in offline rumor diffusion \cite{Shibutani.1966}, conformity should be more important for how users share online as opposed to offline rumors. 

\subsection{Online Diffusion on Social Media}

% relevance and context

Social media has become a prevalent platform for consuming and sharing information online and shapes the political discourse \cite{Allcott.2017,Stieglitz.2013b,Solovev.2022,Bar.2022}. As any user can contribute on social media, the responsibility of quality control over content is shifted from trained journalists (as in the case of traditional media) to end-users. However, when in a hedonic mindset, users primarily seek pleasure from interacting with content, rather than engaging in extensive reasoning to determine its believability \citep{Lutz.2020}. This offers an explanation of why social media is especially prone to the dissemination of false information. 

% characteristics: retweeting

Social media allows users within a social network to share content with their follower base. For example, in the case of Twitter, content is presented in units referred to as \textquote{tweets,} while sharing someone's else content is called \textquote{retweeting.} Interactions occur primarily through existing connections within the underlying network \cite{Myers.2012}, thereby forming so-called cascades. The inherent nature of resharing can give birth to cascades that go \textquote{viral} \cite{Goel.2016}, whereby a post from a single broadcast grows into a large cascade due to extensive resharing. Even though all posts could potentially reach the complete user base, viral cascades are fairly rare \cite{Goel.2016}.

% understanding: information diffusion

Prior literature has covered information diffusion across a variety of social media platforms such as Twitter \cite{Jung.2018,Myers.2012,Stieglitz.2012b,Stieglitz.2013,Drolsbach.2023}, Flickr \cite{Cha.2009}, Last.fm \cite{Garg.2011}, and Facebook \cite{Bakshy.2015}. Some works study the different structural properties that explain the dynamics behind online diffusion \cite{Centola.2010,Garg.2011,Leskovec.2007}. A summary of this stream of research is provided in the following. 

% descriptives of casacdes

% sharing behavior

The sharing behavior of users has been studied in order to identify drivers of online diffusion, specifically by estimating both the volume and speed with which content has been shared. Both dimensions define sharing behavior \cite{Stieglitz.2013,Zaman.2014}. In terms of volume, a larger reshare count suggests that a message is shared more extensively and it is thus viewed as a \textquote{measure of virality} \cite{Han.2020}. Another important dimension of online diffusion is the speed with which content travels \cite{Yang.2010b,Stieglitz.2013}. This is measured by the response time, that is, the time difference between a user's reshare and the parent reshare. The response time indicates a user's interest in a given piece of content and thus represents an important feature for promoting information diffusion \cite{Stieglitz.2013}. Across both dimensions, sharing behavior varies across users. Some users have a large social influence, as documented by having more followers or by following a larger number of other users \cite{Han.2020,Kwak.2010,Zaman.2014}. This corresponds to a large number of users that can be influenced to disseminate content, implying that virality is the result of influential users, \ie, how many incoming or outgoing ties a user has \cite{Kwak.2010,Lerman.2010,Myers.2014}. Having accounts labeled as \textquote{verified} can also make users more influential. Furthermore, users differ in their past engagement levels, as reflected in the number of previous shares, likes, or posts. Hence, we later control for the number of followers and followees, past engagement levels, and a \textquote{verified} badge, in our analysis. 

Besides user-level characteristics, important determinants of sharing behavior are the lifetime and depth of a resharing cascade. (1)~{Lifetime} is relevant as novel information has a higher likelihood of being shared. To account for this, previous literature on modeling online diffusion takes into consideration the lifetime of a cascade \cite{Macskassy.2011}. For instance, on YouTube, the lifetime of content (\ie, the age of videos) is associated with the overall number of user interactions, thereby confirming lifetime as an antecedent of virality \cite{Susarla.2012}. (2)~{cascade depth} refers to the number of generations or hops in the diffusion tree. It indicates how far the original message has spread from the root user. Hence, cascade depth can be seen as a measure of the past sharing behavior of the crowd, where higher values imply a higher chance of the rumor reaching more follower bases and social communities \cite{Weng.2013}. In the case of Twitter, the past history of tweets is prominently displayed, including the author, the retweeting user (\ie, the parent in the resharing cascade), and statistics on past sharing activities. Past sharing behavior has the ability to influence future resharing behavior, as has been previously confirmed. Hence, when modeling resharing, previous works have controlled for cascade depth \cite{Zaman.2014}.

\subsection{Online Diffusion of Rumors}

% examples

A vast number of social media users are exposed to online rumors comprising false information. This has been observed, for instance, in the case of electoral debates \cite{Allcott.2017,Bakshy.2015} and scientific conspiracy theories \cite{Bessi.2015,DelVicario.2016,Domenico.2013}. A prominent example of an online rumor with false content is the so-called Pizzagate conspiracy theory, which wrongly accused Hillary Clinton of running a child sex ring \cite{Kim.2019}. Besides politics, rumors can also damage the reputations of companies and organizations, such as when they are erroneously accused of malpractice. False rumors can have serious downsides, even financially. For instance, a rumor once claimed that President Obama had been killed in bomb attack, thus causing stock markets to plummet \cite{Forbes.2017}. 

% online diffusion

Prior research has compared the diffusion dynamics of rumors to those of non-rumors \cite[\eg,][]{Lee.2015}. One possibility is to simulate the spread of rumors in a social network via a contagion process \cite[\eg,][]{Nekovee.2007}. Other works compare summary statistics. At the user level, this involves, for instance, the follower count \cite[\eg,][]{Castillo.2011,DelVicario.2016}. At the cascade level, it comprises lifetime \cite[\eg,][]{Castillo.2011,DelVicario.2016} and reshares \cite{Friggeri.2014}. Differences in summary statistics can eventually be leveraged by machine learning in order to distinguish rumors from other social media posts \cite{Castillo.2011,Kwon.2017,Kwon.2013,Vosoughi.2017}. However, all of the aforementioned works focus on a comparison of rumors vs. non-rumors but not on true vs. false rumors.

% by veracity

Only a few works have studied differences in online diffusion across true and false rumors. In \cite{Friggeri.2014}, around 4,000 rumors were collected from Facebook and then classified according to veracity. Based on this data, the authors studied upload and deletion rates, yet not whether rumor veracity is linked to different diffusion dynamics. In \cite{Vosoughi.2018}, the authors characterize differences between true and false rumor cascades on Twitter. Differences are observed in a variety of metrics across true and false rumors, such as the overall depth and breadth of cascades, as well as their lifetime and the number of involved users. For false rumors, cascades appear to be wider and deeper. The authors also find that user level characteristics such as the number of followers, account age, and engagement levels are linked to rumor sharing. Our work expands upon \cite{Vosoughi.2018} by studying the role of lifetime and crowd effects in rumor diffusion. In particular, we seek to discern differences in the diffusion of true vs. false rumors and thereby contribute to the understanding of why false rumors are more likely to go viral. For this purpose, we explain differences in the volume and speed with which true vs. false rumors are shared, depending on the lifetime and depth of resharing cascades. 

\section{Hypotheses Development}
\label{sec:hypotheses}

As shown in the background section, true and false rumors are characterized by different diffusion patterns. Motivated by this finding, we now seek to uncover possible reasons for these differences in the diffusion of true and false rumors. This would then allow us to better understand the viral effects of false rumors. In the following, we first describe how we quantify volume and speed in online diffusion, and then derive hypotheses concerning the role of lifetime and crowd effects as underlying mechanisms in user sharing behavior.    

\subsection{Characterizing Online Rumor Diffusion}

% two DVs

Sharing behavior on social media has been previously characterized along two dimensions, namely \emph{volume} and \emph{speed} \cite{Stieglitz.2013,Zaman.2014}. The former, volume, refers to the number of reshares and thus reflects how frequently content is shared by users. A larger reshare count characterizes virality \cite{Han.2020}. Besides volume, speed is another important dimension of online diffusion that captures how interested users are in a message \cite{Stieglitz.2013}. It indicates how quickly content spreads through social media. In social media, speed is measured by the response time, that is, the time lag between the previous reshare in the cascade and the current one. A high diffusion speed characterizes content that quickly reaches a larger audience and is thus viewed as a measure of how much attention a social media post received \cite{Susarla.2012}. In particular, a high diffusion speed is needed for content to become viral \cite{Stieglitz.2013}. In the context of rumor theory, a high diffusion speed is further relevant as a prerequisite for outpacing verification from official channels \cite{Shibutani.1966}. Previous studies have shown that both volume and speed describe sharing behavior in the context of online diffusion \cite{Yang.2010b}.

% A user's probability of retweeting rumors is higher for false than for true rumors. (b)~Users share false rumors faster than true rumors.

In this paper, we thus follow the work of \cite{Stieglitz.2013} and base our analysis on both the volume and speed of user sharing behavior. Both represent the dependent variables in our later analysis. These variables allow us to explain differences in the diffusion dynamics of true vs. false rumors. Specifically, they allow us to discern the role of \emph{lifetime} and \emph{cascade depth} in the more viral resharing of false as opposed to true rumors. Outside the scope of rumors, lifetime and cascade depth have been demonstrated as important antecedents of virality \cite{Macskassy.2011,Crane.2008}. Lifetime is the temporal difference between the initial source post and the point in time at which users reshare, while cascade depth is the distance (in hops) between the initial source post and the current reshare \cite{Goel.2012,Zaman.2014}. 

\begin{figure}[H]
	\captionsetup[subfloat]{position=above,labelformat=empty,skip=0pt}%, labelfont=bf,textfont=normalfont,singlelinecheck=off,justification=raggedright
	%\FIGURE
	%\centering
	\caption{\centering Differences Between the Concepts of Lifetime and cascade Depth in Online Rumor Diffusion.}
	{
		\subfloat{{\includegraphics[width=3.9cm]{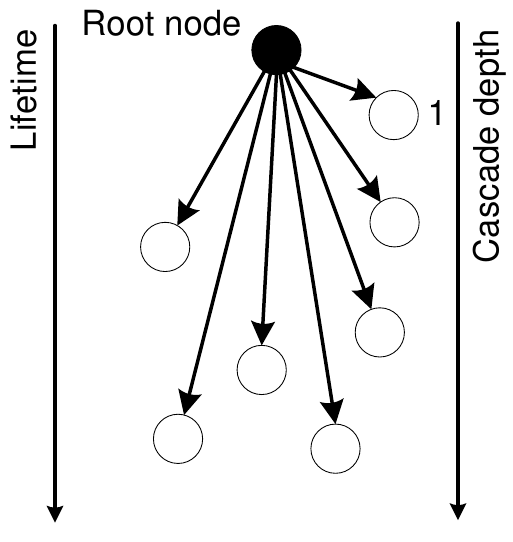}}}
		\quad\quad\quad
		\subfloat{{\includegraphics[width=3.9cm]{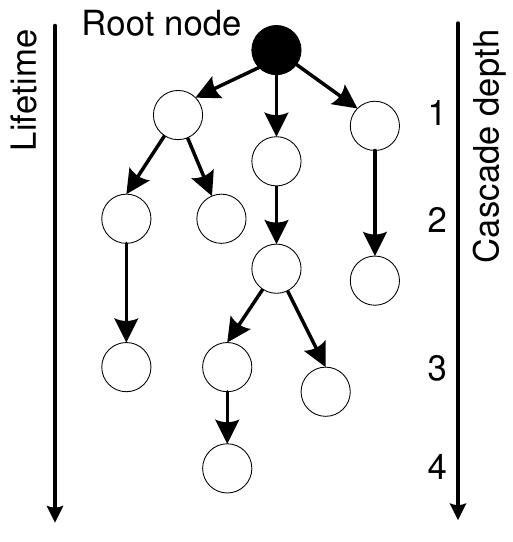}}}%
	}
	\label{fig:diffusion_examples}
%	{}
\end{figure}

Importantly, lifetime and cascade depth are inherently different concepts (\eg, \cite{Zaman.2014}). One might think that a longer lifetime corresponds to a larger cascade depth (and vice versa), yet each describes a different property of diffusion (see \Cref{fig:diffusion_examples}). Importantly, in our later empirical analysis, we find that lifetime and cascade depth have a correlation coefficient of only $-0.016$ and, therefore, are largely unrelated. On the one hand, a cascade might originate from an initial broadcast that is viral and attracts a large audience. Such a cascade can have a long lifetime but, if it is not reshared further, it has a small cascade depth. In this case, users receive the message directly from the root user and the chances that those users come from different social communities is low \cite{Weng.2013}. On the other hand, a cascade can have a long lifetime but also a large cascade depth. Here the diffusion has many intermediaries and the post is more likely to reach different follower bases with heterogeneous characteristics \cite{Weng.2013}. In this light, it is important to remind that the user base can be highly heterogeneous. For example, a rumor may be started in an echo chamber where users are supportive of the content and thus respond quickly without fact-checking, leading to a fast speed of resharing. However, rumormongers may have a smaller follower base than regular users, because of which the rumor may eventually reach only a smaller audience. Hence, as seen in this example, posts may vary greatly with respect to their speed vs. volume of resharing.

%(interaction terms?)

\subsection{The Role of Lifetime Effects}

% lifetime

In the case of virtually all social media content, user interest declines over its lifetime as the underlying information becomes less novel (or is perceived as such). For instance, virality on YouTube is negatively associated with the lifetime of content \cite{Susarla.2012}. Lifetime also plays an important role in offline rumor diffusion \cite{Shibutani.1966}. Rumormongers have been theorized as having a {short attention} span, resulting in a less viral spread of rumors over time. Hence, lifetime can be regarded as a natural barrier to the viral dissemination of rumors. We thus expect a longer lifetime to be negatively associated with a rumor's reshare volume and speed. 

% rumor theory: novelty

While lifetime is expected to curb the spread of both true and false rumors, we hypothesize that the (negative) effect is stronger for false than for true rumors. Our rationale is that rumors have only a short time span during which verification of veracity is absent. Oftentimes, such verification is not available or perhaps not feasible in the beginning, when a rumor is spawned, but has a lag of several hours or days \cite{Shao.2016,Micallef.2022}. Rumors that are false thus need to outpace verification. Rumor theory from offline settings \cite{Allport.1947,Knapp.1944,Shibutani.1966} suggests that the authors of false rumors usually have the intention to deceive users and, hence, want their rumor to attract attention. False rumors are thus often designed to be especially contagious. Therefore, in the early stages of diffusion, false rumors are expected to exhibit particularly viral diffusion dynamics (\ie, spread with high volume and speed). Over time, the uncertainty associated with rumors gradually reduces. In offline settings, it has been repeatedly emphasized that rumors thrive on uncertainty \citep[\eg,][]{Rosnow.1988}. If a rumor turns out to be false, there should be a vested interest among individuals in preventing it from being passed on to other individuals \cite{Rosnow.1988}. Given the decreasing uncertainty associated with rumors over time, false veracity should thus become visible in the form of a diminished tendency among users to reshare false rumors. We expect this tendency to manifest in differences in the lifetime effect for true vs. false rumors, with respect to both the volume and speed of user sharing. Specifically, we propose that the lifetime reduces the reshare volume more strongly for false than for true rumors and, likewise, in the case of a longer lifetime, the response time should be longer for false than for true rumors. The following hypotheses formulate the role of lifetime as a mechanism for user sharing behavior:

%\vspace{0.3cm}
\begin{quote}
 \emph{\textbf{Hypothesis~H1.}} \emph{The volume of user sharing is more {sharply} reduced with a longer lifetime for false rumors than for true rumors.} 

\noindent
 \emph{\textbf{Hypothesis~H2.}} \emph{The response time is more {sharply} reduced with a longer lifetime for false rumors than for true rumors.} 
\end{quote}
%\vspace{0.3cm}

\subsection{The Role of Crowd Effects}

% rumors are social

Propagating rumors is regarded as an inherently sociological phenomenon \cite{Rosnow.1991}: \textquote{rumor is a collaborative process, selection is necessarily collective [\ldots] The result is a collective product to which each participant has contributed in some way} \citep[pp.\,177--178]{Shibutani.1966}. Hence, when propagating rumors, individuals are interdependent: they form a group with a collective orientation towards a rumor that is progressively refined while the underlying hypotheses regarding its veracity are tested and revised, until verification is available \cite{Shibutani.1966}. Altogether, this emphasizes the importance of crowd behavior in (offline) rumor diffusion.  

% roles

% In offline settings, 

Individuals are known to follow different roles in rumor propagation. On the one hand, a common assumption is that individuals will recognize false rumors and refrain from sharing them. This assumption is motivated by evidence suggesting that some individuals adopt verification-seeking behavior, which allows them to evaluate a rumor's veracity and thereby falsify existing rumors \cite{Rosnow.1991}. In this case, crowd behavior in the context of rumor diffusion should entail the debunking of false rumors, thereby reflecting \emph{collective intelligence.} On the other hand, research has also suggested an opposing view, whereby individuals conform to the behavior of the group and thus propagate rumors despite their veracity being unverified or false \cite{Shibutani.1966}. In this scenario, crowd behavior would be driven by \emph{herd behavior.} In the following, we delineate these opposite manifestations of crowd behavior in online rumor diffusion and, on this basis, formulate two competing hypotheses.

\subsubsection{collective intelligence in Rumor Sharing?}%Wisdom of Crowds 

% rumor sharing / intelligence

The spread of rumors can be understood as a collective sense-making process \cite{Bordia.2004}. Individuals often have critical dispositions towards rumors, which helps cultivate a collective perspicacity that can eventually lead to greater intelligence among the group \cite{Pendleton.1998}. Shibutani (1966) \cite{Shibutani.1966} describes the sociological mechanism of rumor sharing as \textquote{communication through which men (sic) caught together in an ambiguous situation attempt to construct a meaningful interpretation of it by pooling their intellectual resources} (p.\,17). Similarly, when a group is introduced to uncertain information requiring interpretation, such as an unverified rumor, \textquote{group members typically pool their knowledge} \citep[p.\,35]{Rosnow.1991}. Through the pooling of knowledge, crowd behavior in rumor sharing should act as a form of quality control that prevents false rumors from becoming viral.  

Prior research has used the term \emph{collective intelligence} to characterize the work of online crowds to identify, verify, and amplify truthful information \cite[\eg,][]{Starbird.2013,Vieweg.2008,Zeng.2019}. Collective intelligence refers to seemingly intelligent behavior in large-scale interaction \cite{Levy.1997}. Different from the concept of the ``wisdom of crowds,'' which focuses on aggregating isolated inputs (\ie, assumes the evaluations of group members to be independent), collective intelligence entails the possibility for information transfer between group members \cite{Kameda.2022}. In some situations, information flow between group members can aid group performance, particularly when group members share their unique information \cite{Kameda.2022}. The notion of collective intelligence may be especially applicable in online environments due to the ease with which users can access additional information, including verifications from official channels. Evidence for collective intelligence has been observed in a wide range of applications and settings \cite[\eg,][]{Han.2021,Hong.2004,Prollochs.2022}. For instance, online health communities enable socially connected crowds to offer accurate diagnostic recommendations \cite{Dissanayake.2019}. 

%Situations in which collective behavior leads to more effective decision-making have been subsumed under the term \textquote{wisdom of crowds} \cite{Hong.2004}. Crowds are \textquote{wise} when a group pools their knowledge, so that their aggregate estimate is closer to the true value and yields a better result than relying upon the estimates of individuals, especially when the individual estimates are fairly dispersed \cite{Lorenz.2011}. The notion of the wisdom of crowds may be especially applicable in online environments due to the ease with which users can access additional information, including verifications from official channels. Wisdom of crowds effects have been observed in a wide range of applications and settings \cite[\eg,][]{Han.2021,Hong.2004,Prollochs.2022}. For instance, online health communities enable socially connected crowds to offer accurate diagnostic recommendations \cite{Dissanayake.2019}. 

In keeping with the concept of collective intelligence, we hypothesize that the crowd in online rumor sharing acts as a form of quality control and diminishes the spread of false rumors. Specifically, we propose that the sharing of false rumors is less pronounced at higher depths of a resharing cascade. The rationale is that resharers at higher cascade depths have a higher chance of belonging to different communities \cite{Liang.2018}. The integration of diverse perspectives from groups with different pools of knowledge increases the likelihood of finding solutions to complex problems \cite{Dissanayake.2019} and is a key condition for the emergence of collective intelligence \cite{Mann.2017}. Exposure to different communities also makes groups less susceptible to simply mirroring the resharing behavior of the root user \cite{Weng.2013}. Previous research has further shown that potential sharers who are far from the seed user (\ie, the initial poster) tend to be less familiar with the ideological positions of the seed user \cite{Liang.2018}. As a result, ideological identity may be less salient in guiding their behavior. Existing theory implies that higher levels of user heterogeneity promote collective intelligence \cite{Lorenz.2011}. We therefore expect collective intelligence to become increasingly evident with a larger cascade depth (\ie, after the rumor has spread over multiple generations and reached more and more follower bases) and thus reduce the spread of false rumors. Put differently, we expect a comparatively lower volume and longer response times for false rumors at higher cascade depths. The collective intelligence notion leads us to hypothesize the following:

%\vspace{0.3cm}
\begin{quote}
 \emph{\textbf{Hypothesis~H3a.}} \emph{The volume of user sharing is more {sharply} reduced with higher cascade depths for false rumors than for true rumors.} 

\noindent
 \emph{\textbf{Hypothesis~H4a.}} \emph{The response time is more {sharply} reduced with higher cascade depths for false rumors than for true rumors.}  %& (\ie, \textquote{wisdom of crowds})

\end{quote}
%\vspace{0.3cm}

\subsubsection{Herd Behavior in Rumor Sharing?}

% herd behavior

Collective intelligence can be undermined by herd behavior \cite{Mann.2017,Lorenz.2011}. Specifically, herd tendencies may induce individuals to revise their estimates. This can happen for various reasons \cite{Lorenz.2011}. For instance, individuals may suspect that others have better information or they may be driven by conformity to imitate peers. By imitating others, individuals engage in herding, by which they \textquote{observe[s] others and make[s] the same decisions or choices that the others have done} \citep[p.\,115]{Sun.2013}. Herding results in information cascades whereby signals are passed throughout the crowd, from the source to followers, often without new information being added. The literature suggests that herd behavior occurs under two primary conditions: uncertainty and the observation of others' behavior \cite{Sun.2013}. In the context of Twitter rumors, both conditions are fulfilled since (1)~unverified content is often published and (2)~the past history of tweets is prominently displayed. One example is Groupon where consumption is characterized by herding \cite{Li.2018}.

% conformity in rumor sharing

The importance of conformity has also been acknowledged in rumor theory. It \textquote{is the underlying concept that helps to explain uncertainty reduction in group cohesiveness [\ldots] and is one motivator for transmitting rumors} \citep[p.\,80]{Pendleton.1998}. Content that lacks verification, such as rumors, produces conformity and leads individuals to follow their peers \cite{Pendleton.1998}. Therefore, rumor sharing should exhibit characteristics of herd behavior.

The sharing of offline and online rumors is fundamentally different, which implies that online rumors (vis-à-vis offline rumors) are characterized by conformity. Online rumors are shared in a straightforward manner by users through a simple click on a reshare button. Conformity due to new technologies has also been observed in other online settings because of which users follow herd behavior \cite[\cf][]{Sun.2013}. In the case of online rumors, conformity might be further exacerbated by the fact that users of social media are in a hedonic mindset and avoid cognitive reasoning such as verifying the veracity of a rumor \citep{Lutz.2020}. Hence, one would expect that users sharing online rumors are especially prone to act in conformity (as opposed to offline rumors where sharing requires extensive cognitive reasoning and should thus be linked to increased verification behavior). 

While herd behavior might -- at least partially -- undermine collective intelligence for both true and false rumors, differences in the perceived level of uncertainty lead us to expect that the effects will be more pronounced for false rumors. Previous research has shown that false rumors tend to be more surprising and novel \cite{Vosoughi.2018}. Since novelty creates uncertainty \cite[\eg,][]{Kagan.2009}, the level of uncertainty of a social media user when confronted with a post should, on average, be higher for false than for true rumors. Herding theory implies that individuals are more likely to go with the herd when they are uncertain about the decision to be made \cite{Sun.2013}. We thus expect that users are more likely to conform to the resharing behavior of other users in the case of false rumors. In this case, users would not curtail false rumors after multiple generations of a resharing cascade but rather conform to the resharing behavior of their peers (by continuing to spread the false rumor). Put differently, the volume of user sharing would be less reduced with higher cascade depths for false rumors than for true rumors. Previous research also suggests that, in herd situations, users tend to spend less time on their decisions \cite{Thies.2016}. We therefore hypothesize a comparatively shorter response time at higher cascade depths for false rumors. In the presence of herd behavior, we propose the following:

\vspace{0.3cm}
\begin{quote}
 \emph{\textbf{Hypothesis~H3b.}} \emph{The volume of user sharing is less reduced with higher cascade depths for false rumors than for true rumors.}  % (\ie, \textquote{herd behavior})

\noindent
 \emph{\textbf{Hypothesis~H4b.}} \emph{The response time is less reduced with higher cascade depths for false than for true rumors.} % (\ie, \textquote{herd behavior})
\end{quote}
%\vspace{0.3cm}

\section{Data and Methodology}
\label{sec:dataset}

\subsection{Data}

% why Twitter

This work studies the spread of false information based on rumor cascades from Twitter. We chose Twitter as it is a leading social media platform that enjoys widespread popularity \cite{Pew.2016}. Currently, Twitter has around 330 million active users \cite{Statista.2020}. It has recently received critical coverage in the public discourse due to its impact on public opinion, especially with regard to electoral campaigns \cite{Allcott.2017}, and the risk that some of its content has a deceptive effect on society \cite{Lazer.2018,Vosoughi.2018}.

% terminology

As is common in online rumor theory, we distinguish between \textquote{rumors} and \textquote{cascades.} Each \emph{rumor} in our sample involves one or more \emph{cascades}. A rumor has more than one cascade if it exhibits multiple independent resharing (\ie, retweeting) chains started by different users but pertaining to the same story. Put simply, multiple cascades occur when the same rumor is broadcast multiple times. To account for this, we later control for heterogeneity at the cascade-level.

% coverage

Our empirical analysis builds upon a comprehensive and representative dataset of Twitter rumors from its founding in 2006 through 2017 \cite{Vosoughi.2018}. Our choice of data provides a real-world, cross-sectional sample and is thus subject to considerable heterogeneity. In sum, our data amounts to $N =$\,\num[group-separator={,},group-minimum-digits=1]{126301} cascades corresponding to \num{2448} rumors. These include more than 4.5 million reshares (\ie, retweets) by around 3 million different users. Permission to process this dataset for the purpose of our study was obtained from Twitter. Approval for this research was received from the institutional review board at (Blinded). 

% fact-checking

For all rumor cascades, a veracity label was obtained from six independent fact-checking organizations, namely factcheck.org, hoax-slayer.com, politifact.com, snoopes.com, truthorfiction.com, and urbanlegends.about.com. This yielded labels classifying the veracity of rumors. As a result, each rumor cascade belongs to one of three different groups, depending on whether its veracity is classified as \textquote{true} or \textquote{false.} Otherwise, it is assigned to a third category named \textquote{else.} Out of the \num[group-separator={,},group-minimum-digits=1]{126301} rumor cascades in our data, \num{19287} were classified as true and \num{82605} as false (for a discussion of the distribution, we refer to \cite{Vosoughi.2018}). In addition, \num{24409} rumors fell into the \textquote{else} category, \ie, there was no clear assignment to positive or negative veracity. We utilize rumors that belong to true and false veracity in our main analysis. We later repeat the analysis with the additional rumors from the \textquote{else} category as part of robustness checks (\ie, to control for rumors of mixed veracity and to control for the possibility that fact-checking organization selected a stratified sample for verification). The results confirm the findings from the main analysis. The fact-checking labels from the different organizations attain a fairly high pairwise agreement, amounting to values between \SI{95}{\percent} and \SI{98}{\percent}. Rumors classified as \textquote{true} or \textquote{false} even show a perfect pairwise agreement of \SI{100}{\percent}.

% Filtering

Our data were processed analogous to \cite{Vosoughi.2018} in that we study rumor cascades started by humans and thus exclude rumor cascades started by bots. This leaves the actual retweet chain intact and allows us to model and understand the role of user behavior in rumor dissemination. The resulting data consist of $N=\num[group-separator={,},group-minimum-digits=1]{3724197}$ retweets (\ie, observations), which represent the unit of analysis in the following.

\subsection{Model Variables}

% structure

\begin{figure}
\centering
\caption{\centering Structure of an Example Cascade.}
\includegraphics[width=.35\textwidth]{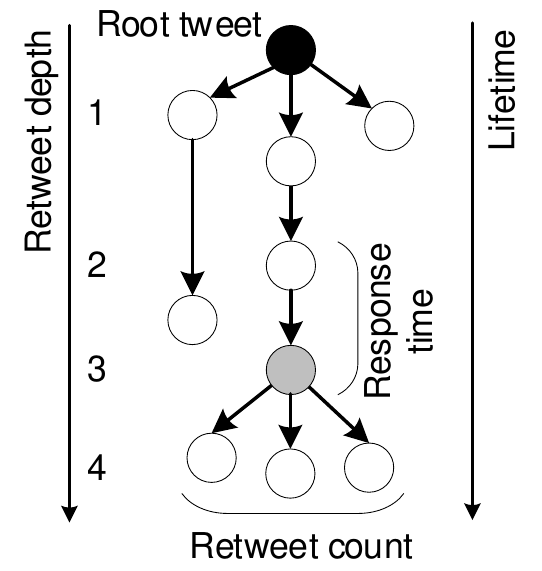}
\label{fig:cascade_example}
\vspace{-0.3cm}
\end{figure}

Recall that a rumor consists of multiple cascades. Each cascade originates from a separate broadcast of the same story and has a tree structure, in which the root node represents the original source tweet with the content of the rumor that is disseminated.  An example of a cascade on Twitter is shown in \Cref{fig:cascade_example}. Each node represents an event whereby a user has shared the tweet, whereas the edges indicate the dependency between reshares, that is, which tweet is a retweet from another tweet.

In the following, we introduce the model variables as later used in our empirical analysis. Note that our unit of analysis are individual retweets. Hence, all model variables are collected at the retweet level and, thereby, we study social media sharing behavior based on the unfolded retweet cascades where each node in the cascade tree represents one observation.

% dependent variables

\subsubsection*{\textbf{Dependent Variables.}} Building upon earlier research, we characterize retweet behavior according to two {variables}, namely (1)~retweet count and (2)~response time. Retweet count quantifies how many users share content \cite{Stieglitz.2013} and is thus used \textquote{as the measure of virality} \cite{Han.2020}. Response time examines the speed of the underlying diffusion \cite{Stieglitz.2013,Zaman.2014}. Response time sheds light on whether rumors of a certain veracity are more interesting to users and thus spread faster.

% structural variables

%\subsubsection*{\emph{Diffusion Variables.}}
\subsubsection*{\textbf{Rumor-Theoretic Dissemination.}} In our analysis, we are interested in the differences in the dissemination of true vs. false cascades. To this end, we include the veracity of a rumor according to the fact-checking label, that is, a binary label indicating whether or not it conveys false information. 

In order to test our hypotheses, we build upon two additional variables that characterize online rumor diffusion, namely the lifetime and the retweet depth within a cascade. Lifetime is the time difference between the current retweet under consideration and the initial source tweet that represents the root within the cascade. Retweet depth quantifies past sharing activities. Hence, for each retweet, we compute its depth within the overall cascade by counting the number of retweets (\ie, hops) that it is away from the tweet at the root. Importantly, both variables are time-varying in the sense that they are re-computed for each retweet, so that they refer to the lifetime/depth of a cascade up to the time at which a specific retweet occurred (and not of the complete, ex~post cascade). Both also capture different concepts (see \cite{Zaman.2014} or the summary statistics later).

Reassuringly, we emphasize that retweet depth does \emph{not} provide a direct measure of collective intelligence or herd behavior. Instead, we utilize retweet depth to measure the number of generations of a retweet cascade, where larger values imply that a rumor has reached larger follower bases and thus larger social communities. Motivated by crowd behavior \cite{Lorenz.2011}, we expect users to exhibit a different sharing behavior at higher retweet depths. If users continue to spread false rumors at higher retweet depths, then users exhibit herding tendencies. If users refrain from retweeting rumors at higher retweet depths, then they are characterized by \textquote{intelligent} behavior.

%% User-level variables

\subsubsection*{\textbf{User-Specific Variables.}}
At the user level, we consider variables that could potentially explain retweet behavior due to differences in both social influence and engagement levels These are (1)~the number of followers; and (2)~the number of users who are followed. Both are count variables with a heavy-tailed distribution and are therefore subject to log-transformation in the regression analysis. Furthermore, we include (3)~the account age; (4)~user engagement as measured by the past number of interactions on Twitter (\ie, tweets, shares, replies, and likes) relative to the account age \cite[\cf][]{Vosoughi.2018}; and (5)~whether the user account was verified by Twitter. Verified accounts are highlighted with a blue badge next to the user's name in order to verify the authenticity of accounts of public interest (\eg, celebrities, politicians).\footnote{\footnotesize Twitter: \emph{About verified accounts}. URL: \url{https://help.twitter.com/en/managing-your-account/about-twitter-verified-accounts}} The aforementioned variables are time-varying in the sense that they provide a snapshot at the time of the retweet (\eg, the account age at the time of the retweet).

All variable definitions are summarized in \Cref{tbl:variables}. In addition, we control for heterogeneity among rumors and cascades (\eg, to account for different topics or for the fact that some cascades were started by a user with a greater social influence). This is detailed later as part of the model specification. 

\begin{table}[H]
%\TABLE
\caption{Variable Definitions.}
\label{tbl:variables}
{
%\OneAndAHalfSpacedXI
\footnotesize
\begin{tabular}{l p{11.0cm}}
\toprule
\textbf{Variable} & \textbf{Description}  \\
\midrule
\multicolumn{2}{c}{\textsc{Dependent variables}} \\
\midrule
	Retweet Count & The number of retweets (\ie, the number of child nodes in the cascade tree) \\
  Response Time & Time difference between a user's retweet and the direct parent tweet (in hours) \\
\midrule
\multicolumn{2}{c}{\textsc{Independent variables}} \\
\midrule
\multicolumn{2}{l}{\underline{Rumor-theoretic dissemination}} \\
	Falsehood & The falsehood of the rumor ($=1$ if false, otherwise 0) \\ 
  Lifetime & Time difference (in hours) between the original broadcast representing the root of the cascade and the current retweet \\ 
	Retweet Depth & Depth of the current retweet in the cascade (\ie, the number of hops from the root to the retweet) \\ 
	\addlinespace
\multicolumn{2}{l}{\underline{User-specific variables}} \\
	Followers & Number of people who follow the user on Twitter (in \num{1000}s) \\ 
  Followees & Number of people whom the user follows on Twitter (in \num{1000}s) \\ 
  Account Age & The age of the user's account (in years) \\ 
	User Engagement & Average daily number of interactions (tweets, retweets, replies, likes) of a user on Twitter\\% a  relative to the account age \\
  Verified & Whether the user's account has been officially verified by Twitter \\ 
\bottomrule
\multicolumn{2}{l}{Unit of analysis: retweet level ($N=\,$\num[group-separator={,},group-minimum-digits=1]{3724197})}
\end{tabular}
}
{}
\end{table}

\subsection{Summary Statistics}

% summary statistics: DV

\begin{figure}[htb]
	\captionsetup[subfloat]{position=below,labelformat=empty,skip=0pt}%, labelfont=bf,textfont=normalfont,singlelinecheck=off,justification=raggedright
	\caption{Complementary Cumulative Distribution Functions (CCDFs) for Cascades.\label{fig:ccdf}}
%	\FIGURE
	{
		\subfloat{{\includegraphics[width=5.5cm]{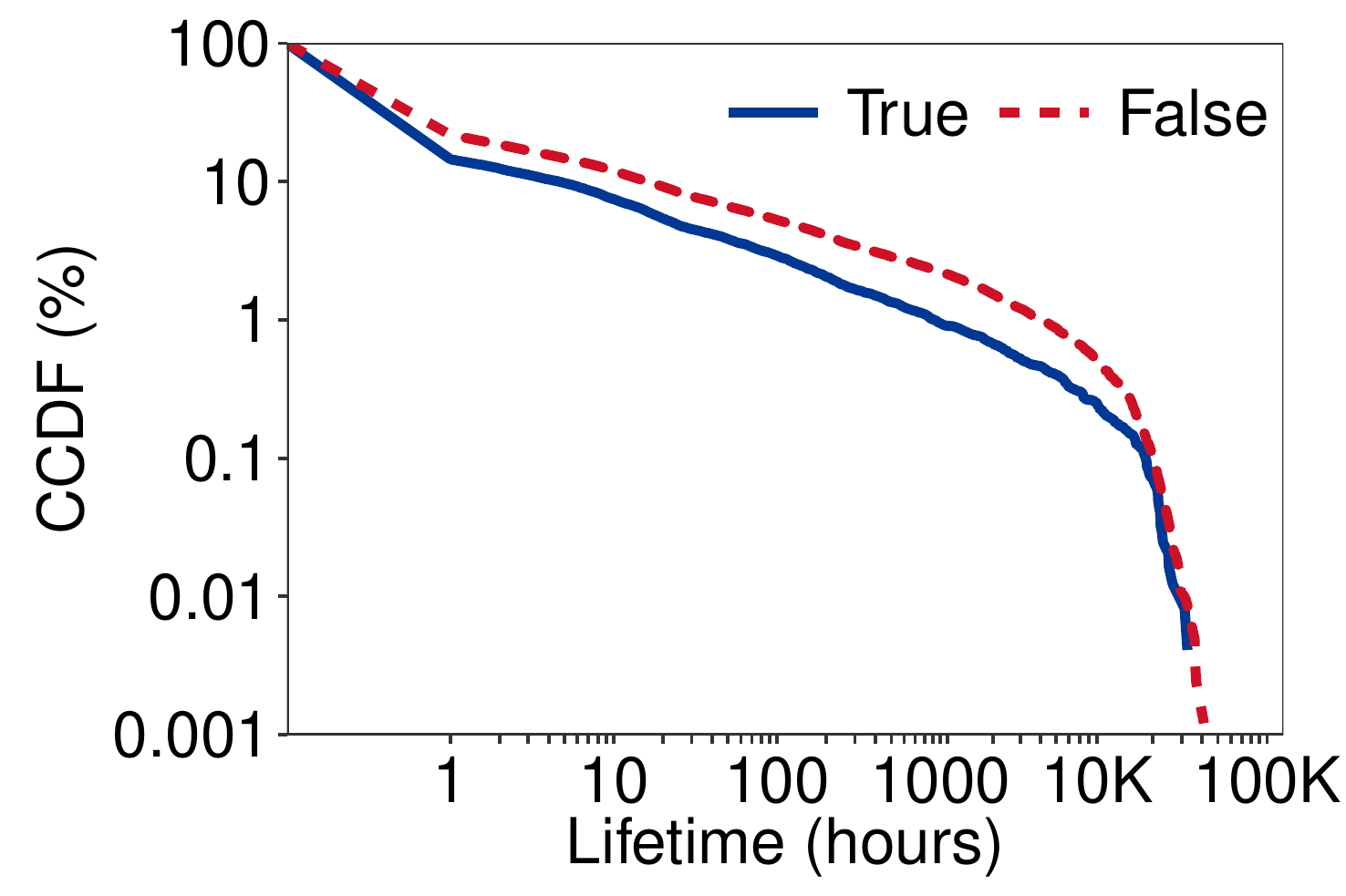}}}
		\quad\quad
		\subfloat{{\includegraphics[width=5.5cm]{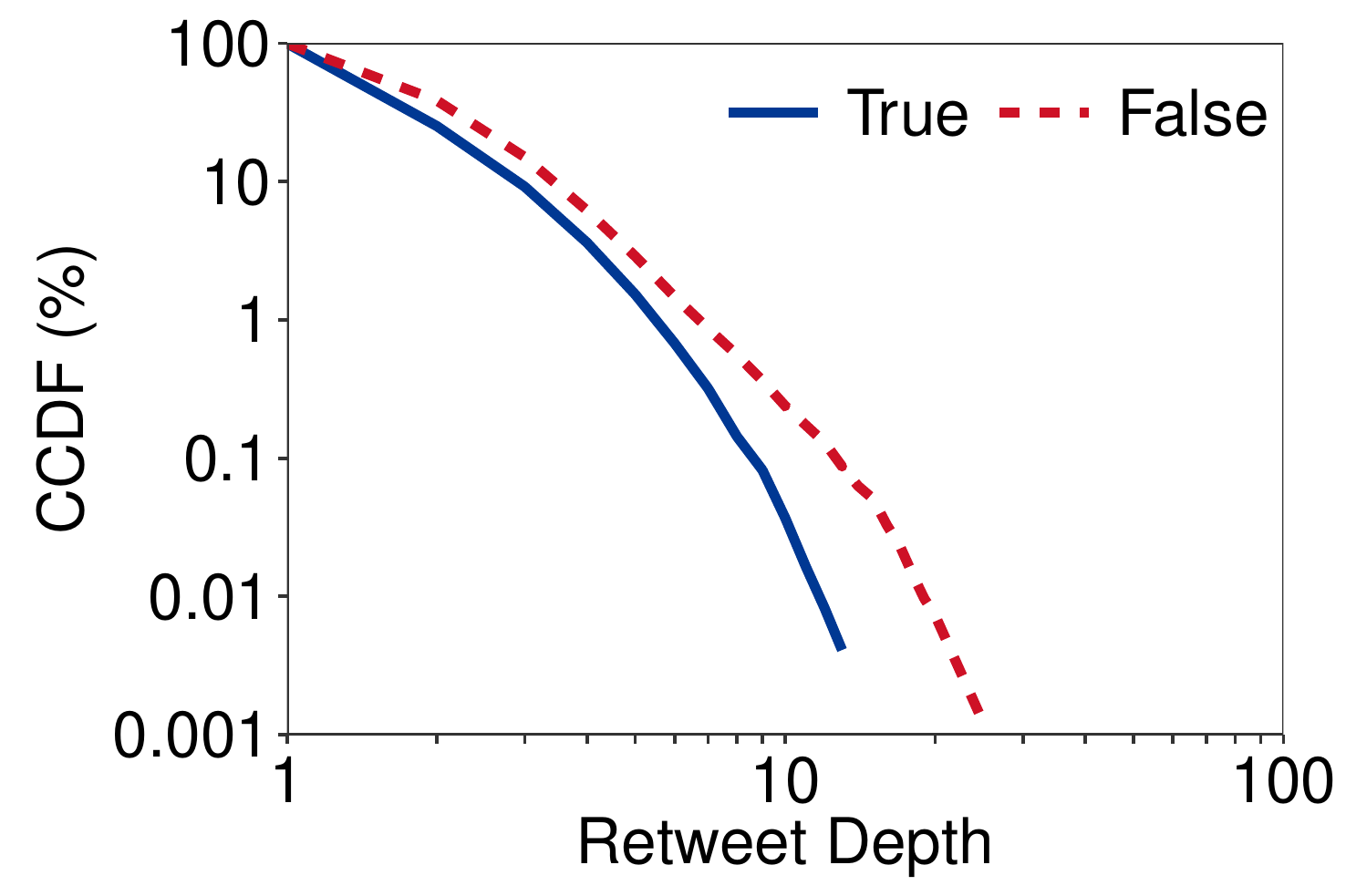}}}%
	}
	{}
	%}
\end{figure}

Summary statistics of the model variables are reported in \Cref{tbl:descriptives}. We observe a higher retweet volume for false rumors as compared to true rumors. Specifically, the average retweet count amounts to 0.977 for false rumors and to 0.879 for true rumors. \Cref{fig:ccdf} compares the diffusion based on the complementary cumulative distribution function (CCDF). Here the CCDF measures the fraction of cascades across true and false rumors with a given lifetime (left) and depth (right). The CCDF is useful to study how often a random variable is above a particular level, and, hence, a strength of the CCDF is that it focuses specifically on the tail distribution, that is, the few cascades that go viral. Evidently, false rumors are characterized by a longer lifetime and a higher depth across the complete distribution.

% CCDFs

% IV

Both false and true rumors are subject to considerable heterogeneity across almost all model variables (\Cref{tbl:descriptives}). The average lifetime of a retweet cascade spans \num{300.78} hours. This duration is longer for false rumors (\num{309.08} hours) than for true rumors (\num{156.15} hours). Cascades from false rumors also tend to be deeper. For false rumors, the mean retweet depth is \num{1.78}, while, on average, the retweet depth amounts to only \num{1.28} for true rumors. Furthermore, false rumors are more frequently tweeted by users from accounts that are not verified and younger in age. False rumors are also more frequently observed among users with a lower engagement level, users who follow a smaller number of other users, and users who themselves have a smaller number of followers. Each difference in means is statistically significant at common levels when performing a two-sided Welch $t$-test.

\begin{table}[H]
	%\TABLE
	\caption{Descriptive Statistics.} 
	\label{tbl:descriptives}
	{
		%\OneAndAHalfSpacedXI
		\footnotesize
		\begin{tabular}{lS[table-format=4.2]S[table-format=4.2]S[table-format=4.2]S[table-format=4.2]S[table-format=4.2]S[table-format=4.2]}
			\toprule
			\textbf{Variable} & \multicolumn{2}{c}{\textbf{Mean}} & \multicolumn{2}{c}{\textbf{Median}} & \multicolumn{2}{c}{\textbf{SD}} \\
			\cmidrule(lr){2-3} \cmidrule(lr){4-5} \cmidrule(lr){6-7} 
			& {\textbf{False}} & {\textbf{True}} & {\textbf{False}} & {\textbf{True}}  & {\textbf{False}} & {\textbf{True}} \\
			\midrule
			\multicolumn{7}{c}{\textsc{Dependent variables}} \\
			\midrule
			Retweet Count & 0.977 & 0.879 & 0.000 & 0.000 & 56.470 & 16.329 \\ 
			Response Time & 260.908 & 104.188 & 4.467 & 0.772 & 1079.889 & 919.635 \\ 
			\midrule
			\multicolumn{7}{c}{\textsc{Independent variables}} \\
			\midrule
			\multicolumn{2}{l}{\underline{Rumor-theoretic dissemination}} \\
			Falsehood & 1.000 & 0.000 & 1.000 & 0.000 & 0.000 & 0.000 \\  
			Lifetime & 309.081 & 156.146 & 12.118 & 1.466 & 1184.721 & 1162.483 \\ 
			Retweet Depth & 1.780 & 1.280 & 1.000 & 1.000 & 1.496 & 0.919 \\ 
			\addlinespace
			\multicolumn{2}{l}{\underline{User-specific variables}} \\
			Followers & 2.234 & 5.240 & 0.410 & 0.466 & 83.146 & 178.604 \\ 	
			Followees & 1.002 & 1.707 & 0.383 & 0.509 & 4.588 & 6.847 \\ 
			Account Age & 2.939 & 3.478 & 2.690 & 3.326 & 1.878 & 2.152 \\  
			User Engagement & 19.698 & 24.655 & 9.518 & 9.541 & 34.715 & 45.257 \\ 
			Verified & 0.002 & 0.006 & 0.000 & 0.000 & 0.043 & 0.079 \\ 
			\bottomrule
			\multicolumn{7}{l}{Unit of analysis: retweet level ($N=\,$\num[group-separator={,},group-minimum-digits=1]{3724197})}
		\end{tabular}
	}
	{}
\end{table}

\subsection{Cross-Correlations}

The cross-correlations among the dependent variables are close to zero (see \Cref{tbl:correlation_DV}). The Pearson correlation coefficient between retweet count and response time amounts to $-0.004$. This can be expected as speed and volume of retweeting are different concepts. Rumormongers, for example, are likely users that are particularly engaged and thus have a fast response time but have a smaller follower base than regular users, which implies a lower retweet count. Recall that we study social media behavior based on the unfolded retweet cascades where each node in the cascade tree represents one observation (\ie, at the retweet level). A higher correlation between retweet count and response time would be expected when studying rumor spreading based on the folded retweet cascades (\ie, at the cascade level). Reassuringly, this confirms earlier findings suggesting that retweet count and response time reflect different concepts \cite{Stieglitz.2013}, thus motivating our analysis based on both the volume and speed of diffusion.

\begin{table}[H]
%\TABLE
\caption{Cross-Correlations Among Dependent Variables.}
\label{tbl:correlation_DV}
{
%\OneAndAHalfSpacedXI
\footnotesize
\sisetup{round-mode=places,round-precision=3}
\begin{tabular}{rl *{5}{S[table-format=2.3,table-column-width=.85cm]}S[table-column-width=.85cm, table-format=1.0]}
\toprule
&& \multicolumn{2}{c}{\textbf{All}} & \multicolumn{2}{c}{\textbf{False}} & \multicolumn{2}{c}{\textbf{True}}\\
\cmidrule(lr){3-4} \cmidrule(lr){5-6} \cmidrule(lr){7-8} 
%\midrule 
& \textbf{Variable} & \mc{$1$} & \mc{$2$}  & \mc{$1$} & \mc{$2$} & \mc{$1$} & \mc{$2$} \\
  \midrule
1 & Retweet Count &	{1} &  & {1} &  & {1} &  \\ 
2 & Response Time &  -0.004 & {1} & -0.004 & {1} & -0.004 & {1} \\
\bottomrule
\multicolumn{8}{p{9.5cm}}{Stated: Pearson correlation coefficient; unit of analysis: retweet level ($N=\,$\num[group-separator={,},group-minimum-digits=1]{3724197})}%, $^*p< 0.05$, $^{**}p< 0.01$, $^{***}p< 0.001$.}
\end{tabular}
}
{
}
\end{table}

\Cref{tbl:correlation_IV} reports the cross-correlations among the independent variables. We observe a strong positive correlation between the number of followers and the number of followees, which is consistent with previous works. The table further shows that the level user engagement is positively correlated with the number of followees and that verified users tend to have more followers. All remaining correlations are fairly small. In particular, the correlation between lifetime and retweet depth of cascades is only \num{-0.016}. Hence, a large retweet depth does \emph{not} necessarily coincide with a long lifetime and vice versa. This demonstrates that both variables relate to different constructs. It also motivates our research, in which we distinguish the role of lifetime and retweet depth during rumor dissemination. 

\begin{table}[H]
\caption{Cross-Correlations Among Independent Variables.}
\label{tbl:correlation_IV}
{
\footnotesize
\sisetup{round-mode=places,round-precision=3}
\begin{tabular}{rl *{7}{S[table-format=2.3,table-column-width=.85cm]}S[table-column-width=.85cm, table-format=1.0]}
\toprule
& \textbf{Variable} & \mc{$1$} & \mc{$2$}  & \mc{$3$} & \mc{$4$} & \mc{$5$} & \mc{$6$} & \mc{$7$} & \mc{$8$} \\% & \mc{$9$} & \mc{$10$} & \mc{$11$}\\
\midrule 
1	& Falsehood & {1} &  &  &  &  &  &  &  \\ 
2	& Lifetime & -0.029 & {1} &  &  &  &  &  &  \\ 
3	& Retweet Depth & -0.077 & -0.016 & {1} &  &  &  &  &  \\ 
4	& Followers &  0.007 & -0.002 & -0.014 & {1} &  &  &  &  \\ 
5	& Followees &  0.034 &  0.012 & -0.042 &  0.127 & {1} &  &  &  \\ 
6	& Account Age &  0.064 & -0.022 &  0.062 &  0.020 &  0.085 & {1} &  &  \\ 
7	& User Engagement &  0.032 &  0.008 & -0.014 &  0.025 &  0.236 & -0.019 & {1} &  \\ 
8	& Verified &  0.022 & -0.008 & -0.012 &  0.179 &  0.022 &  0.051 & -0.006 & {1} \\ 
\bottomrule
\multicolumn{10}{l}{Stated: Pearson correlation coefficient; unit of analysis: retweet level ($N=\,$\num[group-separator={,},group-minimum-digits=1]{3724197})}%, $^*p< 0.05$, $^{**}p< 0.01$, $^{***}p< 0.001$.}
\end{tabular}
}
{
}
\end{table}

\subsection{Model Specification}

% overview

In order to test our hypotheses, we analyze the role of lifetime and crowd effects as mechanisms for user sharing behavior. For this, we examine differences in sharing behavior of true vs. false rumors at the retweet level. Specifically, we focus on the volume and speed of sharing \cite{Stieglitz.2013,Zaman.2014} as given by (1)~retweet count and (2)~response time. We specify two regression models with retweet count and response time as dependent variables. We explain both using lifetime and retweet depth as independent variables, but allow for potential variation in both across true vs. false rumors. This is done through an interaction term between veracity and lifetime (and between veracity and retweet depth) in order to quantify whether false rumors are shared by users more extensively (and more quickly). Thereby, we discern differences in the diffusion dynamics (\ie, volume and speed) of true vs. false rumors.  

% Random effects

In our model specification, we must control for heterogeneity among rumors. For example, some rumors cover topics that are more appealing than others and are thus characterized by a more pronounced diffusion. There might also be heterogeneity among the cascades that belong to a single rumor. For example, some cascades within a rumor have been broadcast at the root by users with a large social influence (\eg, due to their \textquote{verified} badge) and the cascade is thus characterized by a more pronounced diffusion.\footnote{\footnotesize To gain quantitative insight, we offer additional estimation result, including an analysis of the social influence of the root user, in Supplement~\ref{appendix:root_social_influence}.} Likewise, cascades might differ for novel and recurring rumors. For the purpose of capturing such heterogeneity, a model specification with cascade-specific random effects is needed. Thereby, we account for variation across groups but also for the fact that veracity does not vary within a group, as is the case for cascades. Importantly, different from fixed effects, a random effects specification still allows us interpret between-group variation within a single rumor \cite[\cf][]{Townsend.2013}. Hence, we control for heterogeneity among rumor cascades via cascade-specific random effects throughout all analyses. 

\subsubsection{Volume of Rumor Diffusion.}

% retweet count

In the first regression model, the dependent variable is given by the retweet count. Retweet count is a non-negative count variable, and its variance is larger than the mean. To adjust for overdispersion, we draw upon a negative binomial regression to model the retweet count, analogous to \cite{Stieglitz.2013}. In a negative binomial regression, a log-transformation of the conditional expectation is modeled as the dependent variable. 

The independent variables are as follows. Rumor theory stipulates differences in the diffusion of true vs. false rumors. A prerequisite for false rumors to be shared is that they must be composed in a way that attracts attention \cite{Allport.1947}, and therefore, false rumors have a higher baseline propensity of being shared \cite{Vosoughi.2018}. Consistent with this, we control for $\mathit{Falsehood}$ (\ie, $=1$ if the rumor veracity is classified as false and otherwise $=0$). To test hypotheses H1 and H3, we include interactions $\mathit{Falsehood}$ $\times$ $\mathit{Lifetime}_t$ and $\mathit{Falsehood}$ $\times$ $\mathit{RetweetDepth}_t$, respectively (as well as the corresponding individual covariates for $\mathit{Lifetime}_t$ and $\mathit{RetweetDepth}_t$). Since we apply a negative binomial regression, the interpretation of the effects requires an exponential transformation of the coefficients. {We later adhere to \cite{Buis.2010} and \cite{Ai.2003} when interpreting interactions in negative binomial regressions.} If the exponent of the sum of $\mathit{RetweetDepth}_t$ and $\mathit{Falsehood}$ $\times$ $\mathit{RetweetDepth}_t$ is smaller than the exponent of $\mathit{RetweetDepth}_t$, false rumors are shared less extensively at higher retweet depths (by some multiplying factor). This finding would indicate that additional generations of a retweet cascade promote collective intelligence such that users refrain from retweeting false rumors. This would support H3a. If the exponent of the sum of $\mathit{RetweetDepth}_t$ and $\mathit{Falsehood}$ $\times$ $\mathit{RetweetDepth}_t$ is larger than the exponent of $\mathit{RetweetDepth}_t$, false rumors are instead shared more extensively at higher depths. In this case, our result would suggests that users do not curtail false rumors after multiple generations of a retweet cascade but rather conform to the retweeting behavior of their peers (by continuing to spread the rumor). This would support H3b. As previously mentioned, the explanatory variables in our model are dynamic variables that change over time. Specifically, these variables refer to the lifetime/depth of a cascade up to the time $t$ at which a specific retweet occurred. In addition, we control for heterogeneity in user characteristics by including the variables $\mathit{Followers}_t$, $\mathit{Followees}_t$, $\mathit{AccountAge}_t$, $\mathit{UserEngagement}_t$, and $\mathit{Verified}_t$. 

The resulting model is

\vspace{-.7\baselineskip}

{\footnotesize
\begin{align}
& \log({\mathup{E}(RetweetCount \,\mid\, ^*)}) = \, \gamma + \beta_{1} \, \mathit{Falsehood} + \underbrace{\beta_{2} \, \mathit{Falsehood} \times \mathit{Lifetime}_t}_{\text{H1}}  \nonumber\\
& \qquad + \underbrace{\beta_{3} \, \mathit{Falsehood} \times \mathit{RetweetDepth}_t}_{\text{H3a/b}} + \beta_{4} \, \mathit{Lifetime}_t + \beta_{5} \, \mathit{RetweetDepth}_t 
\label{eq:regression_volume} \\
& \qquad + \beta_{6} \, \mathit{Followers}_t + \beta_{7} \, \mathit{Followees}_t  + \beta_{8} \, \mathit{Account Age}_t + \beta_{9} \, \mathit{User Engagement}_t +  \beta_{10} \, \mathit{Verified}_t + u_\text{cascade} \nonumber 
\end{align} 
}%
\normalsize
with intercept $\gamma$, and cascade-specific random effect $u_\text{cascade}$.

\subsubsection{Speed of Rumor Diffusion.}

% speed

In the second regression model, we explain differences in the diffusion speed of true vs. false rumors. The dependent variable is given by the response time, \ie, the time lag between the previous retweet and the current one. Again we control for the veracity of rumors (\ie, $\mathit{Falsehood}$). This is consistent with rumor theory \cite{Shibutani.1966}, which stipulates a short attention window for unverified rumors to go viral and, hence, a prerequisite to achieving virality would be a faster reaction time among users. Again we include interactions to test our hypotheses concerning the role of lifetime (H2) and retweet depth (H4). The interactions test whether the underlying mechanisms by which users share false as opposed to true rumors are different. A positive interaction $\mathit{Falsehood} \times \mathit{Lifetime}_t$ would imply that the response time declines more sharply for false rumors than for true rumors, thus supporting H2. A negative interaction $\mathit{Falsehood} \times \mathit{RetweetDepth}_t$ would provide support for H4a, while a positive interaction would instead provide support for H4b. All other variables are analogous to the previous model. This yields 

\vspace{-.7\baselineskip}

{\footnotesize
\begin{align}
& \log({ResponseTime}) = \, \gamma + \beta_{1} \, \mathit{Falsehood} + \underbrace{\beta_{2} \, \mathit{Falsehood} \times \mathit{Lifetime}_t}_{\text{H2}}  \nonumber\\
& \qquad + \underbrace{\beta_{3} \, \mathit{Falsehood} \times \mathit{RetweetDepth}_t}_{\text{H4a/b}} + \beta_{4} \, \mathit{Lifetime}_t + \beta_{5} \, \mathit{RetweetDepth}_t 
\label{eq:regression_response_time} \\
& \qquad + \beta_{6} \, \mathit{Followers}_t + \beta_{7} \, \mathit{Followees}_t  + \beta_{8} \, \mathit{Account Age}_t + \beta_{9} \, \mathit{User Engagement}_t +  \beta_{10} \, \mathit{Verified}_t + u_\text{cascade} \nonumber 
\end{align} 
}%
\normalsize
with intercept $\gamma$, and cascade-specific random effect $u_\text{cascade}$.

Prior research has found response times on social media to be log-normally distributed \cite{Zaman.2014}. Accordingly, we log-transform $\mathit{ResponseTime}$. Results of the Shapiro-Wilk test for normality as applied to the log-transformed variable suggest that the null hypothesis of normal distribution cannot be rejected. We thus estimate the model using ordinary least squares~(OLS). As a robustness check, we also estimated a censored Tobit model. The results are qualitatively identical.

% focus: effect size

Note that our analyses in the next section primarily focus on the estimated effect sizes. This allows us to determine how strongly diffusion is linked to a rumor's veracity. We emphasize that traditional hypothesis testing will largely return $p$-values that are statistically significant at common thresholds, since our analysis is based on a large-scale dataset with $N=\,\num[group-separator={,},group-minimum-digits=1]{3724197}$ observations. We refer to \cite{Lin.2013} for a discussion of $p$-values in large sample sizes.

\section{Empirical Analysis}
\label{sec:results}

In this section, we test our hypotheses. Specifically, we study the role of lifetime and retweet depth in explaining the volume and speed of retweet behavior, that is, retweet count and response time. In addition, we perform an extensive series of checks in order to validate the robustness of our findings. 

\subsection{Analysis of Retweet Count in Rumor Sharing Behavior (H1 \& H3)}
%\subsection{Analysis of Retweet Count}

% Model

We now analyze determinants that underlie the retweet count in user sharing behavior. For this purpose, we draw upon a negative binomial regression with retweet count as the dependent variable; see \Cref{tbl:regression_retweet_count}. All variables are standardized in order to facilitate interpretability. Given the large sample size of approximately \SI{3.7} million observations, all $p$-values can be expected to be fairly low and, hence, our analysis primarily focuses on the estimated effect sizes. We find that user-specific variables are associated with retweet counts at statistically significant levels. More retweets occur when rumors are shared by users with a larger number of followers. This implies that users with more followers have a larger social influence. In contrast, a higher number of followees is negatively associated with retweet count. Furthermore, more retweets are observed for accounts that are younger in age, have a lower engagement level, and are verified. Here we note that, by using random effects, we capture other sources of heterogeneity among rumor cascades.

%% Veracity

Prior literature has suggested that users retweet false rumors more frequently than true rumors. The results in \Cref{tbl:regression_retweet_count} show a positive and statistically significant coefficient for $\mathit{Falsehood}$. Yet, because we estimate a negative binomial regression model with interaction terms, the coefficients cannot be interpreted as the change in the mean of the dependent variable for a one unit (\ie, standard deviation) increase in the respective predictor variable, with all other predictors remaining constant. Instead, there are two viable options to interpret coefficients in nonlinear regression models with interaction terms (see \cite{Buis.2010} for details): (i) on a multiplicative scale by calculating the incidence rate ratio (IRR), which is equal to the exponent of the coefficient of the respective variable \cite{Buis.2010}. Here the coefficients can be interpreted as the natural logarithm of a multiplying factor by which the predicted number of retweets changes, given a one-unit increase in the predictor variable, holding all other predictor variables constant \cite{Buis.2010}. (ii) In terms of marginal effects, which are nonlinear functions of the coefficients and the levels of the explanatory variables \cite{Ai.2003, Buis.2010}. In the following, we start by interpreting the coefficient estimates in terms of IRR, while we later also evaluate marginal effects. Both variants yield consistent findings. The coefficient for $\mathit{Falsehood}$ amounts to \num{0.196}, which corresponds to an IRR of $e^{0.196} \approx 1.22$. This implies that a false rumor is expected to have \SI{21.65}{\percent} more retweets than a true rumor.

\begin{table}[H]
\caption{Regression Results for Retweet Count.\label{tbl:regression_retweet_count}}
{
\footnotesize
\begin{tabularx}{\textwidth}{@{\hspace{\tabcolsep}\extracolsep{\fill}}l *{4}{S} } 
\toprule
\multicolumn{5}{l}{Dependent Variable: Number of Retweets ($\mathit{RetweetCount})$}\\
\midrule
 & {\textbf{Model (1)}} & {\textbf{Model (2)}} & {\textbf{Model (3)}} & {\textbf{Model (4)}}\\ 
\midrule
Falsehood                         &              &              & 0.067^{***}  & 0.196^{***}  \\
                                  &              &              & (0.005)      & (0.006)      \\
Falsehood $\times$ Lifetime ({\textbf{H1}}) &              &              &              & -0.040^{***} \\
                                  &              &              &              & (0.004)      \\
Falsehood $\times$ Retweet Depth ({\textbf{H3a/b}}) &              &              &              & 0.191^{***}  \\
                                  &              &              &              & (0.005)      \\
Lifetime                    &              & -0.080^{***} & -0.080^{***} & -0.041^{***} \\
                                  &              & (0.001)      & (0.001)      & (0.004)      \\
Retweet Depth                     &              & -0.128^{***} & -0.128^{***} & -0.316^{***} \\
                                  &              & (0.001)      & (0.001)      & (0.005)      \\
Followers                        & 0.543^{***}  & 0.529^{***}  & 0.529^{***}  & 0.528^{***}  \\
                                  & (0.001)      & (0.001)      & (0.001)      & (0.001)      \\
Followees                        & -0.124^{***} & -0.120^{***} & -0.120^{***} & -0.119^{***} \\
                                  & (0.001)      & (0.001)      & (0.001)      & (0.001)      \\
Account Age                       & -0.051^{***} & -0.046^{***} & -0.046^{***} & -0.046^{***} \\
                                  & (0.001)      & (0.001)      & (0.001)      & (0.001)      \\
User Engagement                   & -0.081^{***} & -0.078^{***} & -0.078^{***} & -0.078^{***} \\
                                  & (0.001)      & (0.001)      & (0.001)      & (0.001)      \\
Verified                  & 0.050^{***}  & 0.058^{***}  & 0.058^{***}  & 0.052^{***}  \\
                                  & (0.007)      & (0.007)      & (0.007)      & (0.007)      \\
Intercept                         & 0.133^{***}  & 0.035^{***}  & -0.020^{***} & -0.150^{***} \\
                                  & (0.002)      & (0.002)      & (0.004)      & (0.005)      \\
Random effects (cascade level) & {Yes} & {Yes} & {Yes} & {Yes} \\
\midrule
AIC                                 & {\num{10728836}}   & {\num{10669093}}   & {\num{10668888}}   & {\num{10666825}}   \\
Observations                        & {\num{3724197}}      & {\num{3724197}}      & {\num{3724197}}      & {\num{3724197}}      \\
\bottomrule
\multicolumn{5}{r}{Significance levels: $^*p< 0.05$, $^{**}p< 0.01$, $^{***}p< 0.001$; standard errors in parentheses} \\ \\
\end{tabularx}
}
\subcaption*{\emph{Note:} Negative binomial regression explains the number of retweets. Unit of analysis is the 
retweet level ($N=\,$\num[group-separator={,},group-minimum-digits=1]{3724197}). Cascade-specific random effects are included.
}
\end{table}

%% Lifetime

Next, we assess how the lifetime is associated with the number of retweets for true vs. false rumors~(H1). The coefficient of the interaction between $\mathit{Falsehood}$ and $\mathit{Lifetime}$ is negative and statistically significant ($\beta=-0.040$, $p < 0.001$); the coefficient of $\mathit{Lifetime}$ is also negative and statistically significant ($\beta=-0.041$, $p < 0.001$). For true rumors, a one standard deviation increase in the lifetime of a cascade is expected to decrease the retweet count by \SI{4.02}{\percent}. Here the IRR for false rumors is calculated via the exponent of the sum of the coefficients of $\mathit{Lifetime}$ and $\mathit{Falsehood}$ $\times$ $\mathit{Lifetime}$. The resulting IRR is 0.922, which implies that a one standard deviation increase in the lifetime of a cascade reduces the predicted retweet count by \SI{7.78}{\percent}. The reduction for false rumors is \num{1.9} times larger than the reduction for true rumors. Put differently, the expected number of retweets declines more quickly over the lifetime of the cascade for false rumors. H1 is thus supported.

%% Depth

We now test how the retweet depth is associated with the number of retweets for true vs. false rumors~(H3). The coefficient of the interaction between $\mathit{Falsehood}$ and $\mathit{RetweetDepth}$ is positive and significant ($\beta=0.191$, $p < 0.001$), whereas the coefficient of $\mathit{RetweetDepth}$ is negative and significant ($\beta=-0.316$, $p < 0.001$). This implies that a one standard deviation increase in retweet depth decreases the expected number of retweets by \SI{27.09}{\percent} for true rumors and \SI{11.75}{\percent} for false rumors. The decrease in retweet count for false rumors is only \SI{43.37}{\percent} of the decrease for true rumors. Users thus share false rumors at higher depths more frequently. We must thus reject H3a and, instead, find support for H3b. 

%% Model fit 

According to the Akaike information criterion (AIC), \Cref{tbl:regression_retweet_count} suggests that the variables for lifetime and retweet depth should be included in the model. For each, the difference in AIC is greater than ten, indicating strong support for the corresponding candidate models \cite{Burnham.2004}. Hence, both lifetime and retweet depth are relevant for explaining how often rumors are retweeted.

%% Plot

\begin{figure}%[H]
	\captionsetup[subfloat]{position=bottom,labelformat=empty}% %, labelfont=bf,textfont=normalfont,singlelinecheck=off,justification=raggedright
	\caption{\centering Predicted Marginal Means of Retweet Count.}
	\label{fig:ggpredict_retweet_quantity}
	%\centering
	{
		\subfloat{{\includegraphics[width=5.5cm]{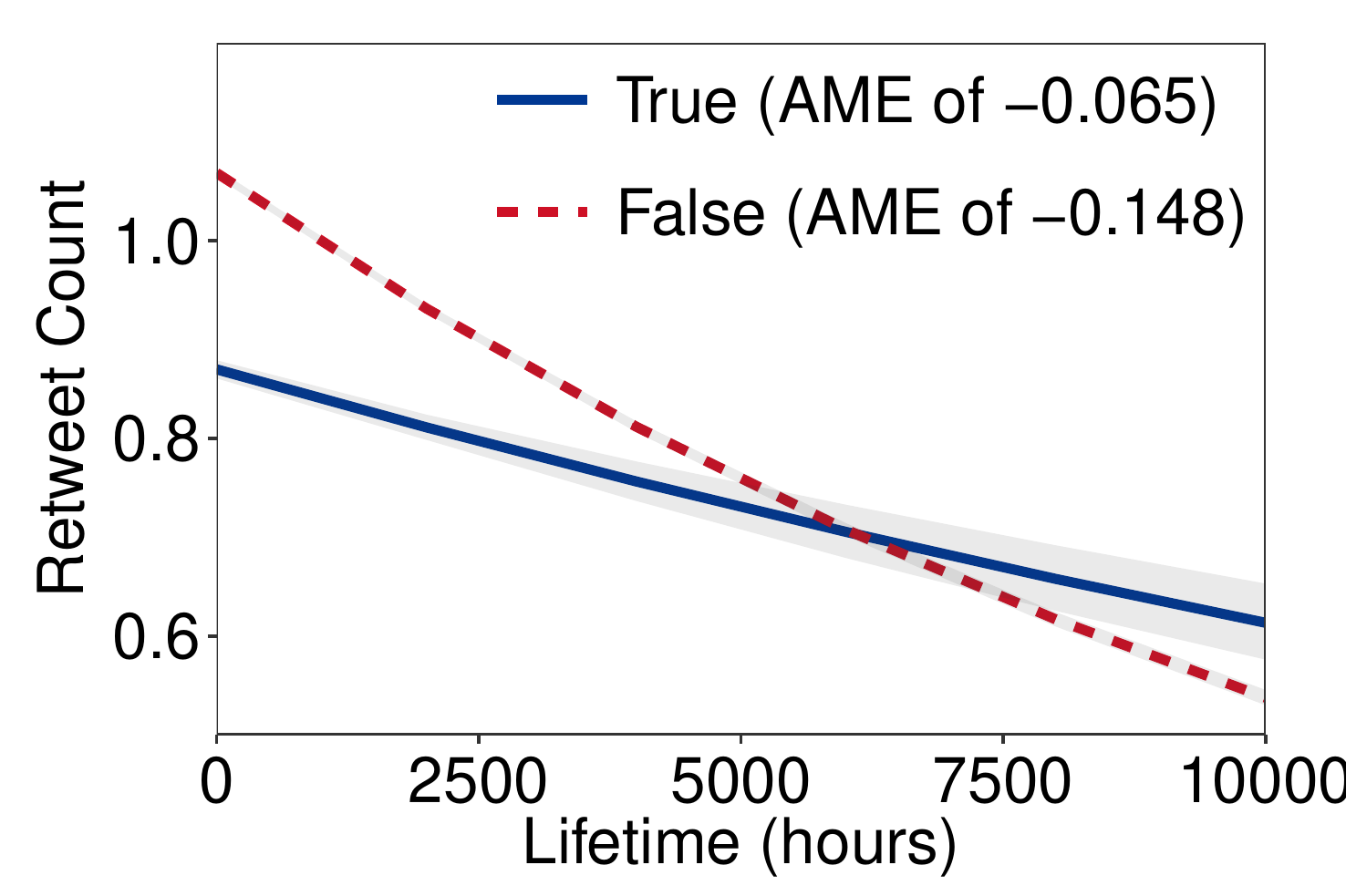}}}
		\quad\quad
		\subfloat{{\includegraphics[width=5.5cm]{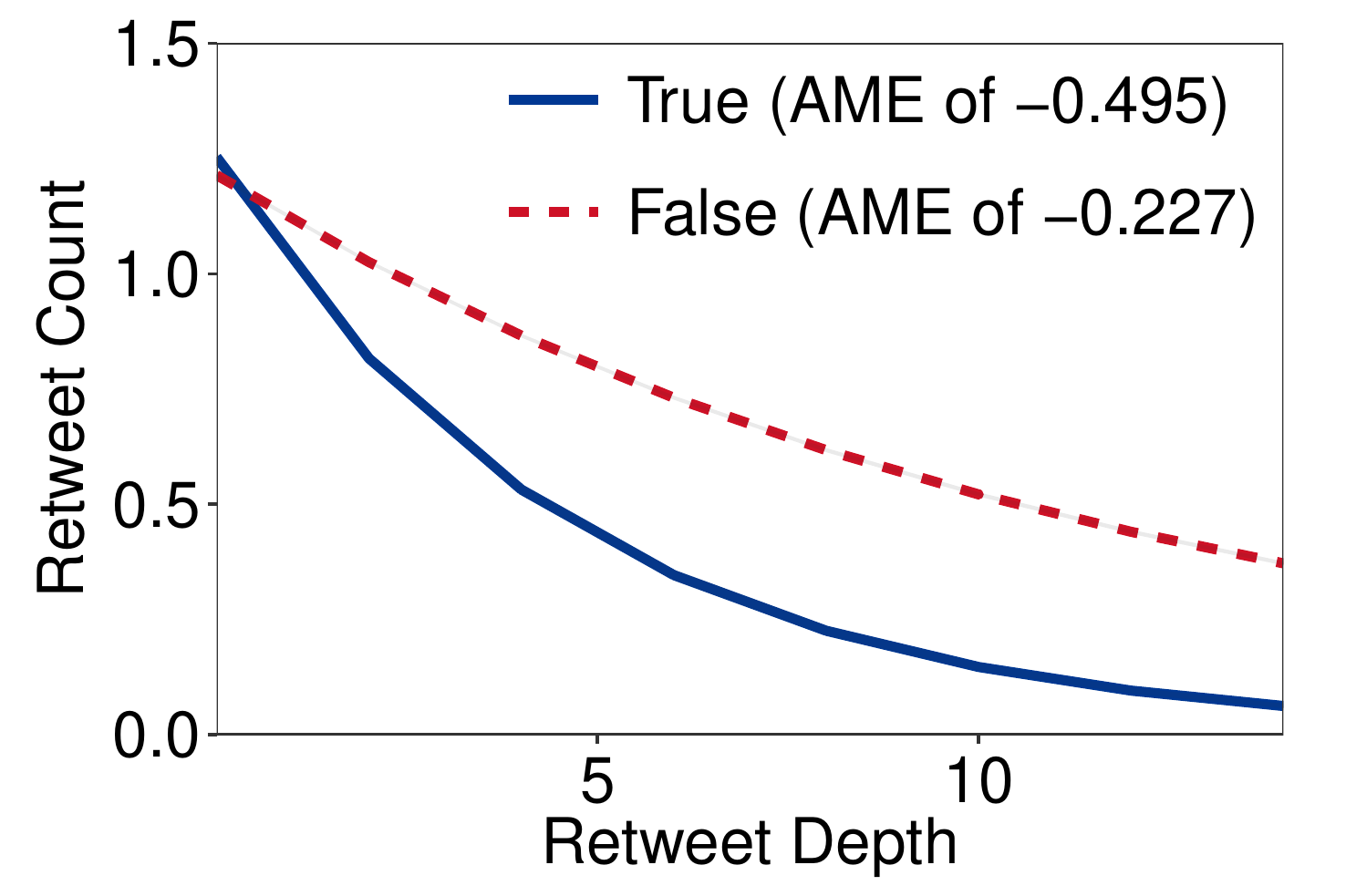}}}%
	}
	\subcaption*{\emph{Note:} The figures show the predicted marginal means of retweet count for different values of lifetime (left) and retweet depth (right). The \SI{95}{\percent} confidence intervals are highlighted in gray (but are comparatively small due to the large sample size). Cascade-specific random effects are included in the analysis. }
	%}
\end{figure}

To better examine the interactions, \Cref{fig:ggpredict_retweet_quantity} compares the predicted marginal means for false and true rumors. The plots thus show that (1)~false rumors receive more retweets than true rumors if the lifetime is low and fewer retweets than true rumors if the lifetime is high; and (2)~false rumors receive more retweets at higher depths than true rumors. In addition, we compare the average marginal effect~(AME) \cite{Buis.2010}. The AME denotes how a one standard deviation increase in lifetime (or retweet depth) changes retweet count on average. The AME of lifetime on the number of retweets is 2.3 times larger for false rumors (AME of $-0.148$) than for true rumors (AME of $-0.065$). In contrast, the AME of retweet depth for false rumors (AME of $-0.227$) is only \SI{48.48}{\percent} of the AME for true rumors (AME of $-0.495$). Altogether, we find strong support in favor of our hypotheses. In particular, the collected evidence suggests pronounced and statistically significant differences regarding how often users share true vs. false rumors depending on the lifetime and depth of a retweet cascade.

%\newpage
\subsection{Analysis of Response Time in Rumor Sharing Behavior (H2 \& H4)}
%\subsubsection{Analysis of Response Time}

The regression results in \Cref{tbl:regression_response_time} explain the response time between a user's tweet and the parent tweet. We again $z$-standardize all explanatory variables.

The regression results indicate a longer response time for false than for true rumors. In particular, false rumors have a \SI{12.41}{\percent} ($e^{0.117} \approx 1.12$) longer response time than true rumors, as demonstrated by the positive coefficient for $\mathit{Falsehood}$ ($\beta=0.117$, $p < 0.001$). Evidently, users share false rumors at a lower speed compared to true rumors. 

%Hypothesis H1c must thus be rejected. 

\begin{table}[H]
	\caption{Regression Results for Response Time.\label{tbl:regression_response_time}}
	{
		
		\footnotesize
		\begin{tabularx}{\textwidth}{@{\hspace{\tabcolsep}\extracolsep{\fill}}l *{4}{S} } 
			\toprule
			\multicolumn{5}{l}{Dependent Variable: Response Time~(log)}\\
			\midrule
			& {\textbf{Model (1)}} & {\textbf{Model (2)}} & {\textbf{Model (3)}} & {\textbf{Model (4)}}\\ 
			\midrule
			Falsehood                           &              &              & 0.162^{***}  & 0.117^{***}  \\
			&              &              & (0.010)      & (0.012)      \\
			Falsehood $\times$ Lifetime ({\textbf{H2}}) &              &              &              & 0.218^{***}  \\
			&              &              &              & (0.006)      \\
			Falsehood $\times$ Retweet Depth ({\textbf{H4a/b}})    &              &              &              & -0.109^{***} \\
			&              &              &              & (0.007)      \\
			Lifetime                    &              & 0.871^{***}  & 0.871^{***}  & 0.661^{***}  \\
			&              & (0.001)      & (0.001)      & (0.006)      \\
			Retweet Depth                       &              & -0.345^{***} & -0.346^{***} & -0.238^{***} \\
			&              & (0.001)      & (0.001)      & (0.007)      \\
			Followers                     & -0.048^{***} & -0.044^{***} & -0.044^{***} & -0.044^{***} \\
			& (0.002)      & (0.002)      & (0.002)      & (0.002)      \\
			Followees                     & 0.152^{***}  & 0.153^{***}  & 0.153^{***}  & 0.153^{***}  \\
			& (0.002)      & (0.001)      & (0.001)      & (0.001)      \\
			Account Age                         & -0.065^{***} & -0.064^{***} & -0.064^{***} & -0.064^{***} \\
			& (0.001)      & (0.001)      & (0.001)      & (0.001)      \\
			User Engagement                     & -0.021^{***} & -0.029^{***} & -0.029^{***} & -0.029^{***} \\
			& (0.001)      & (0.001)      & (0.001)      & (0.001)      \\
			Verified                     & -0.305^{***} & -0.383^{***} & -0.382^{***} & -0.379^{***} \\
			& (0.021)      & (0.020)      & (0.020)      & (0.020)      \\
			Intercept                           & 0.486^{***}  & 0.407^{***}  & 0.277^{***}  & 0.323^{***}  \\
			& (0.004)      & (0.004)      & (0.009)      & (0.011)      \\
			Random effects (cascade level) & {Yes} & {Yes} & {Yes} & {Yes} \\													
			\midrule
			AIC                                & {\num{14799531}}      &  {\num{14204368}}  & {\num{14204126}}       & {\num{14202754}}        \\
			Observations                        & {\num{3724197}}      & {\num{3724197}}      & {\num{3724197}}      & {\num{3724197}}      \\
			\bottomrule
			\multicolumn{5}{r}{Significance levels: $^*p< 0.05$, $^{**}p< 0.01$, $^{***}p< 0.001$; standard errors in parentheses} \\
		\end{tabularx}
	}
	\subcaption*{\emph{Note:} OLS regression explains the time difference between a user's tweet and the parent tweet (in log). Unit of analysis is the retweet level. {Cascade-specific random effects are included.} 
	}
\end{table}

%% Depth & lifetime

We again observe statistically significant differences for the $\mathit{Lifetime}$ variable across true vs. false rumors (H2). A longer lifetime of a cascade is associated with longer response times for false rumors vs. true rumors. An increase of one standard deviation in the lifetime of false rumors increases the response time by an additional \SI{24.36}{\percent} as compared to true rumors. We thus find support for H2. 

Furthermore, our results provide support for H4b, which states that false rumors have a shorter response time at higher depths than true rumors. For true rumors, a one standard deviation increase in retweet depth is expected to result in a \SI{21.18}{\percent} shorter response time. For false rumors, the estimated effect size is larger, as indicated by a negative coefficient for the interaction between $\mathit{Falsehood}$ and $\mathit{RetweetDepth}$ ($\beta=-0.109$, $p < 0.001$). Ceteris paribus, an increase of one standard deviation in retweet depth is expected to result in an additional \SI{10.33}{\percent} lower response time for false rumors than for true rumors. In other words, users located at a higher retweet depth share false rumors more quickly than true rumors. Hence, we must reject H4a and instead find support for H4b. 

Again, the AIC favors the inclusion of both lifetime and retweet depth in the model and thus establishes both as important determinants for explaining response time.

%% Plot

\begin{figure}[h]%[H]
	\caption{\centering Predicted Marginal Means of Response Time.}
	\label{fig:ggpredict_response_time}
	\captionsetup[subfloat]{position=bottom,labelformat=empty,skip=0pt}%%, labelfont=bf,textfont=normalfont,singlelinecheck=off,justification=raggedright
	%\FIGURE
	%\centering
	{
		\subfloat{{\includegraphics[width=5.5cm]{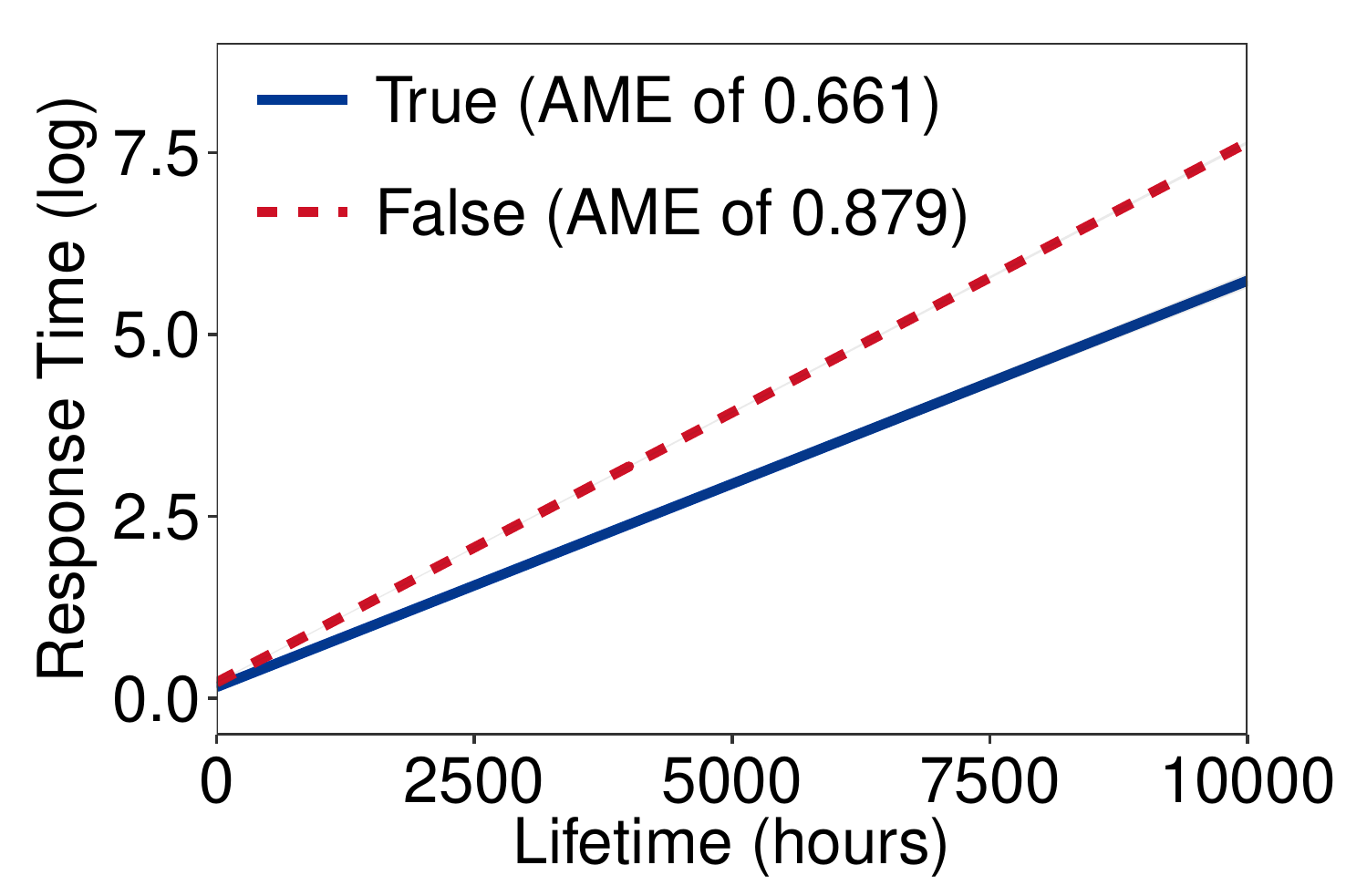}}}
		\quad\quad
		\subfloat{{\includegraphics[width=5.5cm]{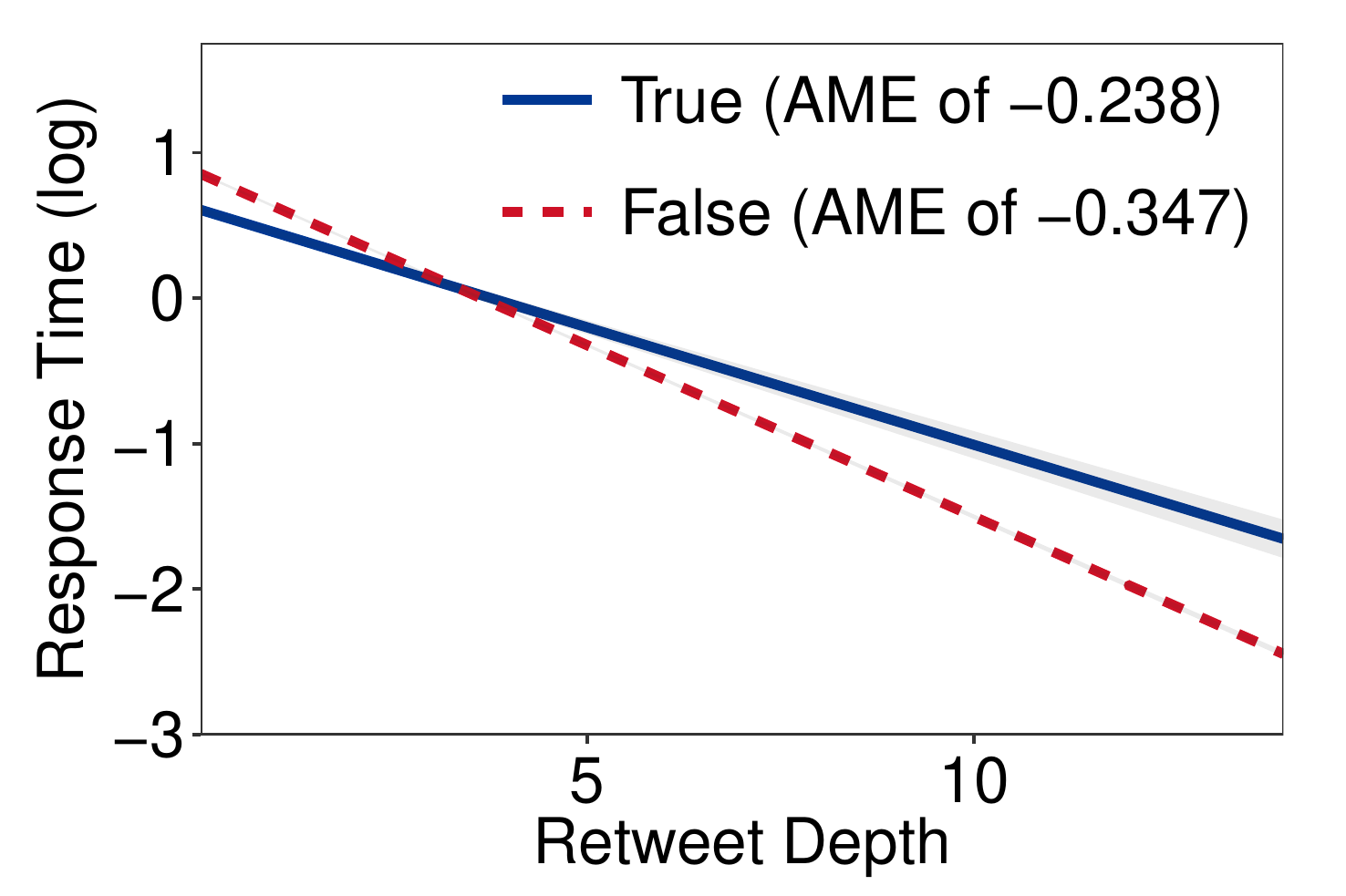}}}%
	}
	\subcaption*{\emph{Note:} The figures show the predicted marginal means of response time (log scale) for different values of lifetime (left) and retweet depth (right). The \SI{95}{\percent} confidence intervals are marked in gray. Cascade-specific random effects are included in the analysis.}
	%}
\end{figure}

\Cref{fig:ggpredict_response_time} shows the predicted marginal means in terms of response for false and true rumors. The left figure visualizes the predictions for lifetime, whereas the right figure shows the predictions for retweet depth. The plots establish the following: (1)~the (positive) marginal effect of lifetime on response time is stronger for false rumors than for true rumors; and (2)~false rumors have a shorter response time at higher depths.

\subsection{Sensitivity Analyses}

In offline settings, it was found that individuals propagate rumors differently depending on the covered topic and the embedded emotions. For instance, \cite{Stieglitz.2013} found that sharing behavior differs across political vs. non-political communication. Similarly, \cite{Knapp.1944} noted a higher chance of dissemination for negative rumors than for positive ones. A theoretical explanation is that negative information is perceived as more unexpected and thus more informative, which, in turn, drives diffusion. 

We therefore performed a sensitivity analysis concerning the role of lifetime and retweet depth. Specifically, we were interested in understanding whether lifetime and retweet depth influence sharing behavior differently across topics and emotions. For this purpose, we repeated the above analysis for (1)~rumors of a political nature and (2)~rumors conveying positive and negative emotions. The results are reported in Appendix~\ref{appendix:sensitivity}. Overall, we confirm both lifetime and retweet depth as important determinants in user sharing. We find a strong influence of lifetime for both false rumors of a political nature and rumors embedding negative emotions. This is in line with rumor theory \cite{Knapp.1944}, which suggests that both draw attention among individuals. We further observe that the extent to which false vs. true rumors spread differently depending on the retweet depth is particularly pronounced for political rumors and for rumors embedding positive emotions. Altogether, our results again confirm that both lifetime and retweet depth are important determinants with regard to the underlying mechanism by which users share rumors and also help to explain why false rumors go viral more frequently than true rumors.

\subsection{Robustness Checks}

We conducted an extensive set of checks that yield consistent findings. Details are relegated to the supplements, while we provide a summary in the following. 

% veracity

Our previous analysis compares the underlying mechanism of user sharing behavior across true vs. false rumors, yet we acknowledge that not all rumors can be clearly sorted according to a binary veracity label. For this reason, we repeated the above regressions and additionally included all rumors for which the fact-checking label was neither positive nor negative and were assigned to the \textquote{else} veracity category. The regression results support the above findings. 

% rumor structure

Cascades can originate either from novel rumors throughout the Twitter network or from rumors that have previously been disseminated. Theoretically, novel vs. recurring rumors might have different diffusion properties as rumormongers compete for attention. We controlled for this through cascade-specific random effects. In addition, we restricted our analysis to rumor cascades that introduce a novel story (\ie, one that has not received previous coverage). This resulted in findings which were qualitatively similar. The conclusions were also supported when we instead utilized a dummy variable reflecting whether a rumor is novel or recurring. We find that novel rumors are linked to more prolific and faster retweeting, while the findings from our hypotheses remain consistent. 

% source

In our analysis, we included user-level characteristics to control for the social influence of retweeting users; however, there is also variability among the initial broadcasters of a rumor. This variability is captured by our use of random effects. Nevertheless, we provide further quantitative insights concerning the role of the root (see Appendix~\ref{appendix:root_social_influence}). For instance, we find that content creators with a \textquote{verified} badge are linked to reduced dissemination of false rumors. The conclusions with regard to our hypotheses remain the same.

% RE vs. FE

We also experimented with the use of cascade-specific fixed effects. Here a benefit is that it allows for larger between-cascade heterogeneity. However, one can no longer retrieve proper estimates of interaction terms. The reason is that veracity varies at the cascade level and, if veracity \emph{and} cascade-level fixed effects are included at the same, the model is over-specified. Hence, in our above analysis, we opted for a random effect specification as suggested in \cite{Townsend.2013}, which is desirable for the objective of our study as it allows us to interpret between-group variation within a single rumor. Nevertheless, we repeated the above analysis with cascade-specific fixed effects, but where we excluded interaction terms. As a result, we find that the coefficients referring to direct effects in our model specification remain qualitatively similar. 

%% Cascade-size FE

Furthermore, previous research observed that structural differences at the cascade level (\ie, for folded retweet cascades) between real and false rumors can be largely explained by differences in the cascade size \cite{Juul.2021}. In our regression analyses, such effects are already accounted for by our random-effects specification, which implicitly controls for the overall size of the cascade. Notwithstanding, we also repeated our analysis with cascade-size fixed effects to control for the cascade size explicitly (see Appendix~\ref{appendix:size_fe}). We find that all our results remain robust.

% estimators

Our estimation routine was subject to robustness checks with consistent findings: (1)~We repeated our analysis with rumor lifetime instead of cascade lifetime; (2)~we controlled for outliers in the dependent variables; (3)~we ran separate regressions for true vs. false rumors; (4)~we calculated variance inflation factors for all independent variables and found that all remain below the critical threshold of four; (5)~we found consistent results when repeating our analysis with alternative estimators such as zero-inflated Poisson regression on the retweet count and Tobit regression on response time; (6)~we utilized Bayesian estimation techniques due to their theoretical advantages in terms of how multicollinearity is addressed; (7)~we incorporated quadratic effects; (8)~we included additional interactions between user-specific variables and veracity; (9)~we used clustered standard errors at annual level and repeated the analysis for different time periods in order to ensure robustness across the complete time period of the study; and (10)~we replaced retweet depth with the overall exposure as measured by the number of nodes in the cascade until a time $t$. The latter is highly correlated with lifetime, indicating that the two are not necessarily measuring different concepts (as opposed to lifetime and retweet depth). (11)~We experimented with alternative measures for social influence, namely degree centrality and so-called structural virality \cite{Goel.2016}. (12)~We also controlled for tie strength as in \cite{Crandall.2008}, which quantifies user similarity and thus accounts for homophily \cite{McPherson.2001,Kwon.2017b}. (13)~We repeated the analysis with rumors where there was no clear assignment to true or false veracity, and were thus the rumor was assigned to the ``else'' category. (14)~We expanded our findings concerning retweet count by performing an additional analysis based on the retweet probability. This should quantify the propensity of users to share rumors and yield the same conclusions (Appendix~\ref{sec:analysis_retweet_probability}). In all of the aforementioned checks, our findings for lifetime and retweet depth are supported. 

In our above analyses, we assumed that the spreading process is characterized by a uniform crowd behavior (\ie, herd behavior or collective intelligence) throughout a rumor. However, in principle, it could be possible that both herd behavior and collective intelligence co-exist within a rumor. Hence, we performed a robustness check in which we relaxed the assumption, so that one cascade can be characterized by both collective intelligence and herd behavior. Formally, we adopt the idea of varying-slopes model (\ie, we no longer assume a global coefficient $\beta_3$ for the entire sample but replace it with a coefficient $\beta_{3,i}$ that is allowed to vary across cascades). For details on varying-slope models, we refer to the introduction in \cite{Gelman.2014}. As a result, we find that a significant part of the probability mass confirms H3b/H4b. Hence, we must reject the premise that both collective intelligence and herd behavior are equally present; on the contrary, user sharing behavior appears to be driven predominantly by herding behavior. 

Lastly, the timing of when a rumor was subject to fact-checking could potentially influence the spreading behavior \cite{Shao.2016}. In theory, this should already be captured by the random effects in our regression model. Nevertheless, we repeated the above analysis by additionally controlling for the different time lags with which rumors were subject to fact-checking. Our results remain consistent. Likewise, we yield robust results regardless of whether we run our analysis with the retweet chain until the different time lags of fact-checking or with the complete retweet cascade.

\section{Discussion}
\label{sec:discussion}

\subsection{Summary of Findings}

% mechanism

This study is motivated by the substantial viral effects of false rumors. In order to understand why false rumors have more pronounced diffusion dynamics than true rumors, we investigate the underlying mechanism of user sharing behavior through the lens of lifetime and crowd behavior. Both are identified as important factors for explaining the differences in how users share true vs. false rumors. 

% H1: lifetime 

We hypothesized that the lifetime of a cascade reduces the resharing volume more sharply for false than for true rumors~(H1). This is supported by our results: ceteris paribus, false rumors receive more reshares than true rumors if the lifetime is low; however, they receive fewer reshares if the lifetime is high. Users are thus \emph{less} likely to share false rumors that have already been around for a long time. We also find support for our hypothesis that users share false rumors more slowly if the lifetime of the cascade is high~(H2). In sum, these hypotheses suggest that the interest in false rumors diminishes more quickly over time, a mechanism by which further spread is naturally inhibited. The lifetime effect is more pronounced for false than for true rumors because false rumors rely upon the short time window during which verification from official channels is absent.

%% H2: Crowd behavior

Our empirical evidence further suggests that the past resharing behavior of the crowd plays an important role in shaping individual resharing behavior. Based on our hypotheses, we distinguish two opposing functions of crowd behavior in online rumor diffusion. On the one hand, we hypothesized that the collective behavior of users would constitute a confirmation of truthful information. As a consequence, every generation of a resharing cascade would promote collective intelligence, which would reduce the diffusion of false rumors~(H3a/H4a). This would imply that users can recognize false rumors as such and refrain from resharing. However, we must reject hypothesis~H3a/H4a. Instead, our results establish the opposite (H3b/H4b): with a greater cascade depth, false rumors are subject to more and faster sharing than true rumors. The user base exhibits characteristics that resemble herd behavior. Thus, users do not curtail false rumors after multiple generations of a resharing cascade but rather conform to the resharing behavior of their peers (by continuing to spread the rumor).

\subsection{Implications}

In this work, we add by studying two important determinants that explain differences in the resharing behavior of true vs. false rumors, yet that have previously not received attention: lifetime and crowd effects. Both explain user sharing behavior. We find that both lifetime and cascade depth imply a different diffusion of true vs. false content. Thereby, our work contributes an explanation for the pronounced viral effects of false rumors. 

Our work further contributes to the literature on crowd behavior \cite[\eg,][]{Hong.2004,Hong.2014,Lorenz.2011,Li.2018}. Crowds can benefit from collective intelligence \cite[\eg,][]{Starbird.2013,Vieweg.2008,Zeng.2019} in that intelligent behavior allows crowds through large-scale interaction to transfer information and thus aid group performance \cite{Kameda.2022}. Yet, collective intelligence can be undermined due to herd behavior. Herd behavior provides an explanation for our findings and thus sheds light on why users are prone to resharing false rumors. One key determinant for collective intelligence is the independence of individuals' evaluations. Yet on social media, collective opinions are rarely simple aggregations of independent minds \cite{Kameda.2022}. Instead, users may act in the interest of conformity, thereby following the resharing behavior of previous users in the cascade. Given the relatively high level of uncertainty associated with false rumors, people may thus partially bypass their own beliefs and instead base their adoption decisions on the observations of other users. 

Our findings suggest the presence of herding tendencies in the spread of online rumors, while we are not aware of similar observations from offline rumors. Hence, a question is why online rumors (rather than offline rumors) are prone to herding tendencies. A possible explanation is located in the nature of herd behavior \cite{Lorenz.2011}. Recall that conformity is theorized as one reason for why individuals imitate peers and thus herd \cite{Lorenz.2011}. In online environments, it is especially easy for users to act in conformity \cite[\cf][]{Sun.2013}, since resharing allows users to share online information with a single click, while comparatively more effort is needed for the dissemination of offline rumors. Moreover, social media user are oftentimes in a hedonic mindset where they avoid cognitive reasoning \citep{Lutz.2020}, and, hence, are likely to share online rumors without verification. In comparison, the dissemination of offline rumors involves extensive cognitive processes, which help in reducing the risk that offline rumors are passed on without verification. Altogether, this provides an explanation behind why online rumors in particular are characterized by herd behavior. 

%\subsection{Implications}

% danger

From a practical perspective, false rumors can lead to the misallocation of resources during social crises, misinformation in the public discourse, and poor investments for companies. Hence, there is a need among managers and policymakers to better understand the viral effects of false rumors. In this regard, our work reveals a number of alarming findings for practice. False rumors go viral not only through the efforts of influential users but also due to their lifetime and crowd behavior. Hence, in order to curb the proliferation of false rumors, both lifetime and crowd behavior, as the underlying drivers of user behavior, need to be managed effectively.

Our findings suggest that in order to limit the spread of false rumors, social media platforms should pay close attention to the role of lifetime. As we have seen, the spread of false rumors is naturally curtailed by lifetime. Users tend to become less interested in rumors as time passes. However, rather than waiting for this process to take its course, the spread of false rumors could be more effectively managed in the early phases of dissemination. This would allow official channels to release a clarification regarding false information before a rumor proliferates. 

% implications for IS design (Wisdom of crowds)

Social media platforms cannot rely upon collective intelligence to combat the proliferation of false rumors. This is contrary to the widespread belief that users recognize false rumors and refrain from resharing them. Instead, we find signs of herd behavior: with a greater cascade depth, false rumors are subject to more and faster sharing. Social media platforms would thus need to change their design such that herd behavior is explicitly discouraged. For instance, displaying past sharing activity less prominently for potential false rumors might reduce the risk of users emulating the behavior of their peers. Herd tendencies might also be mitigated if the uncertainty associated with false rumors is effectively reduced. Examples of strategies to achieve this goal include, among others, group moderation and user ratings where false content is flagged. 

Our findings are also of value for the design of better detection mechanisms. In practice, a platform could prioritize rumors with a short lifetime and considerable cascade depth for fact-checking, as these diffusion dynamics are characteristic of falsehood. Similarly, our findings suggest that lifetime and crowd variables could be used to distinguish true vs. false rumors. Through the use of machine learning, social media platforms could detect false rumors based on the resharing dynamics and thus provide an early warning system. Importantly, warnings of false rumors could even be generated before official fact-checkers step in, which typically suffer from a severe time lag of more than 24 hours \cite{Shao.2016}. 

\subsection{Future Research and Limitations}

% limitations

Our work points to opportunities for future research. First, we draw upon a large-scale dataset with more than \num{3.7} million observations and an extensive set of control variables (\eg, followers, account age), analogous to earlier research \cite[\eg,][]{Stieglitz.2012b,Stieglitz.2013}. The choice of variables was the result of data processing agreements with Twitter that ensure user privacy. We further controlled for unobserved heterogeneity at the cascade level through the use of random effects. Nevertheless, future research could extend our work by developing and validating additional hypotheses regarding the role of user behavior or the underlying network structure. Another rewarding direction would be to enrich our findings with evidence on how social media users cognitively process social media posts. For instance, it remains an open research question how users revise their beliefs depending on the resharing behavior of others \cite[\cf][]{Sun.2013}. Previous research suggests that herding is more likely to undermine collective intelligence if the individuals are less confident in their decisions \cite{Lorenz.2011}.

Further limitations are related to the analyzed dataset. First, fact-checking efforts primarily focus on rumors which eventually may turn out as false. This may potentially lead to a sampling bias in the dataset. In our analysis, we nevertheless control for this by including a veracity dummy. Second, in recent years, Twitter and other platforms have increased their content moderation efforts (but, generally, after our dataset was constructed). This may have implications for when new data are collected to repeat our analysis as it may be biased as certain false rumors may be removed in line with platform guidelines. Third, our dataset is based on the preprocessing strategy from Vosoughi et al. (2018) \cite{Vosoughi.2018}, and, in line with that, rumor cascades started by bots are excluded. This does not affect our analysis as it leaves the retweet chain intact; however, our results must be interpreted in the light of human-created rumors. Notwithstanding, there are other works that analyze the spread of low-credibility content that was created by bots \cite{Shao.2018,Geissler.2022}.

\section{Conclusion}
\label{sec:conclusion}

% relevance & what we do

False rumors influence the formation of public opinion and can yield severe negative consequences: individuals might be erroneously accused of wrongdoing, companies can see their reputation harmed, and society runs the risk of political discourse and even election outcomes being manipulated. We hypothesized that lifetime and crowd effects explain why false rumors are more apt to go viral than true rumors. In order to provide empirical evidence, we collected \num[group-separator={,},group-minimum-digits=1]{126301} rumors from Twitter that were reshared more than 3.7 million times. Overall, false and true rumors entail distinctive diffusion patterns: lifetime is an important determinant explaining why false rumors die out. We further hypothesized that false rumors are shared less frequently due to ``intelligent'' behavior among users, yet no evidence was found for this. On the contrary, false rumors are more viral at greater cascade depths, thus implying herd tendencies. Hence, in order to diminish the proliferation of false rumors, both lifetime and crowd behavior, as underlying drivers of user behavior, need to be managed effectively by social media platforms.

\section{Acknowledgments}
This research was supported by a grant from the German Research Foundation (DFG grant 455368471).

\bibliographystyle{ACM-Reference-Format-no-doi-abbrv}
%\balance 
\bibliography{literature}


\begin{thebibliography}{95}

%%% the Web address.

\ifx \showCODEN    \undefined \def \showCODEN     #1{\unskip}     \fi
\ifx \showDOI      \undefined \def \showDOI       #1{#1}\fi
\ifx \showISBNx    \undefined \def \showISBNx     #1{\unskip}     \fi
\ifx \showISBNxiii \undefined \def \showISBNxiii  #1{\unskip}     \fi
\ifx \showISSN     \undefined \def \showISSN      #1{\unskip}     \fi
\ifx \showLCCN     \undefined \def \showLCCN      #1{\unskip}     \fi
\ifx \shownote     \undefined \def \shownote      #1{#1}          \fi
\ifx \showarticletitle \undefined \def \showarticletitle #1{#1}   \fi
\ifx \showURL      \undefined \def \showURL       {\relax}        \fi
% invisible to TeX
\providecommand\bibfield[2]{#2}
\providecommand\bibinfo[2]{#2}
\providecommand\natexlab[1]{#1}
\providecommand\showeprint[2][]{arXiv:#2}

\bibitem[\protect\citeauthoryear{Ai and Norton}{Ai and Norton}{2003}]%
        {Ai.2003}
\bibfield{author}{\bibinfo{person}{Chunrong Ai} {and}
  \bibinfo{person}{Edward~C. Norton}.} \bibinfo{year}{2003}\natexlab{}.
\newblock \showarticletitle{Interaction terms in logit and probit models}.
\newblock \bibinfo{journal}{\emph{Economics Letters}} \bibinfo{volume}{80},
  \bibinfo{number}{1} (\bibinfo{year}{2003}), \bibinfo{pages}{123--129}.
\newblock
\showISSN{01651765}


\bibitem[\protect\citeauthoryear{Allcott and Gentzkow}{Allcott and
  Gentzkow}{2017}]%
        {Allcott.2017}
\bibfield{author}{\bibinfo{person}{Hunt Allcott} {and} \bibinfo{person}{Matthew
  Gentzkow}.} \bibinfo{year}{2017}\natexlab{}.
\newblock \showarticletitle{Social media and fake news in the 2016 election}.
\newblock \bibinfo{journal}{\emph{Journal of Economic Perspectives}}
  \bibinfo{volume}{31}, \bibinfo{number}{2} (\bibinfo{year}{2017}),
  \bibinfo{pages}{211--236}.
\newblock
\showISSN{0895-3309}


\bibitem[\protect\citeauthoryear{Allport and Postman}{Allport and
  Postman}{1947}]%
        {Allport.1947}
\bibfield{author}{\bibinfo{person}{Gordon~Willard Allport} {and}
  \bibinfo{person}{Leo Postman}.} \bibinfo{year}{1947}\natexlab{}.
\newblock \bibinfo{booktitle}{\emph{The Psychology of Rumor}}.
\newblock \bibinfo{publisher}{{Henry Holt}}, \bibinfo{address}{New York, NJ}.
\newblock


\bibitem[\protect\citeauthoryear{Bakshy, Messing, and Adamic}{Bakshy
  et~al\mbox{.}}{2015}]%
        {Bakshy.2015}
\bibfield{author}{\bibinfo{person}{Eytan Bakshy}, \bibinfo{person}{Solomon
  Messing}, {and} \bibinfo{person}{Lada~A. Adamic}.}
  \bibinfo{year}{2015}\natexlab{}.
\newblock \showarticletitle{Exposure to ideologically diverse news and opinion
  on {F}acebook}.
\newblock \bibinfo{journal}{\emph{Science}} \bibinfo{volume}{348},
  \bibinfo{number}{6239} (\bibinfo{year}{2015}), \bibinfo{pages}{1130--1132}.
\newblock


\bibitem[\protect\citeauthoryear{B{\"a}r, Pr{\"o}llochs, and
  Feuerriegel}{B{\"a}r et~al\mbox{.}}{2023a}]%
        {Bar.2022}
\bibfield{author}{\bibinfo{person}{Dominik B{\"a}r}, \bibinfo{person}{Nicolas
  Pr{\"o}llochs}, {and} \bibinfo{person}{Stefan Feuerriegel}.}
  \bibinfo{year}{2023}\natexlab{a}.
\newblock \showarticletitle{Finding Qs: Profiling QAnon supporters on Parler}.
  In \emph{ICWSM}.
\newblock


\bibitem[\protect\citeauthoryear{B{\"a}r, Pr{\"o}llochs, and
  Feuerriegel}{B{\"a}r et~al\mbox{.}}{2023b}]%
        {Bar.2023}
\bibfield{author}{\bibinfo{person}{Dominik B{\"a}r}, \bibinfo{person}{Nicolas
  Pr{\"o}llochs}, {and} \bibinfo{person}{Stefan Feuerriegel}.}
  \bibinfo{year}{2023}\natexlab{b}.
\newblock \showarticletitle{New threats to society from free-speech social
  media platforms}.
\newblock \bibinfo{journal}{\emph{Commun. ACM}}  \bibinfo{volume}{Forthcoming}
  (\bibinfo{year}{2023}).
\newblock


\bibitem[\protect\citeauthoryear{Bessi, Coletto, Davidescu, Scala, Caldarelli,
  and Quattrociocchi}{Bessi et~al\mbox{.}}{2015}]%
        {Bessi.2015}
\bibfield{author}{\bibinfo{person}{Alessandro Bessi}, \bibinfo{person}{Mauro
  Coletto}, \bibinfo{person}{George~Alexandru Davidescu},
  \bibinfo{person}{Antonio Scala}, \bibinfo{person}{Guido Caldarelli}, {and}
  \bibinfo{person}{Walter Quattrociocchi}.} \bibinfo{year}{2015}\natexlab{}.
\newblock \showarticletitle{Science vs conspiracy: Collective narratives in the
  age of misinformation}.
\newblock \bibinfo{journal}{\emph{PLOS ONE}} \bibinfo{volume}{10},
  \bibinfo{number}{2} (\bibinfo{year}{2015}), \bibinfo{pages}{e0118093}.
\newblock
\showISSN{1932-6203}


\bibitem[\protect\citeauthoryear{Bordia and DiFonzo}{Bordia and
  DiFonzo}{2004}]%
        {Bordia.2004}
\bibfield{author}{\bibinfo{person}{Prashant Bordia} {and}
  \bibinfo{person}{Nicholas DiFonzo}.} \bibinfo{year}{2004}\natexlab{}.
\newblock \showarticletitle{Problem solving in social interactions on the
  Internet: Rumor as social cognition}.
\newblock \bibinfo{journal}{\emph{Social Psychology Quarterly}}
  \bibinfo{volume}{67}, \bibinfo{number}{1} (\bibinfo{year}{2004}),
  \bibinfo{pages}{33--49}.
\newblock


\bibitem[\protect\citeauthoryear{Buis}{Buis}{2010}]%
        {Buis.2010}
\bibfield{author}{\bibinfo{person}{Maarten~L. Buis}.}
  \bibinfo{year}{2010}\natexlab{}.
\newblock \showarticletitle{Stata tip 87: Interpretation of interactions in
  nonlinear models}.
\newblock \bibinfo{journal}{\emph{The Stata Journal}} \bibinfo{volume}{10},
  \bibinfo{number}{2} (\bibinfo{year}{2010}), \bibinfo{pages}{305--308}.
\newblock


\bibitem[\protect\citeauthoryear{Burnham and Anderson}{Burnham and
  Anderson}{2004}]%
        {Burnham.2004}
\bibfield{author}{\bibinfo{person}{Kenneth~P. Burnham} {and}
  \bibinfo{person}{David~R. Anderson}.} \bibinfo{year}{2004}\natexlab{}.
\newblock \showarticletitle{Multimodel inference: Understanding {AIC} and {BIC}
  in model selection}.
\newblock \bibinfo{journal}{\emph{Sociological Methods {\&} Research}}
  \bibinfo{volume}{33}, \bibinfo{number}{2} (\bibinfo{year}{2004}),
  \bibinfo{pages}{261--304}.
\newblock


\bibitem[\protect\citeauthoryear{Castillo, Mendoza, and Poblete}{Castillo
  et~al\mbox{.}}{2011}]%
        {Castillo.2011}
\bibfield{author}{\bibinfo{person}{Carlos Castillo}, \bibinfo{person}{Marcelo
  Mendoza}, {and} \bibinfo{person}{Barbara Poblete}.}
  \bibinfo{year}{2011}\natexlab{}.
\newblock \showarticletitle{Information credibility on {T}witter}. In
  \emph{WWW}.
\newblock


\bibitem[\protect\citeauthoryear{Centola}{Centola}{2010}]%
        {Centola.2010}
\bibfield{author}{\bibinfo{person}{Damon Centola}.}
  \bibinfo{year}{2010}\natexlab{}.
\newblock \showarticletitle{The spread of behavior in an online social network
  experiment}.
\newblock \bibinfo{journal}{\emph{Science}} \bibinfo{volume}{329},
  \bibinfo{number}{5996} (\bibinfo{year}{2010}), \bibinfo{pages}{1194--1197}.
\newblock


\bibitem[\protect\citeauthoryear{Cha, Mislove, and Gummadi}{Cha
  et~al\mbox{.}}{2009}]%
        {Cha.2009}
\bibfield{author}{\bibinfo{person}{Meeyoung Cha}, \bibinfo{person}{Alan
  Mislove}, {and} \bibinfo{person}{Krishna~P. Gummadi}.}
  \bibinfo{year}{2009}\natexlab{}.
\newblock \showarticletitle{A measurement-driven analysis of information
  propagation in the {F}lickr social network}. In \emph{WWW}.
\newblock


\bibitem[\protect\citeauthoryear{Crandall, Cosley, Huttenlocher, Kleinberg, and
  Suri}{Crandall et~al\mbox{.}}{2008}]%
        {Crandall.2008}
\bibfield{author}{\bibinfo{person}{David Crandall}, \bibinfo{person}{Dan
  Cosley}, \bibinfo{person}{Daniel Huttenlocher}, \bibinfo{person}{Jon
  Kleinberg}, {and} \bibinfo{person}{Siddharth Suri}.}
  \bibinfo{year}{2008}\natexlab{}.
\newblock \showarticletitle{Feedback effects between similarity and social
  influence in online communities}. In \emph{KDD}.
\newblock


\bibitem[\protect\citeauthoryear{Crane and Sornette}{Crane and
  Sornette}{2008}]%
        {Crane.2008}
\bibfield{author}{\bibinfo{person}{Riley Crane} {and} \bibinfo{person}{Didier
  Sornette}.} \bibinfo{year}{2008}\natexlab{}.
\newblock \showarticletitle{Robust dynamic classes revealed by measuring the
  response function of a social system}.
\newblock \bibinfo{journal}{\emph{PNAS}} \bibinfo{volume}{105},
  \bibinfo{number}{41} (\bibinfo{year}{2008}), \bibinfo{pages}{15649--15653}.
\newblock


\bibitem[\protect\citeauthoryear{de~Domenico, Lima, Mougel, and
  Musolesi}{de~Domenico et~al\mbox{.}}{2013}]%
        {Domenico.2013}
\bibfield{author}{\bibinfo{person}{Manilo de Domenico},
  \bibinfo{person}{Antonio Lima}, \bibinfo{person}{Paul Mougel}, {and}
  \bibinfo{person}{Mirco Musolesi}.} \bibinfo{year}{2013}\natexlab{}.
\newblock \showarticletitle{The anatomy of a scientific rumor}.
\newblock \bibinfo{journal}{\emph{Scientific Reports}} \bibinfo{volume}{3},
  \bibinfo{number}{2980} (\bibinfo{year}{2013}).
\newblock


\bibitem[\protect\citeauthoryear{{Del Vicario}, Bessi, Zollo, Petroni, Scala,
  Caldarelli, Stanley, and Quattrociocchi}{{Del Vicario} et~al\mbox{.}}{2016}]%
        {DelVicario.2016}
\bibfield{author}{\bibinfo{person}{Michela {Del Vicario}},
  \bibinfo{person}{Alessandro Bessi}, \bibinfo{person}{Fabiana Zollo},
  \bibinfo{person}{Fabio Petroni}, \bibinfo{person}{Antonio Scala},
  \bibinfo{person}{Guido Caldarelli}, \bibinfo{person}{H.~Eugene Stanley},
  {and} \bibinfo{person}{Walter Quattrociocchi}.}
  \bibinfo{year}{2016}\natexlab{}.
\newblock \showarticletitle{The spreading of misinformation online}.
\newblock \bibinfo{journal}{\emph{PNAS}} \bibinfo{volume}{113},
  \bibinfo{number}{3} (\bibinfo{year}{2016}), \bibinfo{pages}{554--559}.
\newblock


\bibitem[\protect\citeauthoryear{Dissanayake, Nerur, Singh, and
  Lee}{Dissanayake et~al\mbox{.}}{2019}]%
        {Dissanayake.2019}
\bibfield{author}{\bibinfo{person}{Indika Dissanayake},
  \bibinfo{person}{Sridhar Nerur}, \bibinfo{person}{Rahul Singh}, {and}
  \bibinfo{person}{Yang Lee}.} \bibinfo{year}{2019}\natexlab{}.
\newblock \showarticletitle{Medical Crowdsourcing: Harnessing the ``Wisdom of
  the Crowd'' to Solve Medical Mysteries}.
\newblock \bibinfo{journal}{\emph{Journal of the Association for Information
  Systems}} \bibinfo{volume}{20}, \bibinfo{number}{11} (\bibinfo{year}{2019}),
  \bibinfo{pages}{4}.
\newblock


\bibitem[\protect\citeauthoryear{Drolsbach and Pr{\"o}llochs}{Drolsbach and
  Pr{\"o}llochs}{2023}]%
        {Drolsbach.2023}
\bibfield{author}{\bibinfo{person}{Chiara~P Drolsbach} {and}
  \bibinfo{person}{Nicolas Pr{\"o}llochs}.} \bibinfo{year}{2023}\natexlab{}.
\newblock \showarticletitle{Believability and Harmfulness Shape the Virality of
  Misleading Social Media Posts}. In \emph{WWW}.
\newblock


\bibitem[\protect\citeauthoryear{Ducci, Kraus, and Feuerriegel}{Ducci
  et~al\mbox{.}}{2020}]%
        {Ducci.2020}
\bibfield{author}{\bibinfo{person}{Francesco Ducci}, \bibinfo{person}{Mathias
  Kraus}, {and} \bibinfo{person}{Stefan Feuerriegel}.}
  \bibinfo{year}{2020}\natexlab{}.
\newblock \showarticletitle{Cascade-{LSTM}: A tree-structured neural classifier
  for detecting misinformation cascades}. In \emph{KDD}.
\newblock


\bibitem[\protect\citeauthoryear{Forbes}{Forbes}{2017}]%
        {Forbes.2017}
\bibfield{author}{\bibinfo{person}{Forbes}.} \bibinfo{year}{2017}\natexlab{}.
\newblock \bibinfo{title}{Can 'fake news'~impact the stock market?}
\newblock
\newblock
\urldef\tempurl%
\url{https://www.forbes.com/sites/kenrapoza/2017/02/26/can-fake-news-impact-the-stock-market/}
\showURL{%
\tempurl}


\bibitem[\protect\citeauthoryear{Friggeri, Adamic, Eckles, and Cheng}{Friggeri
  et~al\mbox{.}}{2014}]%
        {Friggeri.2014}
\bibfield{author}{\bibinfo{person}{Adrien Friggeri}, \bibinfo{person}{Lada~A
  Adamic}, \bibinfo{person}{Dean Eckles}, {and} \bibinfo{person}{Justin
  Cheng}.} \bibinfo{year}{2014}\natexlab{}.
\newblock \showarticletitle{Rumor cascades}. In \emph{ICWSM}.
\newblock


\bibitem[\protect\citeauthoryear{Garg, Smith, and Telang}{Garg
  et~al\mbox{.}}{2011}]%
        {Garg.2011}
\bibfield{author}{\bibinfo{person}{Rajiv Garg}, \bibinfo{person}{Michael~D.
  Smith}, {and} \bibinfo{person}{Rahul Telang}.}
  \bibinfo{year}{2011}\natexlab{}.
\newblock \showarticletitle{Measuring information diffusion in an online
  community}.
\newblock \bibinfo{journal}{\emph{Journal of Management Information Systems}}
  \bibinfo{volume}{28}, \bibinfo{number}{2} (\bibinfo{year}{2011}),
  \bibinfo{pages}{11--38}.
\newblock
\showISSN{0742-1222}


\bibitem[\protect\citeauthoryear{Geissler, B{\"a}r, Pr{\"o}llochs, and
  Feuerriegel}{Geissler et~al\mbox{.}}{2022}]%
        {Geissler.2022}
\bibfield{author}{\bibinfo{person}{Dominique Geissler},
  \bibinfo{person}{Dominik B{\"a}r}, \bibinfo{person}{Nicolas Pr{\"o}llochs},
  {and} \bibinfo{person}{Stefan Feuerriegel}.} \bibinfo{year}{2022}\natexlab{}.
\newblock \showarticletitle{Russian propaganda on social media during the 2022
  invasion of Ukraine}.
\newblock \bibinfo{journal}{\emph{arXiv:2211.04154}} (\bibinfo{year}{2022}).
\newblock


\bibitem[\protect\citeauthoryear{Gelman, Carlin, Stern, Dunson, Vehtari, and
  Rubin}{Gelman et~al\mbox{.}}{2014}]%
        {Gelman.2014}
\bibfield{author}{\bibinfo{person}{Andrew Gelman}, \bibinfo{person}{John~B.
  Carlin}, \bibinfo{person}{Hal~S. Stern}, \bibinfo{person}{David~B. Dunson},
  \bibinfo{person}{Aki Vehtari}, {and} \bibinfo{person}{Donald~B. Rubin}.}
  \bibinfo{year}{2014}\natexlab{}.
\newblock \bibinfo{booktitle}{\emph{Bayesian Data Analysis}}.
\newblock \bibinfo{publisher}{{CRC Press}}, \bibinfo{address}{Boca Raton, FL}.
\newblock


\bibitem[\protect\citeauthoryear{Goel, Anderson, Hofman, and Watts}{Goel
  et~al\mbox{.}}{2016}]%
        {Goel.2016}
\bibfield{author}{\bibinfo{person}{Sharad Goel}, \bibinfo{person}{Ashton
  Anderson}, \bibinfo{person}{Jake Hofman}, {and} \bibinfo{person}{Duncan~J.
  Watts}.} \bibinfo{year}{2016}\natexlab{}.
\newblock \showarticletitle{The structural virality of online diffusion}.
\newblock \bibinfo{journal}{\emph{Management Science}} \bibinfo{volume}{62},
  \bibinfo{number}{1} (\bibinfo{year}{2016}), \bibinfo{pages}{180--196}.
\newblock
\showISSN{0025-1909}


\bibitem[\protect\citeauthoryear{Goel, Watts, and Goldstein}{Goel
  et~al\mbox{.}}{2012}]%
        {Goel.2012}
\bibfield{author}{\bibinfo{person}{Sharad Goel}, \bibinfo{person}{Duncan~J.
  Watts}, {and} \bibinfo{person}{Daniel~G. Goldstein}.}
  \bibinfo{year}{2012}\natexlab{}.
\newblock \showarticletitle{The structure of online diffusion networks}. In
  \emph{EC}.
\newblock


\bibitem[\protect\citeauthoryear{Han, Lappas, and Sabnis}{Han
  et~al\mbox{.}}{2020a}]%
        {Han.2020}
\bibfield{author}{\bibinfo{person}{Yue Han}, \bibinfo{person}{Theodoros
  Lappas}, {and} \bibinfo{person}{Gaurav Sabnis}.}
  \bibinfo{year}{2020}\natexlab{a}.
\newblock \showarticletitle{The importance of interactions between content
  characteristics and creator characteristics for studying virality in social
  media}.
\newblock \bibinfo{journal}{\emph{Information Systems Research}}
  \bibinfo{volume}{31}, \bibinfo{number}{2} (\bibinfo{year}{2020}),
  \bibinfo{pages}{297--652}.
\newblock
\showISSN{1047-7047}


\bibitem[\protect\citeauthoryear{Han, Ozturk, and Nickerson}{Han
  et~al\mbox{.}}{2020b}]%
        {Han.2021}
\bibfield{author}{\bibinfo{person}{Yue Han}, \bibinfo{person}{Pinar Ozturk},
  {and} \bibinfo{person}{Jeffrey~V. Nickerson}.}
  \bibinfo{year}{2020}\natexlab{b}.
\newblock \showarticletitle{Leveraging the wisdom of crowd to address societal
  challenges: A revisit to the knowledge reuse process for innovation through
  analytics}.
\newblock \bibinfo{journal}{\emph{Journal of the Association for Information
  Systems}} \bibinfo{volume}{21}, \bibinfo{number}{5} (\bibinfo{year}{2020}).
\newblock


\bibitem[\protect\citeauthoryear{Hong and Page}{Hong and Page}{2004}]%
        {Hong.2004}
\bibfield{author}{\bibinfo{person}{Lu Hong} {and} \bibinfo{person}{Scott~E.
  Page}.} \bibinfo{year}{2004}\natexlab{}.
\newblock \showarticletitle{Groups of diverse problem solvers can outperform
  groups of high-ability problem solvers}.
\newblock \bibinfo{journal}{\emph{PNAS}} \bibinfo{volume}{101},
  \bibinfo{number}{46} (\bibinfo{year}{2004}), \bibinfo{pages}{16385--16389}.
\newblock
\showISSN{0027-8424}


\bibitem[\protect\citeauthoryear{Hong and Pavlou}{Hong and Pavlou}{2014}]%
        {Hong.2014}
\bibfield{author}{\bibinfo{person}{Yili Hong} {and} \bibinfo{person}{Paul~A.
  Pavlou}.} \bibinfo{year}{2014}\natexlab{}.
\newblock \showarticletitle{Product fit uncertainty in online markets: Nature,
  effects, and antecedents}.
\newblock \bibinfo{journal}{\emph{Information Systems Research}}
  \bibinfo{volume}{25}, \bibinfo{number}{2} (\bibinfo{year}{2014}),
  \bibinfo{pages}{328--344}.
\newblock
\showISSN{1047-7047}


\bibitem[\protect\citeauthoryear{Jakubik, V{\"o}ssing, B{\"a}r, Pr{\"o}llochs,
  and Feuerriegel}{Jakubik et~al\mbox{.}}{2023}]%
        {Jakubik.2022}
\bibfield{author}{\bibinfo{person}{Johannes Jakubik}, \bibinfo{person}{Michael
  V{\"o}ssing}, \bibinfo{person}{Dominik B{\"a}r}, \bibinfo{person}{Nicolas
  Pr{\"o}llochs}, {and} \bibinfo{person}{Stefan Feuerriegel}.}
  \bibinfo{year}{2023}\natexlab{}.
\newblock \showarticletitle{Online emotions during the storming of the US
  Capitol: Evidence from the social media network Parler}. In \emph{ICWSM}.
\newblock


\bibitem[\protect\citeauthoryear{Jung, Mirbabaie, Ross, Stieglitz, Neuberger,
  and Kapidzic}{Jung et~al\mbox{.}}{2018}]%
        {Jung.2018}
\bibfield{author}{\bibinfo{person}{Anna-Katharina Jung}, \bibinfo{person}{Milad
  Mirbabaie}, \bibinfo{person}{Bj{\"o}rn Ross}, \bibinfo{person}{Stefan
  Stieglitz}, \bibinfo{person}{Christoph Neuberger}, {and}
  \bibinfo{person}{Sanja Kapidzic}.} \bibinfo{year}{2018}\natexlab{}.
\newblock \showarticletitle{Information diffusion between {T}witter and online
  media}. In \emph{International Conference on Information Systems (ICIS)}.
\newblock


\bibitem[\protect\citeauthoryear{Juul and Ugander}{Juul and Ugander}{2021}]%
        {Juul.2021}
\bibfield{author}{\bibinfo{person}{Jonas~L Juul} {and} \bibinfo{person}{Johan
  Ugander}.} \bibinfo{year}{2021}\natexlab{}.
\newblock \showarticletitle{Comparing information diffusion mechanisms by
  matching on cascade size}.
\newblock \bibinfo{journal}{\emph{PNAS}} \bibinfo{volume}{118},
  \bibinfo{number}{46} (\bibinfo{year}{2021}), \bibinfo{pages}{e2100786118}.
\newblock


\bibitem[\protect\citeauthoryear{Kagan}{Kagan}{2009}]%
        {Kagan.2009}
\bibfield{author}{\bibinfo{person}{Jerome Kagan}.}
  \bibinfo{year}{2009}\natexlab{}.
\newblock \showarticletitle{Categories of novelty and states of uncertainty}.
\newblock \bibinfo{journal}{\emph{Review of General Psychology}}
  \bibinfo{volume}{13}, \bibinfo{number}{4} (\bibinfo{year}{2009}),
  \bibinfo{pages}{290--301}.
\newblock


\bibitem[\protect\citeauthoryear{Kameda, Toyokawa, and Tindale}{Kameda
  et~al\mbox{.}}{2022}]%
        {Kameda.2022}
\bibfield{author}{\bibinfo{person}{Tatsuya Kameda}, \bibinfo{person}{Wataru
  Toyokawa}, {and} \bibinfo{person}{R~Scott Tindale}.}
  \bibinfo{year}{2022}\natexlab{}.
\newblock \showarticletitle{Information aggregation and collective intelligence
  beyond the wisdom of crowds}.
\newblock \bibinfo{journal}{\emph{Nature Reviews Psychology}}
  \bibinfo{volume}{1} (\bibinfo{year}{2022}), \bibinfo{pages}{345--357}.
\newblock


\bibitem[\protect\citeauthoryear{Karaca-Mandic, Norton, and Dowd}{Karaca-Mandic
  et~al\mbox{.}}{2012}]%
        {KaracaMandic.2012}
\bibfield{author}{\bibinfo{person}{Pinar Karaca-Mandic},
  \bibinfo{person}{Edward~C. Norton}, {and} \bibinfo{person}{Bryan Dowd}.}
  \bibinfo{year}{2012}\natexlab{}.
\newblock \showarticletitle{Interaction terms in nonlinear models}.
\newblock \bibinfo{journal}{\emph{Health Services Research}}
  \bibinfo{volume}{47}, \bibinfo{number}{1} (\bibinfo{year}{2012}),
  \bibinfo{pages}{255--274}.
\newblock


\bibitem[\protect\citeauthoryear{Kim and Dennis}{Kim and Dennis}{2019}]%
        {Kim.2019}
\bibfield{author}{\bibinfo{person}{Antino Kim} {and} \bibinfo{person}{Alan~R.
  Dennis}.} \bibinfo{year}{2019}\natexlab{}.
\newblock \showarticletitle{Says who? {T}he effects of presentation format and
  source rating on fake news in social media}.
\newblock \bibinfo{journal}{\emph{MIS Quarterly}} \bibinfo{volume}{43},
  \bibinfo{number}{3} (\bibinfo{year}{2019}), \bibinfo{pages}{1025--1039}.
\newblock


\bibitem[\protect\citeauthoryear{Knapp}{Knapp}{1944}]%
        {Knapp.1944}
\bibfield{author}{\bibinfo{person}{Robert~H. Knapp}.}
  \bibinfo{year}{1944}\natexlab{}.
\newblock \showarticletitle{A psychology of rumor}.
\newblock \bibinfo{journal}{\emph{The Public Opinion Quarterly}}
  \bibinfo{volume}{8}, \bibinfo{number}{1} (\bibinfo{year}{1944}),
  \bibinfo{pages}{22--37}.
\newblock


\bibitem[\protect\citeauthoryear{Kratzwald, Ili{\'c}, Kraus, Feuerriegel, and
  Prendinger}{Kratzwald et~al\mbox{.}}{2018}]%
        {Kratzwald.2018}
\bibfield{author}{\bibinfo{person}{Bernhard Kratzwald}, \bibinfo{person}{Suzana
  Ili{\'c}}, \bibinfo{person}{Mathias Kraus}, \bibinfo{person}{Stefan
  Feuerriegel}, {and} \bibinfo{person}{Helmut Prendinger}.}
  \bibinfo{year}{2018}\natexlab{}.
\newblock \showarticletitle{Deep learning for affective computing: Text-based
  emotion recognition in decision support}.
\newblock \bibinfo{journal}{\emph{Decision Support Systems}}
  \bibinfo{volume}{115} (\bibinfo{year}{2018}), \bibinfo{pages}{24--35}.
\newblock
\showISSN{01679236}


\bibitem[\protect\citeauthoryear{Kwak, Lee, Park, and Moon}{Kwak
  et~al\mbox{.}}{2010}]%
        {Kwak.2010}
\bibfield{author}{\bibinfo{person}{Haewoon Kwak}, \bibinfo{person}{Changhyun
  Lee}, \bibinfo{person}{Hosung Park}, {and} \bibinfo{person}{Sue Moon}.}
  \bibinfo{year}{2010}\natexlab{}.
\newblock \showarticletitle{What is {T}witter, a social network or a news
  media?}. In \emph{WWW}.
\newblock


\bibitem[\protect\citeauthoryear{Kwon, Oh, and Kim}{Kwon
  et~al\mbox{.}}{2017b}]%
        {Kwon.2017b}
\bibfield{author}{\bibinfo{person}{Hyeokkoo~Eric Kwon},
  \bibinfo{person}{Wonseok Oh}, {and} \bibinfo{person}{Taekyung Kim}.}
  \bibinfo{year}{2017}\natexlab{b}.
\newblock \showarticletitle{Platform structures, homing preferences, and
  homophilous propensities in online social networks}.
\newblock \bibinfo{journal}{\emph{Journal of Management Information Systems}}
  \bibinfo{volume}{34}, \bibinfo{number}{3} (\bibinfo{year}{2017}),
  \bibinfo{pages}{768--802}.
\newblock
\showISSN{0742-1222}


\bibitem[\protect\citeauthoryear{Kwon, Cha, and Jung}{Kwon
  et~al\mbox{.}}{2017a}]%
        {Kwon.2017}
\bibfield{author}{\bibinfo{person}{Sejeong Kwon}, \bibinfo{person}{Meeyoung
  Cha}, {and} \bibinfo{person}{Kyomin Jung}.} \bibinfo{year}{2017}\natexlab{a}.
\newblock \showarticletitle{Rumor detection over varying time windows}.
\newblock \bibinfo{journal}{\emph{PLOS ONE}} \bibinfo{volume}{12},
  \bibinfo{number}{1} (\bibinfo{year}{2017}), \bibinfo{pages}{e0168344}.
\newblock
\showISSN{1932-6203}


\bibitem[\protect\citeauthoryear{Kwon, Cha, Jung, Chen, and Wang}{Kwon
  et~al\mbox{.}}{2013}]%
        {Kwon.2013}
\bibfield{author}{\bibinfo{person}{Sejeong Kwon}, \bibinfo{person}{Meeyoung
  Cha}, \bibinfo{person}{Kyomin Jung}, \bibinfo{person}{Wei Chen}, {and}
  \bibinfo{person}{Yajun Wang}.} \bibinfo{year}{2013}\natexlab{}.
\newblock \showarticletitle{Prominent features of rumor propagation in online
  social media}. In \emph{ICDM}.
\newblock


\bibitem[\protect\citeauthoryear{Lazer, Baum, Benkler, Berinsky, Greenhill,
  Menczer, Metzger, Nyhan, Pennycook, Rothschild, Schudson, Sloman, Sunstein,
  Thorson, Watts, and Zittrain}{Lazer et~al\mbox{.}}{2018}]%
        {Lazer.2018}
\bibfield{author}{\bibinfo{person}{David M.~J. Lazer},
  \bibinfo{person}{Matthew~A. Baum}, \bibinfo{person}{Yochai Benkler},
  \bibinfo{person}{Adam~J. Berinsky}, \bibinfo{person}{Kelly~M. Greenhill},
  \bibinfo{person}{Filippo Menczer}, \bibinfo{person}{Miriam~J. Metzger},
  \bibinfo{person}{Brendan Nyhan}, \bibinfo{person}{Gordon Pennycook},
  \bibinfo{person}{David Rothschild}, \bibinfo{person}{Michael Schudson},
  \bibinfo{person}{Steven~A. Sloman}, \bibinfo{person}{Cass~R. Sunstein},
  \bibinfo{person}{Emily~A. Thorson}, \bibinfo{person}{Duncan~J. Watts}, {and}
  \bibinfo{person}{Jonathan~L. Zittrain}.} \bibinfo{year}{2018}\natexlab{}.
\newblock \showarticletitle{The science of fake news}.
\newblock \bibinfo{journal}{\emph{Science}} \bibinfo{volume}{359},
  \bibinfo{number}{6380} (\bibinfo{year}{2018}), \bibinfo{pages}{1094--1096}.
\newblock


\bibitem[\protect\citeauthoryear{Lee, Agrawal, and Rao}{Lee
  et~al\mbox{.}}{2015}]%
        {Lee.2015}
\bibfield{author}{\bibinfo{person}{Jaeung Lee}, \bibinfo{person}{Manish
  Agrawal}, {and} \bibinfo{person}{H.~R. Rao}.}
  \bibinfo{year}{2015}\natexlab{}.
\newblock \showarticletitle{Message diffusion through social network service:
  The case of rumor and non-rumor related tweets during {B}oston bombing 2013}.
\newblock \bibinfo{journal}{\emph{Information Systems Frontiers}}
  \bibinfo{volume}{17}, \bibinfo{number}{5} (\bibinfo{year}{2015}),
  \bibinfo{pages}{997--1005}.
\newblock


\bibitem[\protect\citeauthoryear{Lerman and Ghosh}{Lerman and Ghosh}{2010}]%
        {Lerman.2010}
\bibfield{author}{\bibinfo{person}{Kristina Lerman} {and} \bibinfo{person}{Rumi
  Ghosh}.} \bibinfo{year}{2010}\natexlab{}.
\newblock \showarticletitle{Information contagion: An empirical study of spread
  of news on {D}igg and {T}witter social networks}. In \emph{ICWSM}.
\newblock


\bibitem[\protect\citeauthoryear{Leskovec, Adamic, and Huberman}{Leskovec
  et~al\mbox{.}}{2007}]%
        {Leskovec.2007}
\bibfield{author}{\bibinfo{person}{Jure Leskovec}, \bibinfo{person}{Lada~A.
  Adamic}, {and} \bibinfo{person}{Bernardo~A. Huberman}.}
  \bibinfo{year}{2007}\natexlab{}.
\newblock \showarticletitle{The dynamics of viral marketing}.
\newblock \bibinfo{journal}{\emph{ACM Transactions on the Web}}
  \bibinfo{volume}{1}, \bibinfo{number}{1} (\bibinfo{year}{2007}),
  \bibinfo{pages}{Article 5}.
\newblock


\bibitem[\protect\citeauthoryear{L{\'e}vy}{L{\'e}vy}{1997}]%
        {Levy.1997}
\bibfield{author}{\bibinfo{person}{Pierre L{\'e}vy}.}
  \bibinfo{year}{1997}\natexlab{}.
\newblock \bibinfo{booktitle}{\emph{Collective Intelligence: Mankind's emerging
  world in cyberspace}}.
\newblock \bibinfo{publisher}{Plenum/Harper Collins, New York, NY}.
\newblock


\bibitem[\protect\citeauthoryear{Li and Wu}{Li and Wu}{2018}]%
        {Li.2018}
\bibfield{author}{\bibinfo{person}{Xitong Li} {and} \bibinfo{person}{Lynn Wu}.}
  \bibinfo{year}{2018}\natexlab{}.
\newblock \showarticletitle{Herding and social media word-of-mouth: Evidence
  from {G}roupon}.
\newblock \bibinfo{journal}{\emph{MIS Quarterly}} \bibinfo{volume}{42},
  \bibinfo{number}{4} (\bibinfo{year}{2018}), \bibinfo{pages}{1331--1351}.
\newblock


\bibitem[\protect\citeauthoryear{Liang}{Liang}{2018}]%
        {Liang.2018}
\bibfield{author}{\bibinfo{person}{Hai Liang}.}
  \bibinfo{year}{2018}\natexlab{}.
\newblock \showarticletitle{Broadcast versus viral spreading: The structure of
  diffusion cascades and selective sharing on social media}.
\newblock \bibinfo{journal}{\emph{Journal of Communication}}
  \bibinfo{volume}{68}, \bibinfo{number}{3} (\bibinfo{year}{2018}),
  \bibinfo{pages}{525--546}.
\newblock


\bibitem[\protect\citeauthoryear{Lin, Lucas, and Shmueli}{Lin
  et~al\mbox{.}}{2013}]%
        {Lin.2013}
\bibfield{author}{\bibinfo{person}{Mingfeng Lin}, \bibinfo{person}{Henry~C.
  Lucas, Jr.}, {and} \bibinfo{person}{Galit Shmueli}.}
  \bibinfo{year}{2013}\natexlab{}.
\newblock \showarticletitle{Too big to fail: Large samples and the p-value
  problem}.
\newblock \bibinfo{journal}{\emph{Information Systems Research}}
  \bibinfo{volume}{24}, \bibinfo{number}{4} (\bibinfo{year}{2013}),
  \bibinfo{pages}{906--917}.
\newblock
\showISSN{1047-7047}


\bibitem[\protect\citeauthoryear{Lorenz, Rauhut, Schweitzer, and
  Helbing}{Lorenz et~al\mbox{.}}{2011}]%
        {Lorenz.2011}
\bibfield{author}{\bibinfo{person}{Jan Lorenz}, \bibinfo{person}{Heiko Rauhut},
  \bibinfo{person}{Frank Schweitzer}, {and} \bibinfo{person}{Dirk Helbing}.}
  \bibinfo{year}{2011}\natexlab{}.
\newblock \showarticletitle{How social influence can undermine the wisdom of
  crowd effect}.
\newblock \bibinfo{journal}{\emph{PNAS}} \bibinfo{volume}{108},
  \bibinfo{number}{22} (\bibinfo{year}{2011}), \bibinfo{pages}{9020--9025}.
\newblock
\showISSN{0027-8424}


\bibitem[\protect\citeauthoryear{Love and Hirschheim}{Love and
  Hirschheim}{2017}]%
        {Love.2017}
\bibfield{author}{\bibinfo{person}{James Love} {and} \bibinfo{person}{Rudy
  Hirschheim}.} \bibinfo{year}{2017}\natexlab{}.
\newblock \showarticletitle{Crowdsourcing of information systems research}.
\newblock \bibinfo{journal}{\emph{European Journal of Information Systems}}
  \bibinfo{volume}{26}, \bibinfo{number}{3} (\bibinfo{year}{2017}),
  \bibinfo{pages}{315--332}.
\newblock


\bibitem[\protect\citeauthoryear{Lutz, Adam, Feuerriegel, Pr{\"o}llochs, and
  Neumann}{Lutz et~al\mbox{.}}{2020}]%
        {Lutz.2020}
\bibfield{author}{\bibinfo{person}{Bernhard Lutz}, \bibinfo{person}{Marc T.~P.
  Adam}, \bibinfo{person}{Stefan Feuerriegel}, \bibinfo{person}{Nicolas
  Pr{\"o}llochs}, {and} \bibinfo{person}{Dirk Neumann}.}
  \bibinfo{year}{2020}\natexlab{}.
\newblock \showarticletitle{Affective information processing of fake news:
  Evidence from {NeuroIS}}.
\newblock \bibinfo{series}{Lecture Notes in Information Systems and
  Organisation}, Vol.~\bibinfo{volume}{32}. \bibinfo{publisher}{Springer},
  \bibinfo{pages}{121--128}.
\newblock


\bibitem[\protect\citeauthoryear{Macskassy and Michelson}{Macskassy and
  Michelson}{2011}]%
        {Macskassy.2011}
\bibfield{author}{\bibinfo{person}{Sofus~A Macskassy} {and}
  \bibinfo{person}{Matthew Michelson}.} \bibinfo{year}{2011}\natexlab{}.
\newblock \showarticletitle{Why do people retweet? {A}nti-homophily wins the
  day!}. In \emph{ICWSM}.
\newblock


\bibitem[\protect\citeauthoryear{Mann and Helbing}{Mann and Helbing}{2017}]%
        {Mann.2017}
\bibfield{author}{\bibinfo{person}{Richard~P Mann} {and} \bibinfo{person}{Dirk
  Helbing}.} \bibinfo{year}{2017}\natexlab{}.
\newblock \showarticletitle{Optimal incentives for collective intelligence}.
\newblock \bibinfo{journal}{\emph{PNAS}} \bibinfo{volume}{114},
  \bibinfo{number}{20} (\bibinfo{year}{2017}), \bibinfo{pages}{5077--5082}.
\newblock


\bibitem[\protect\citeauthoryear{McPherson, Smith-Lovin, and Cook}{McPherson
  et~al\mbox{.}}{2001}]%
        {McPherson.2001}
\bibfield{author}{\bibinfo{person}{Miller McPherson}, \bibinfo{person}{Lynn
  Smith-Lovin}, {and} \bibinfo{person}{James~M Cook}.}
  \bibinfo{year}{2001}\natexlab{}.
\newblock \showarticletitle{Birds of a feather: Homophily in social networks}.
\newblock \bibinfo{journal}{\emph{Annual Review of Sociology}}
  \bibinfo{volume}{27} (\bibinfo{year}{2001}), \bibinfo{pages}{415--444}.
\newblock


\bibitem[\protect\citeauthoryear{Micallef, Armacost, Memon, and Patil}{Micallef
  et~al\mbox{.}}{2022}]%
        {Micallef.2022}
\bibfield{author}{\bibinfo{person}{Nicholas Micallef},
  \bibinfo{person}{Vivienne Armacost}, \bibinfo{person}{Nasir Memon}, {and}
  \bibinfo{person}{Sameer Patil}.} \bibinfo{year}{2022}\natexlab{}.
\newblock \showarticletitle{True or false: Studying the work practices of
  professional fact-checkers}. In \emph{CSCW}.
\newblock


\bibitem[\protect\citeauthoryear{Myers and Leskovec}{Myers and
  Leskovec}{2014}]%
        {Myers.2014}
\bibfield{author}{\bibinfo{person}{Seth~A. Myers} {and} \bibinfo{person}{Jure
  Leskovec}.} \bibinfo{year}{2014}\natexlab{}.
\newblock \showarticletitle{The bursty dynamics of the Twitter information
  network}. In \emph{WWW}.
\newblock


\bibitem[\protect\citeauthoryear{Myers, Zhu, and Leskovec}{Myers
  et~al\mbox{.}}{2012}]%
        {Myers.2012}
\bibfield{author}{\bibinfo{person}{Seth~A. Myers}, \bibinfo{person}{Chenguang
  Zhu}, {and} \bibinfo{person}{Jure Leskovec}.}
  \bibinfo{year}{2012}\natexlab{}.
\newblock \showarticletitle{Information diffusion and external influence in
  networks}. In \emph{KDD}.
\newblock


\bibitem[\protect\citeauthoryear{Naumzik and Feuerriegel}{Naumzik and
  Feuerriegel}{2022}]%
        {Naumzik.2022}
\bibfield{author}{\bibinfo{person}{Christof Naumzik} {and}
  \bibinfo{person}{Stefan Feuerriegel}.} \bibinfo{year}{2022}\natexlab{}.
\newblock \showarticletitle{Detecting false rumors from retweet dynamics on
  social media}. In \emph{WWW}.
\newblock


\bibitem[\protect\citeauthoryear{Nekovee, Moreno, Bianconi, and
  Marsili}{Nekovee et~al\mbox{.}}{2007}]%
        {Nekovee.2007}
\bibfield{author}{\bibinfo{person}{Maziar Nekovee}, \bibinfo{person}{Yamir
  Moreno}, \bibinfo{person}{Ginestra Bianconi}, {and} \bibinfo{person}{Matteo
  Marsili}.} \bibinfo{year}{2007}\natexlab{}.
\newblock \showarticletitle{Theory of rumor spreading in complex social
  networks}.
\newblock \bibinfo{journal}{\emph{Physica A: Statistical Mechanics and its
  Applications}} \bibinfo{volume}{374}, \bibinfo{number}{1}
  (\bibinfo{year}{2007}), \bibinfo{pages}{457--470}.
\newblock
\showISSN{03784371}


\bibitem[\protect\citeauthoryear{Pendleton}{Pendleton}{1998}]%
        {Pendleton.1998}
\bibfield{author}{\bibinfo{person}{Susan~Coppess Pendleton}.}
  \bibinfo{year}{1998}\natexlab{}.
\newblock \showarticletitle{Rumor research revisited and expanded}.
\newblock \bibinfo{journal}{\emph{Language {\&} Communication}}
  \bibinfo{volume}{18}, \bibinfo{number}{1} (\bibinfo{year}{1998}),
  \bibinfo{pages}{69--86}.
\newblock
\showISSN{02715309}


\bibitem[\protect\citeauthoryear{{Pew Research Center}}{{Pew Research
  Center}}{2016}]%
        {Pew.2016}
\bibfield{author}{\bibinfo{person}{{Pew Research Center}}.}
  \bibinfo{year}{2016}\natexlab{}.
\newblock \bibinfo{title}{News use across social media platforms 2016}.
\newblock
\newblock
\urldef\tempurl%
\url{https://www.journalism.org/2016/05/26/news-use-across-social-media-platforms-2016/}
\showURL{%
\tempurl}


\bibitem[\protect\citeauthoryear{Pr{\"o}llochs}{Pr{\"o}llochs}{2022}]%
        {Prollochs.2022}
\bibfield{author}{\bibinfo{person}{Nicolas Pr{\"o}llochs}.}
  \bibinfo{year}{2022}\natexlab{}.
\newblock \showarticletitle{Community-based fact-checking on Twitter's
  Birdwatch platform}. In \emph{ICWSM}.
\newblock


\bibitem[\protect\citeauthoryear{Pr{\"o}llochs, B{\"a}r, and
  Feuerriegel}{Pr{\"o}llochs et~al\mbox{.}}{2021a}]%
        {SciRep.2021}
\bibfield{author}{\bibinfo{person}{Nicolas Pr{\"o}llochs},
  \bibinfo{person}{Dominik B{\"a}r}, {and} \bibinfo{person}{Stefan
  Feuerriegel}.} \bibinfo{year}{2021}\natexlab{a}.
\newblock \showarticletitle{Emotions explain differences in the diffusion of
  true vs. false social media rumors}.
\newblock \bibinfo{journal}{\emph{Scientific Reports}} \bibinfo{volume}{11},
  \bibinfo{number}{1} (\bibinfo{year}{2021}), \bibinfo{pages}{22721}.
\newblock


\bibitem[\protect\citeauthoryear{Pr{\"o}llochs, B{\"a}r, and
  Feuerriegel}{Pr{\"o}llochs et~al\mbox{.}}{2021b}]%
        {EPJ.2021}
\bibfield{author}{\bibinfo{person}{Nicolas Pr{\"o}llochs},
  \bibinfo{person}{Dominik B{\"a}r}, {and} \bibinfo{person}{Stefan
  Feuerriegel}.} \bibinfo{year}{2021}\natexlab{b}.
\newblock \showarticletitle{Emotions in online rumor diffusion}.
\newblock \bibinfo{journal}{\emph{EPJ Data Science}} \bibinfo{volume}{10},
  \bibinfo{number}{1} (\bibinfo{year}{2021}), \bibinfo{pages}{1--17}.
\newblock


\bibitem[\protect\citeauthoryear{Robertson, Pr{\"o}llochs, Schwarzenegger,
  Parnamets, Van~Bavel, and Feuerriegel}{Robertson et~al\mbox{.}}{2023}]%
        {Upworthy.2023}
\bibfield{author}{\bibinfo{person}{Claire~E Robertson},
  \bibinfo{person}{Nicolas Pr{\"o}llochs}, \bibinfo{person}{Kaoru
  Schwarzenegger}, \bibinfo{person}{Phillip Parnamets}, \bibinfo{person}{Jay~J
  Van~Bavel}, {and} \bibinfo{person}{Stefan Feuerriegel}.}
  \bibinfo{year}{2023}\natexlab{}.
\newblock \showarticletitle{Negativity drives online news consumption}.
\newblock \bibinfo{journal}{\emph{Nature Human Behaviour}}
  \bibinfo{volume}{Forthcoming} (\bibinfo{year}{2023}).
\newblock


\bibitem[\protect\citeauthoryear{Rosnow}{Rosnow}{1988}]%
        {Rosnow.1988}
\bibfield{author}{\bibinfo{person}{Ralph~L. Rosnow}.}
  \bibinfo{year}{1988}\natexlab{}.
\newblock \showarticletitle{Rumor as communication: A contextualist approach}.
\newblock \bibinfo{journal}{\emph{Journal of Communication}}
  \bibinfo{volume}{38}, \bibinfo{number}{1} (\bibinfo{year}{1988}),
  \bibinfo{pages}{12--28}.
\newblock


\bibitem[\protect\citeauthoryear{Rosnow}{Rosnow}{1991}]%
        {Rosnow.1991}
\bibfield{author}{\bibinfo{person}{Ralph~L Rosnow}.}
  \bibinfo{year}{1991}\natexlab{}.
\newblock \showarticletitle{Inside Rumor: A Personal Journey}.
\newblock \bibinfo{journal}{\emph{American Psychologist}} \bibinfo{volume}{46},
  \bibinfo{number}{5} (\bibinfo{year}{1991}), \bibinfo{pages}{484--496}.
\newblock


\bibitem[\protect\citeauthoryear{Shao, Ciampaglia, Flammini, and Menczer}{Shao
  et~al\mbox{.}}{2016}]%
        {Shao.2016}
\bibfield{author}{\bibinfo{person}{Chengcheng Shao},
  \bibinfo{person}{Giovanni~Luca Ciampaglia}, \bibinfo{person}{Alessandro
  Flammini}, {and} \bibinfo{person}{Filippo Menczer}.}
  \bibinfo{year}{2016}\natexlab{}.
\newblock \showarticletitle{Hoaxy: A platform for tracking online
  misinformation}. In \emph{WWW Companion}.
\newblock


\bibitem[\protect\citeauthoryear{Shao, Ciampaglia, Varol, Yang, Flammini, and
  Menczer}{Shao et~al\mbox{.}}{2018}]%
        {Shao.2018}
\bibfield{author}{\bibinfo{person}{Chengcheng Shao},
  \bibinfo{person}{Giovanni~Luca Ciampaglia}, \bibinfo{person}{Onur Varol},
  \bibinfo{person}{Kai-Cheng Yang}, \bibinfo{person}{Alessandro Flammini},
  {and} \bibinfo{person}{Filippo Menczer}.} \bibinfo{year}{2018}\natexlab{}.
\newblock \showarticletitle{The spread of low-credibility content by social
  bots}.
\newblock \bibinfo{journal}{\emph{Nature Communications}} \bibinfo{volume}{9},
  \bibinfo{number}{1} (\bibinfo{year}{2018}), \bibinfo{pages}{4787}.
\newblock


\bibitem[\protect\citeauthoryear{Shibutani}{Shibutani}{1966}]%
        {Shibutani.1966}
\bibfield{author}{\bibinfo{person}{Tamotsu Shibutani}.}
  \bibinfo{year}{1966}\natexlab{}.
\newblock \bibinfo{booktitle}{\emph{Improvised news: A sociological study of
  rumor}}.
\newblock \bibinfo{publisher}{Bobbs-Merrill}, \bibinfo{address}{Indianapolis,
  IN}.
\newblock


\bibitem[\protect\citeauthoryear{Solovev and Pr{\"o}llochs}{Solovev and
  Pr{\"o}llochs}{2022a}]%
        {Solovev.2022}
\bibfield{author}{\bibinfo{person}{Kirill Solovev} {and}
  \bibinfo{person}{Nicolas Pr{\"o}llochs}.} \bibinfo{year}{2022}\natexlab{a}.
\newblock \showarticletitle{Hate speech in the political discourse on social
  media: Disparities across parties, gender, and ethnicity}. In \emph{WWW}.
\newblock


\bibitem[\protect\citeauthoryear{Solovev and Pr{\"o}llochs}{Solovev and
  Pr{\"o}llochs}{2022b}]%
        {Solovev.2022b}
\bibfield{author}{\bibinfo{person}{Kirill Solovev} {and}
  \bibinfo{person}{Nicolas Pr{\"o}llochs}.} \bibinfo{year}{2022}\natexlab{b}.
\newblock \showarticletitle{Moral emotions shape the virality of COVID-19
  misinformation on social media}. In \emph{WWW}.
\newblock


\bibitem[\protect\citeauthoryear{Starbird}{Starbird}{2013}]%
        {Starbird.2013}
\bibfield{author}{\bibinfo{person}{Kate Starbird}.}
  \bibinfo{year}{2013}\natexlab{}.
\newblock \showarticletitle{Delivering patients to Sacr{\'e} Coeur: Collective
  intelligence in digital volunteer communities}. In \emph{CHI}.
\newblock


\bibitem[\protect\citeauthoryear{Statista}{Statista}{2020}]%
        {Statista.2020}
\bibfield{author}{\bibinfo{person}{Statista}.} \bibinfo{year}{2020}\natexlab{}.
\newblock \bibinfo{title}{Number of monthly active {T}witter users worldwide
  from 1st quarter 2010 to 1st quarter 2019}.
\newblock
\newblock
\urldef\tempurl%
\url{https://www.statista.com/statistics/282087/number-of-monthly-active-twitter-users/}
\showURL{%
\tempurl}


\bibitem[\protect\citeauthoryear{Stieglitz and Dang-Xuan}{Stieglitz and
  Dang-Xuan}{2012}]%
        {Stieglitz.2012b}
\bibfield{author}{\bibinfo{person}{Stefan Stieglitz} {and}
  \bibinfo{person}{Linh Dang-Xuan}.} \bibinfo{year}{2012}\natexlab{}.
\newblock \showarticletitle{Political communication and influence through
  microblogging: An empirical analysis of sentiment in {T}witter messages and
  retweet behavior}. In \emph{HICSS}.
\newblock


\bibitem[\protect\citeauthoryear{Stieglitz and Dang-Xuan}{Stieglitz and
  Dang-Xuan}{2013a}]%
        {Stieglitz.2013}
\bibfield{author}{\bibinfo{person}{Stefan Stieglitz} {and}
  \bibinfo{person}{Linh Dang-Xuan}.} \bibinfo{year}{2013}\natexlab{a}.
\newblock \showarticletitle{Emotions and information diffusion in social media:
  Sentiment of microblogs and sharing behavior}.
\newblock \bibinfo{journal}{\emph{Journal of Management Information Systems}}
  \bibinfo{volume}{29}, \bibinfo{number}{4} (\bibinfo{year}{2013}),
  \bibinfo{pages}{217--248}.
\newblock
\showISSN{0742-1222}


\bibitem[\protect\citeauthoryear{Stieglitz and Dang-Xuan}{Stieglitz and
  Dang-Xuan}{2013b}]%
        {Stieglitz.2013b}
\bibfield{author}{\bibinfo{person}{Stefan Stieglitz} {and}
  \bibinfo{person}{Linh Dang-Xuan}.} \bibinfo{year}{2013}\natexlab{b}.
\newblock \showarticletitle{Social media and political communication: A social
  media analytics framework}.
\newblock \bibinfo{journal}{\emph{Social Network Analysis and Mining}}
  \bibinfo{volume}{3}, \bibinfo{number}{4} (\bibinfo{year}{2013}),
  \bibinfo{pages}{1277--1291}.
\newblock


\bibitem[\protect\citeauthoryear{Sun}{Sun}{2013}]%
        {Sun.2013}
\bibfield{author}{\bibinfo{person}{Heshan Sun}.}
  \bibinfo{year}{2013}\natexlab{}.
\newblock \showarticletitle{A longitudinal study of herd behavior in the
  adoption and continued use of technology}.
\newblock \bibinfo{journal}{\emph{MIS Quarterly}} \bibinfo{volume}{4},
  \bibinfo{number}{37} (\bibinfo{year}{2013}), \bibinfo{pages}{1013--1041}.
\newblock


\bibitem[\protect\citeauthoryear{Susarla, Oh, and Tan}{Susarla
  et~al\mbox{.}}{2012}]%
        {Susarla.2012}
\bibfield{author}{\bibinfo{person}{Anjana Susarla}, \bibinfo{person}{Jeong-Ha
  Oh}, {and} \bibinfo{person}{Yong Tan}.} \bibinfo{year}{2012}\natexlab{}.
\newblock \showarticletitle{Social networks and the diffusion of user-generated
  content: Evidence from {YouTube}}.
\newblock \bibinfo{journal}{\emph{Information Systems Research}}
  \bibinfo{volume}{23}, \bibinfo{number}{1} (\bibinfo{year}{2012}),
  \bibinfo{pages}{23--41}.
\newblock
\showISSN{1047-7047}


\bibitem[\protect\citeauthoryear{Tambuscio, Ruffo, Flammini, and
  Menczer}{Tambuscio et~al\mbox{.}}{2015}]%
        {Tambuscio.2015}
\bibfield{author}{\bibinfo{person}{Marcella Tambuscio},
  \bibinfo{person}{Giancarlo Ruffo}, \bibinfo{person}{Alessandro Flammini},
  {and} \bibinfo{person}{Filippo Menczer}.} \bibinfo{year}{2015}\natexlab{}.
\newblock \showarticletitle{Fact-checking effect on viral hoaxes: A model of
  misinformation spread in social networks}. In \emph{WWW Companion}.
\newblock


\bibitem[\protect\citeauthoryear{{The Economist}}{{The Economist}}{2017}]%
        {TheEconomist.2017}
\bibfield{author}{\bibinfo{person}{{The Economist}}.}
  \bibinfo{year}{2017}\natexlab{}.
\newblock \showarticletitle{How the world was trolled}.
\newblock \bibinfo{journal}{\emph{The Economist}}  \bibinfo{volume}{425}
  (\bibinfo{year}{2017}), \bibinfo{pages}{21--24}.
\newblock


\bibitem[\protect\citeauthoryear{Thies, Wessel, and Benlian}{Thies
  et~al\mbox{.}}{2016}]%
        {Thies.2016}
\bibfield{author}{\bibinfo{person}{Ferdinand Thies}, \bibinfo{person}{Michael
  Wessel}, {and} \bibinfo{person}{Alexander Benlian}.}
  \bibinfo{year}{2016}\natexlab{}.
\newblock \showarticletitle{Effects of social interaction dynamics on
  platforms}.
\newblock \bibinfo{journal}{\emph{Journal of Management Information Systems}}
  \bibinfo{volume}{33}, \bibinfo{number}{3} (\bibinfo{year}{2016}),
  \bibinfo{pages}{843--873}.
\newblock
\showISSN{0742-1222}


\bibitem[\protect\citeauthoryear{Townsend, Buckley, Harada, and Scott}{Townsend
  et~al\mbox{.}}{2013}]%
        {Townsend.2013}
\bibfield{author}{\bibinfo{person}{Zac Townsend}, \bibinfo{person}{Jack
  Buckley}, \bibinfo{person}{Masataka Harada}, {and} \bibinfo{person}{Marc~A.
  Scott}.} \bibinfo{year}{2013}\natexlab{}.
\newblock \showarticletitle{The choice between fixed and random effects}.
\newblock \bibinfo{journal}{\emph{SAGE Handbook of Multilevel Modeling}}
  (\bibinfo{year}{2013}), \bibinfo{pages}{73--88}.
\newblock


\bibitem[\protect\citeauthoryear{Vieweg, Palen, Liu, Hughes, and Sutton}{Vieweg
  et~al\mbox{.}}{2008}]%
        {Vieweg.2008}
\bibfield{author}{\bibinfo{person}{Sarah Vieweg}, \bibinfo{person}{Leysia
  Palen}, \bibinfo{person}{Sophia~B Liu}, \bibinfo{person}{Amanda~L Hughes},
  {and} \bibinfo{person}{Jeannette~N Sutton}.} \bibinfo{year}{2008}\natexlab{}.
\newblock \showarticletitle{Collective intelligence in disaster: Examination of
  the phenomenon in the aftermath of the 2007 Virginia Tech Shooting}. In
  \emph{ISRAM}.
\newblock


\bibitem[\protect\citeauthoryear{Vosoughi, Mohsenvand, and Roy}{Vosoughi
  et~al\mbox{.}}{2017}]%
        {Vosoughi.2017}
\bibfield{author}{\bibinfo{person}{Soroush Vosoughi},
  \bibinfo{person}{Mostafa~`Neo' Mohsenvand}, {and} \bibinfo{person}{Deb Roy}.}
  \bibinfo{year}{2017}\natexlab{}.
\newblock \showarticletitle{Rumor gauge: Predicting the veracity of rumors on
  {T}witter}.
\newblock \bibinfo{journal}{\emph{ACM Transactions on Knowledge Discovery from
  Data}} \bibinfo{volume}{11}, \bibinfo{number}{4} (\bibinfo{year}{2017}),
  \bibinfo{pages}{1--36}.
\newblock


\bibitem[\protect\citeauthoryear{Vosoughi, Roy, and Aral}{Vosoughi
  et~al\mbox{.}}{2018}]%
        {Vosoughi.2018}
\bibfield{author}{\bibinfo{person}{Soroush Vosoughi}, \bibinfo{person}{Deb
  Roy}, {and} \bibinfo{person}{Sinan Aral}.} \bibinfo{year}{2018}\natexlab{}.
\newblock \showarticletitle{The spread of true and false news online}.
\newblock \bibinfo{journal}{\emph{Science}} \bibinfo{volume}{359},
  \bibinfo{number}{6380} (\bibinfo{year}{2018}), \bibinfo{pages}{1146--1151}.
\newblock


\bibitem[\protect\citeauthoryear{Webb, Burnap, Procter, Rana, Stahl, Williams,
  Housley, Edwards, and Jirotka}{Webb et~al\mbox{.}}{2016}]%
        {Webb.2016}
\bibfield{author}{\bibinfo{person}{Helena Webb}, \bibinfo{person}{Pete Burnap},
  \bibinfo{person}{Rob Procter}, \bibinfo{person}{Omer Rana},
  \bibinfo{person}{Bernd~Carsten Stahl}, \bibinfo{person}{Matthew Williams},
  \bibinfo{person}{William Housley}, \bibinfo{person}{Adam Edwards}, {and}
  \bibinfo{person}{Marina Jirotka}.} \bibinfo{year}{2016}\natexlab{}.
\newblock \showarticletitle{Digital wildfires}.
\newblock \bibinfo{journal}{\emph{ACM Transactions on Information Systems}}
  \bibinfo{volume}{34}, \bibinfo{number}{3} (\bibinfo{year}{2016}),
  \bibinfo{pages}{Article 15}.
\newblock


\bibitem[\protect\citeauthoryear{Weng, Menczer, and Ahn}{Weng
  et~al\mbox{.}}{2013}]%
        {Weng.2013}
\bibfield{author}{\bibinfo{person}{Lilian Weng}, \bibinfo{person}{Filippo
  Menczer}, {and} \bibinfo{person}{Yong-Yeol Ahn}.}
  \bibinfo{year}{2013}\natexlab{}.
\newblock \showarticletitle{Virality prediction and community structure in
  social networks}.
\newblock \bibinfo{journal}{\emph{Scientific Reports}}  \bibinfo{volume}{3}
  (\bibinfo{year}{2013}), \bibinfo{pages}{Article 2522}.
\newblock


\bibitem[\protect\citeauthoryear{Yang and Counts}{Yang and Counts}{2010}]%
        {Yang.2010b}
\bibfield{author}{\bibinfo{person}{Jiang Yang} {and} \bibinfo{person}{Scott
  Counts}.} \bibinfo{year}{2010}\natexlab{}.
\newblock \showarticletitle{Predicting the speed, scale, and range of
  information diffusion in Twitter}. In \emph{ICWSM}.
\newblock


\bibitem[\protect\citeauthoryear{Zaman, Fox, and Bradlow}{Zaman
  et~al\mbox{.}}{2014}]%
        {Zaman.2014}
\bibfield{author}{\bibinfo{person}{Tauhid Zaman}, \bibinfo{person}{Emily~B.
  Fox}, {and} \bibinfo{person}{Eric~T. Bradlow}.}
  \bibinfo{year}{2014}\natexlab{}.
\newblock \showarticletitle{A {B}ayesian approach for predicting the popularity
  of tweets}.
\newblock \bibinfo{journal}{\emph{The Annals of Applied Statistics}}
  \bibinfo{volume}{8}, \bibinfo{number}{3} (\bibinfo{year}{2014}),
  \bibinfo{pages}{1583--1611}.
\newblock


\bibitem[\protect\citeauthoryear{Zeng and Zhu}{Zeng and Zhu}{2019}]%
        {Zeng.2019}
\bibfield{author}{\bibinfo{person}{Runxi Zeng} {and} \bibinfo{person}{Di Zhu}.}
  \bibinfo{year}{2019}\natexlab{}.
\newblock \showarticletitle{A model and simulation of the emotional contagion
  of netizens in the process of rumor refutation}.
\newblock \bibinfo{journal}{\emph{Scientific Reports}} \bibinfo{volume}{9},
  \bibinfo{number}{1} (\bibinfo{year}{2019}), \bibinfo{pages}{1--15}.
\newblock


\end{thebibliography}

\appendix

\newpage

\newpage
\vspace{1cm}
\begin{center}
\huge Supplementary Materials
\end{center}
\vspace{1cm}

\section{Overview of Robustness Checks}

We conducted an extensive set of checks and complementary analyses to validate the robustness of our results regarding outliers, additional model features, and modification in variables. We briefly present the main findings in the following.
%% Outliers
%log transformation
%winsorizing
%cut-off times
\subsection{Robustness to Outliers}

In our main analysis, we deliberately refrained from excluding outliers in the explanatory count variables, as this would exclude extreme effects that are of particular interest in the context of rumor spreading. As part of our robustness checks, we analyzed the role of outliers as follows. For this, we winsorized all count variables at the \SI{0.1}{\percent} and \SI{1}{\percent} level. We find qualitatively identical results that continue to support our findings.

\subsection{Analysis of Rumors by Veracity (\ie, ``Else'' Category)}

In our main analysis, we excluded rumors for which there was no clear assignment of the veracity to true or false. These rumors were assigned to a third category named ``else.'' As a robustness check, we additionally considered rumor cascades from the ``else'' category. This is done to control for rumors of mixed veracity and to control for the possibility that fact-checking organization selected a stratified sample. Compared to true rumors, we find that rumors from the ``else'' category are expected to receive more retweets, but the estimated retweet count is lower than that of false rumors. Again, we confirm our previous findings for hypotheses H1, H2, H3b, and H4b (and reject both H3a and H4a), suggesting the presence of lifetime and crowd effects in the diffusion of false rumors.

\subsection{Analysis of Exposure}

We evaluated the role of the overall exposure of a retweet cascade (\ie, cascade size). For each tweet, we calculated the number of inner nodes in a cascade until the time point $t$. This measure thus comprises not only all nodes on the direct path from the root to a retweet but also on all other parts of the cascade (before the time point $t$). We find that overall exposure exhibits a fairly high correlation with lifetime (correlation of \num{0.17}). In our regression models, overall exposure has a negative effect on retweet size and a positive effect on response time. This again suggests that the interest of users declines if the underlying information becomes less novel. All other results (hypotheses H1--H4) still stand with this additional control. 

\subsection{Alternative Model Specifications and Estimators}

We further estimated separate regressions for false and true news. Here we obtain qualitatively identical results that continue to support our findings. In addition, we utilized Bayesian estimation techniques due to their theoretical advantages such as robustness against multicollinearity. Based on the estimated posterior means, the previous findings are confirmed. Reassuringly, we note that the variance inflation factors of all independent variables in our main analysis are below the critical threshold of 4. Hence, there is no evidence that multicollinearity impedes the validity of our findings. 

In April 2015, Twitter introduced a quote feature that allows for external retweeting. To address potentially confounding effects, we tested model variants in which we (1) add a dummy variable that takes the value 1 after the introduction of the quote feature (otherwise $=0$) and (2) omit observations at the root node of rumor cascades. In both cases, our results are robust and continue to support our findings.

\newpage
\section{Analysis for Retweet Probability}
\label{sec:analysis_retweet_probability}

We further study the role of lifetime and crowd effects for retweet probability, that is, the user's propensity to share content. For this, we introduce a binary variable $\mathit{Retweeted}$, indicating whether a tweet was shared by a user. This is later used to model the retweet probability, which reflects a user's propensity for retweeting and answers the question of how likely it is that the tweet will be subsequently shared by other users (\ie, their followers) \citep{Zaman.2014}. All else being equal, a higher retweet probability should then result in a retweet cascade that has a larger reach. The observed sharing probability is larger for false rumors (\SI{18.60}{\percent}) than for true rumors (\SI{13.70}{\percent}). 

In the regression model, we examine determinants that explain a user's propensity to share rumors. Recall that the units of analysis are individual retweets, and therefore, the dependent variable is given by \emph{Retweeted}, which equals $1$ if at least one retweet occurred and otherwise $=0$. All other variables are analogous to the previous model. This yields 

\vspace{-.7\baselineskip}

{\footnotesize
\begin{align}
& \logit({Retweeted}) = \, \gamma + {\beta_{1} \, \mathit{Falsehood}} + \beta_{2} \, \mathit{Falsehood} \times \mathit{Lifetime}_t  \nonumber\\
& \qquad + \beta_{3} \, \mathit{Falsehood} \times \mathit{RetweetDepth}_t + \beta_{4} \, \mathit{Lifetime}_t + \beta_{5} \, \mathit{RetweetDepth}_t 
\label{eq:regression_retweet_probability} \\
& \qquad + \beta_{6} \, \mathit{Followers}_t + \beta_{7} \, \mathit{Followees}_t  + \beta_{8} \, \mathit{Account Age}_t + \beta_{9} \, \mathit{User Engagement}_t +  \beta_{10} \, \mathit{Verified}_t + u_\text{cascade} \nonumber 
\end{align} 
}%
\normalsize
with intercept $\gamma$, and cascade-specific random effect $u_\text{cascade}$. In the aforementioned model, the variable \emph{Retweeted} is binary and, hence, the model is estimated as a logit regression. 

%% Model

The results for the logistic regression explaining a user's propensity to share rumors are as follows (see \Cref{tbl:regression_retweeted}). All variables are again standardized in order to facilitate interpretability.

%%% H1

The coefficient for $\mathit{Falsehood}$ is positive and statistically significant. Hence, false rumors are more likely to be retweeted than true rumors. Yet, because we estimate a logistic regression model with interaction terms, the coefficients cannot be interpreted as the change in the mean of the dependent variable for a one unit (\ie, standard deviation) increase in the respective predictor variable, with all other predictors remaining constant. In logistic regression models with interaction terms, marginal effects are nonlinear functions of the coefficients and the levels of the explanatory variables \citep{Ai.2003, Buis.2010}. However, logit models are linear in log-odds and, hence, the odds ratios (\ie, the exponentiated coefficients) represent the constant effect of a given variable on the likelihood of a retweet. In mixed logistic regression models, interaction effects should be interpreted via the natural logarithm of a multiplying factor by which the predicted odds change, given a one unit increase in the predictor variable, holding all other predictor variables constant \citep{KaracaMandic.2012}. Therefore, we first have to calculate the odds ratio, which is equal to the exponent of the coefficient of the respective variable. Formally, the odds of an event is the ratio $p/(1-p)$, where $p$ is the probability of an event. With the logit link function, $\beta$ is the change in $\log{(\mathrm{odds})}$ and the corresponding change in odds of the event is $e^{\beta}$. The coefficient for $\mathit{Falsehood}$ amounts to \num{0.285}, which corresponds to an odds ratio of $e^{0.285} \approx 1.33$. This implies that the odds of a retweet for a false rumor are \SI{32.98}{\percent} higher than for a true rumor. 

%% lifetime

Next, we assess how the lifetime is associated with the odds of a retweet for true vs. false rumors. The coefficient of the interaction between $\mathit{Falsehood}$ and $\mathit{Lifetime}$ is negative and statistically significant ($\beta=-0.234$, $p < 0.001$); the coefficient of $\mathit{Lifetime}$ is also negative and statistically significant ($\beta=-0.116$, $p < 0.001$). For true rumors, a one standard deviation increase in the lifetime of a cascade decreases the odds by \SI{10.95}{\percent}. Here the odds ratio for false rumors is calculated via the exponent of the sum of the coefficients of $\mathit{Lifetime}$ and $\mathit{Falsehood}$ $\times$ $\mathit{Lifetime}$. The resulting odds ratio is 0.704, which implies that a one standard deviation increase in the lifetime of a cascade reduces the odds of a retweet by \SI{29.53}{\percent}. This is $\sim$\num{2.7} times the reduction for true rumors. Put differently, the propensity of users to share a false rumor declines faster over the lifetime of the cascade.

\begin{table}[H]
	%\TABLE
	\caption{Regression Results for Retweet Probability.\label{tbl:regression_retweeted}}
	{
		%\OneAndAHalfSpacedXI
		\footnotesize
		\begin{tabularx}{\textwidth}{@{\hspace{\tabcolsep}\extracolsep{\fill}}l *{4}{S} } 
			\toprule
			\multicolumn{5}{l}{Dependent Variable: $\mathit{Retweeted}$ ($=1$ if true, $=0$ otherwise)}\\
			\midrule
			& {\textbf{Model (1)}} & {\textbf{Model (2)}} & {\textbf{Model (3)}} & {\textbf{Model (4)}}\\ 
			\midrule
			
			Falsehood                         &              &              & 0.198^{***}  & 0.285^{***}  \\
			&              &              & (0.016)      & (0.018)      \\
			Falsehood $\times$ Lifetime &              &              &              & -0.234^{***} \\
			&              &              &              & (0.014)      \\
			Falsehood $\times$ Retweet Depth  &              &              &              & 0.198^{***}  \\
			&              &              &              & (0.013)      \\
			Lifetime                    &              & -0.340^{***} & -0.340^{***} & -0.116^{***} \\
			&              & (0.003)      & (0.003)      & (0.013)      \\
			Retweet Depth                     &              & -0.061^{***} & -0.061^{***} & -0.256^{***} \\
			&              & (0.002)      & (0.002)      & (0.013)      \\
			Followers                         & 0.794^{***}  & 0.797^{***}  & 0.797^{***}  & 0.797^{***}  \\
			& (0.003)      & (0.003)      & (0.003)      & (0.003)      \\
			Followees                        & -0.127^{***} & -0.124^{***} & -0.125^{***} & -0.125^{***} \\
			& (0.002)      & (0.002)      & (0.002)      & (0.002)      \\
			Account Age                       & 0.049^{***}  & 0.050^{***}  & 0.050^{***}  & 0.050^{***}  \\
			& (0.002)      & (0.002)      & (0.002)      & (0.002)      \\																	
			User Engagement                   & 0.048^{***}  & 0.049^{***}  & 0.049^{***}  & 0.049^{***}  \\
			& (0.002)      & (0.002)      & (0.002)      & (0.002)      \\
			Verified                  & 0.488^{***}  & 0.473^{***}  & 0.476^{***}  & 0.472^{***}  \\
			& (0.034)      & (0.034)      & (0.034)      & (0.034)      \\
			Intercept                         & -2.121^{***} & -2.235^{***} & -2.398^{***} & -2.487^{***} \\
			& (0.006)      & (0.006)      & (0.014)      & (0.017)      \\
			Random effects (cascade level) & {Yes} & {Yes} & {Yes} & {Yes} \\													
			\midrule
			AIC                                 & {\num{3016561}}   & {\num{3001679}}   & {\num{3001490}}   & {\num{3001081}}   \\
			Observations                        & {\num{3724197}}      & {\num{3724197}}      & {\num{3724197}}      & {\num{3724197}}      \\
			\bottomrule
			\multicolumn{5}{r}{Significance levels: $^*p< 0.05$, $^{**}p< 0.01$, $^{***}p< 0.001$; standard errors in parentheses} \\
			%\multicolumn{5}{r}{Logit regression; significance levels: $^*p< 0.05$, $^{**}p< 0.01$, $^{***}p< 0.001$; standard errors in parentheses; unit of analysis: retweet level ($N=\,$\num[group-separator={,},group-minimum-digits=1]{3724197})} \\
		\end{tabularx}
	}
	\subcaption*{\emph{Note:} Logistic regression explains whether or not a tweet was retweeted ($=$1 if true, otherwise 0). Unit of analysis is the retweet level ($N=\,$\num[group-separator={,},group-minimum-digits=1]{3724197}). Cascade-specific random effects are included. 
	}
\end{table}

%% Depth

We now test how the retweet depth is associated with the odds of a retweet for true vs. false rumors. The coefficient of the interaction between $\mathit{Falsehood}$ and $\mathit{RetweetDepth}$ is positive and significant ($\beta=0.198$, $p < 0.001$), whereas the coefficient of $\mathit{RetweetDepth}$ is negative and significant ($\beta=-0.256$, $p < 0.001$). The corresponding odds ratios are 0.774 for true rumors and 0.944 for false rumors. This implies that a one standard deviation increase in retweet depth reduces the odds of a retweet by \SI{22.6}{\percent} for true rumors and by \SI{5.6}{\percent} for false rumors. Accordingly, the odds of a retweet decrease as the retweet depth increases, yet to a much smaller extent for false rumors. Specifically, the decrease in odds for false rumors is only \SI{24.78}{\percent} of the decrease in odds for true rumors. Thus, retweet depth has a stronger negative association with a user's retweet probability for true than for false rumors.

% AIC

According to the AIC, \Cref{tbl:regression_retweeted} suggests that the variables for lifetime and retweet depth should be included in the model. For each, the difference in AIC is greater than ten, indicating strong support for the corresponding candidate models \citep{Burnham.2004}. Hence, both lifetime and retweet depth are relevant for explaining the sharing behavior of true and false rumors.   

%% Interaction plots

\begin{figure}%[H]
	\captionsetup[subfloat]{position=bottom,labelformat=empty}% %, labelfont=bf,textfont=normalfont,singlelinecheck=off,justification=raggedright
	%\FIGURE
	%\centering
	\caption{Predicted marginal means of retweet probability.\label{fig:ggpredict_retweet_probability}}
	{
		\subfloat{{\includegraphics[width=5.5cm]{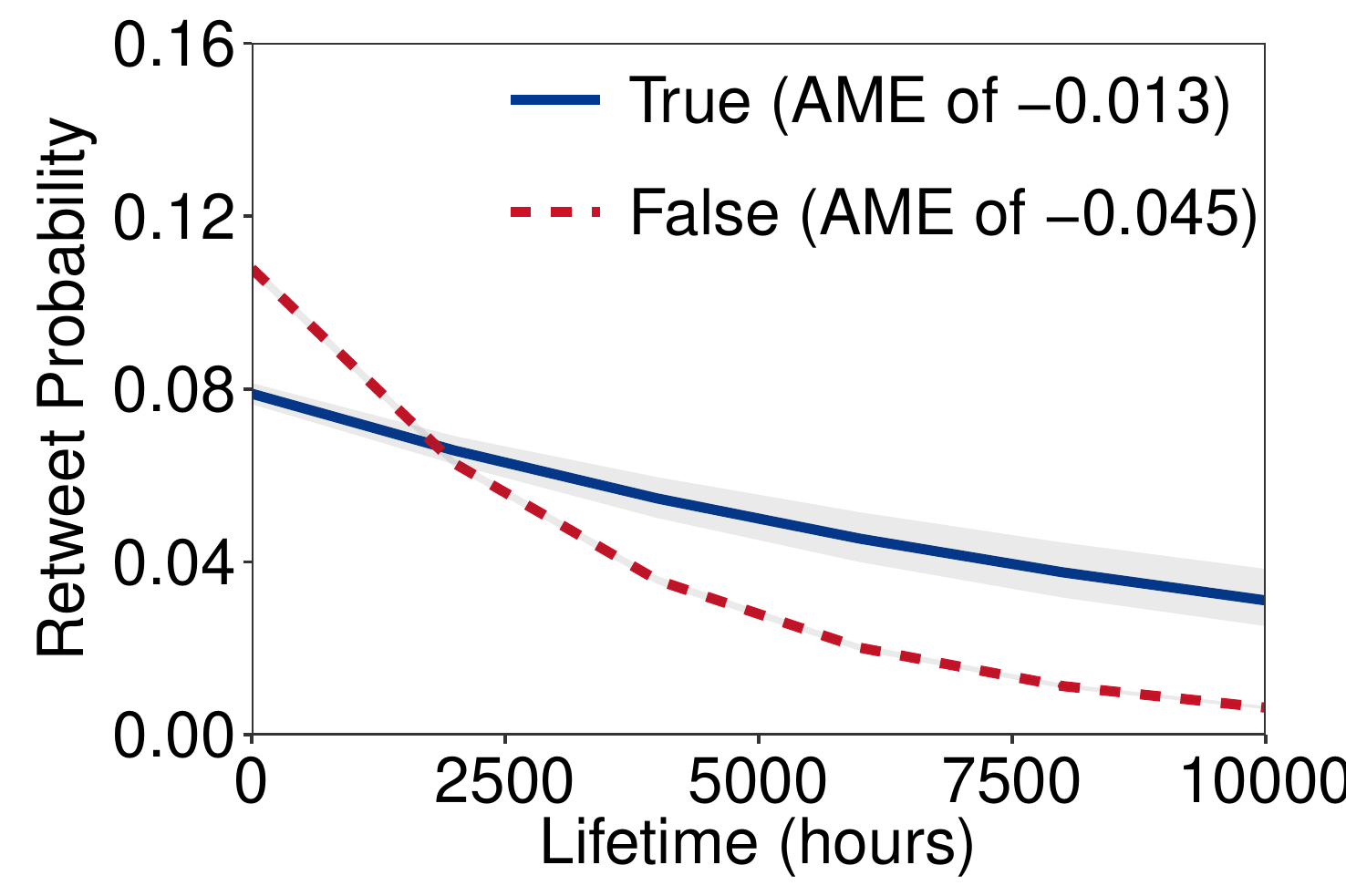}}}
		\quad\quad
		\subfloat{{\includegraphics[width=5.5cm]{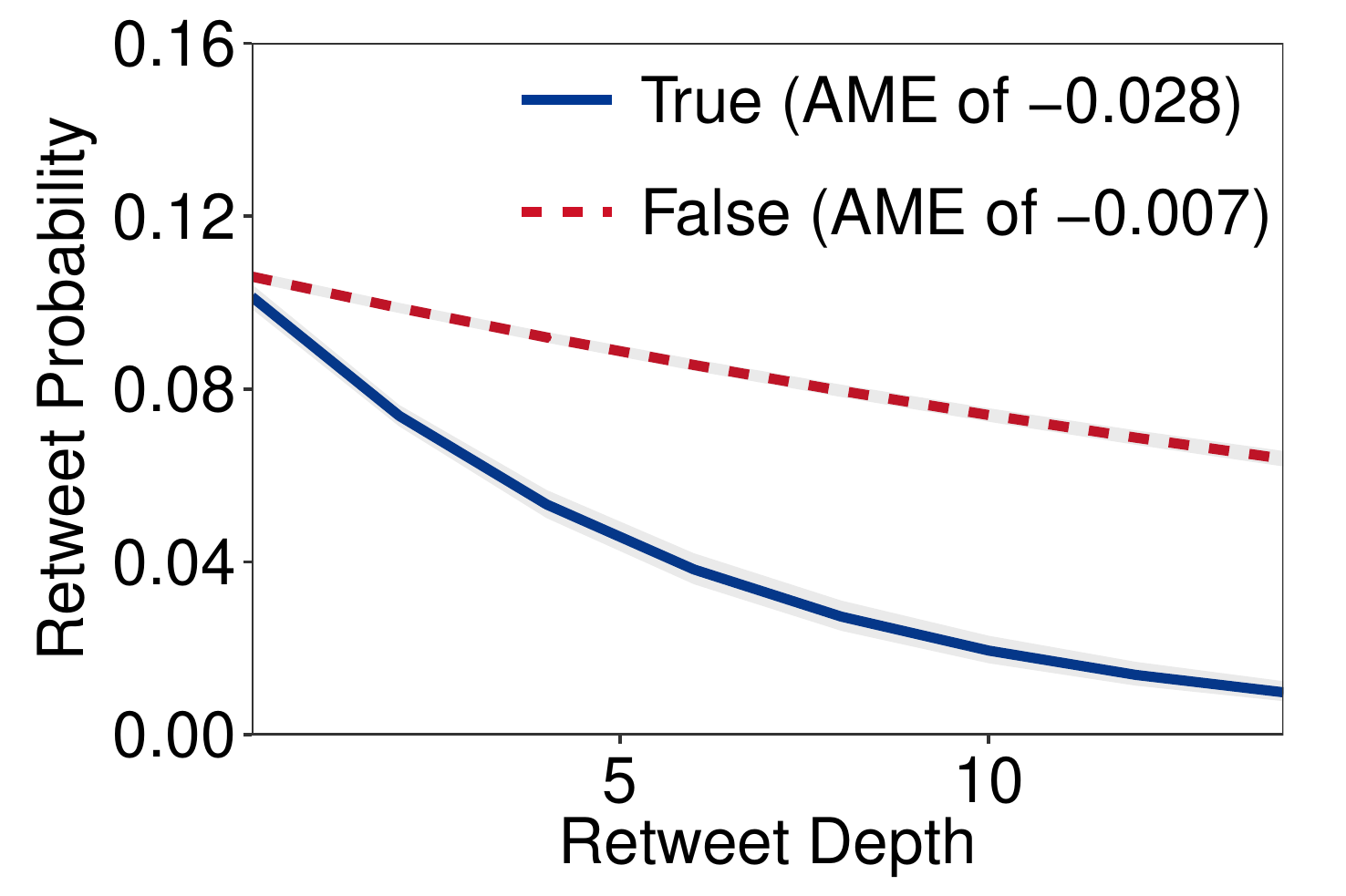}}}%
	}
	
	\subcaption*{\emph{Note:} The figures show the predicted marginal means of retweet probability for different values of lifetime (left) and retweet depth (right). The \SI{95}{\percent} confidence intervals are highlighted in gray (but are comparatively small due to the large sample size). Cascade-specific random effects are included in the analysis.}
	%}
\end{figure}

To shed additional light on the interactions, \Cref{fig:ggpredict_retweet_probability} visualizes the predicted marginal means. The left figure visualizes the predictions as a function of lifetime, whereas the right figure shows the effects as a function of retweet depth. The AME of lifetime is 3.5 times larger for false rumors (AME of $-0.045$) than for true rumors (AME of $-0.013$). In contrast, the AME of retweet depth for false rumors (AME of $-0.007$) is only \SI{26.71}{\percent} of the AME for true rumors (AME of $-0.028$). The plots further show that (1)~false rumors are more likely to be retweeted than true rumors if the lifetime is short and less likely if the lifetime is long, and (2)~false rumors are more likely to be retweeted at higher depths than true rumors. 

Altogether, the above results are consistent with the findings regarding retweet count (\ie, H1 and H3b are confirmed, H3a is rejected).

\newpage
\section{Sensitivity Analyses}
\label{appendix:sensitivity}

Our dataset provides a real-world, cross-sectional sample and is thus subject to considerable heterogeneity. Previous research has shown that the underlying topic (\eg, political vs. non-political content) of a rumor and emotions are important drivers of online behavior \citep[\eg,][]{Stieglitz.2013,SciRep.2021,EPJ.2021,Ducci.2020,Naumzik.2022,Upworthy.2023,Solovev.2022b,Jakubik.2022}. Yet the role of lifetime and crowd behavior in the underlying sharing mechanism is unclear for rumors addressing different topics and comprising different emotions. For this purpose, we examine lifetime and crowd behavior from rumors with (1)~a political topic and (2)~positive and negative emotions. According to previous literature, false rumors entail a higher proportion of replies conveying negative emotions, while true rumors are more likely to be linked with positive emotions. Compared to true rumors, a higher fraction of false rumors are non-political. A labeling of rumors as political vs. non-political and positive vs. negative was obtained from \citep{Vosoughi.2018}. 

\subsection{Political vs. Non-Political Rumor Cascades}

\citet{Stieglitz.2013} have previously confirmed that sharing behavior differs across political vs. non-political communication. The authors argue that political topics are more controversial and thus attract more attention, which itself influences sharing behavior. We thus seek to understand whether the lifetime and crowd effects as identified in our paper differ between political vs. non-political rumors. 

We analyze differences in the lifetime and crowd effects for political vs. non-political rumor cascades as follows. First, a topic label (political vs. non-political) for each rumor cascade was extracted \citep{Vosoughi.2018}. Our topic labels originate from rumor debunking websites. In cases in which the rumor debunking organizations did not provide a topic label, three human annotators were used to assign a topic label. The Fleiss' kappa for the annotators was 0.93. In our dataset, {\num{37120}} rumor cascades are political and {\num{69894}} rumor cascades are non-political. We then use the topic labels to separate political from non-political rumors and estimate two separate regression models for each dependent variable. 

\Cref{tbl:regression_subsets_politics_ame} tabulates the average marginal effects of lifetime and retweet depth for true and false rumors separately. We report the average marginal effects across samples rather than coefficient estimates because coefficient estimates in non-linear models with interaction terms do not reflect marginal effects. Therefore, testing differences in coefficient estimates across samples can be misleading. For completeness, we calculate the differences in the average marginal effects of these variables. This allows us to investigate whether there are significant differences in the magnitude of lifetime and depth effects between political and non-political rumors.

%% Retweet count 

The results for the negative binomial regression with retweet count as the dependent variable confirm lifetime and crowd effects, analogous to the findings in the main paper. 

In addition, the results (see columns (1) and (2)) show that the AME of $\mathit{Lifetime}$ is more negative for false political rumors than for true political rumors; and negative but not significantly different for false non-political rumors vs. true non-political rumors. On average, the lifetime effects are most pronounced for false political rumors. Put differently, older false political rumors tend to receive fewer retweets. For retweet depth, the (positive) offset in the marginal effect between true vs. false rumors is much higher for political (difference of $0.374$, $p<0.001$) than for non-political rumors (difference of $0.149$, $p<0.001$). Political false rumors thus receive relatively more retweets at higher depths.

%% Response time

In columns (3) and (4) of \Cref{tbl:regression_subsets_politics_ame}, we repeat our analysis with the response time as the dependent variable. Also here, we find that the differences in the lifetime and depth effects are higher for rumors that are political vs. rumors that are non-political. The difference in the lifetime effect of true vs. false rumors is approximately 11.6 times larger for political rumors (difference of $0.474$, $p<0.001$) than for non-political rumors (difference of $0.041$, $p<0.001$). This suggests that older false political rumors tend to spread slower. For retweet depth, the difference in the effect size between true vs. false rumors is approximately 1.6 times larger for political rumors (difference of $-0.155$, $p<0.001$) than for non-political rumors (difference of $-0.099$, $p<0.001$). In other words, false political rumors spread faster at higher depths.

\begin{table}[H]
%\TABLE
\caption{{Average Marginal Effects for Political vs. Non-Political Rumor Cascades.}\label{tbl:regression_subsets_politics_ame}}
{
%\OneAndAHalfSpacedXI
\scriptsize
\begin{tabularx}{\textwidth}{@{\hspace{\tabcolsep}\extracolsep{\fill}}l *{4}{S} } 
\toprule
	%	& \multicolumn{2}{c}{{\begin{tabular}{@{}c@{}}Dependent Variable:\\ \emph{Retweeted}\\\end{tabular}}} 
		& \multicolumn{2}{c}{{\begin{tabular}{@{}c@{}}Dependent Variable:\\ \emph{Retweet Count}\\\end{tabular}}}
		& \multicolumn{2}{c}{{\begin{tabular}{@{}c@{}}Dependent Variable:\\\emph{Response Time (log)}\\\end{tabular}}} \\	
	\cmidrule(lr){2-3} \cmidrule(lr){4-5} %\cmidrule(lr){6-7} 
 & {\textbf{Political}} & {\textbf{Non-political}} & {\textbf{Political}} & {\textbf{Non-political}}\\% & {\textbf{Political}} & {\textbf{Non-political}} \\
 & {\textbf{(1)}} & {\textbf{(2)}} & {\textbf{(3)}} & {\textbf{(4)}}\\ 
\midrule
Lifetime, \emph{False Rumors}  & -0.230^{***} & -0.091^{***} & 1.060^{***} & 0.765^{***} \\ 
  & (0.003) & (0.002) & (0.002) & (0.002) \\ 
  Lifetime, \emph{True Rumors} & -0.040^{***} & -0.090^{***} & 0.586^{***} & 0.725^{***} \\ 
  & (0.008) & (0.008) & (0.009) & (0.008) \\ 
  Retweet Depth, \emph{False Rumors}  & -0.223^{***} & -0.236^{***} & -0.272^{***} & -0.490^{***} \\ 
  & (0.002) & (0.002) & (0.001) & (0.002) \\ 
  Retweet Depth, \emph{True Rumors}  & -0.597^{***} & -0.385^{***} & -0.117^{***} & -0.391^{***} \\ 
  & (0.010) & (0.009) & (0.010) & (0.011) \\ 
\addlinespace
\multicolumn{5}{l}{\underline{$\Delta$ between true and false rumors}} \\
%\addlinespace
\quad {Lifetime} & -0.189^{***} & -0.001 & 0.474^{***} & 0.041^{***} \\
%& (0.000) & (0.000) & (0.000) & (0.000) & (0.000) & (0.000)\\
\quad {Retweet Depth} & 0.374^{***} & 0.149^{***} & -0.155^{***} & -0.099^{***}\\
%& (0.000) & (0.000) & (0.000) & (0.000) & (0.000) & (0.000)\\
\bottomrule
\multicolumn{5}{r}{Significance levels: $^*p< 0.05$, $^{**}p< 0.01$, $^{***}p< 0.001$; standard errors in parentheses} \\
\end{tabularx}
}
\subcaption*{\emph{Note:} Average marginal effects (AME) for political vs. non-political rumor cascades.
Columns (1) and (2) report the AME for negative binomial regressions explaining the number of retweets. 
Columns (3) and (4) report the AME for OLS regressions explaining the time difference between a user's tweet and the parent tweet. Control variables and random effects (cascade level) are included. The significance levels for differences in marginal effects are calculated using $t$-tests. 
}
\end{table}

In summary, we established the important role of lifetime and crowd effects in both political and non-political rumors, consistent with the findings from the main paper. Hence, both lifetime and crowd effects are relevant determinants for explaining the underlying mechanism of sharing behavior. Nevertheless, we find some variation. The extent to which false vs. true rumors spread differently depending on the lifetime and depth of the cascade is slightly stronger for political than for non-political rumors. In short, both lifetime and crowd behavior play an important role for political and non-political rumors. 

\subsection{Emotions in Replies to Rumor Cascades}

Rumor theory suggests that rumor spreading is also driven by emotions. In offline settings, \citet{Rosnow.1991} found that negative emotions are usually highly influential in triggering rumormongering. Here the theoretical explanation is again located in a negativity bias according to which negative information is perceived as being more unexpected and thus more informative \citep{Shibutani.1966}. Furthermore, viral effects are particularly pronounced for tweets with emotional content \citep{Stieglitz.2013}. Altogether, this highlights the importance of emotion as an antecedent for diffusion.

We now analyze how the role of lifetime and crowd effects differs between rumor cascades embedding positive vs. negative emotions. For this purpose, a textual analysis of replies to rumor cascades was performed \citep{Kratzwald.2018}. In our dataset, {\num{18641}} false rumor cascades and {\num{3514}} true rumor cascades have replies associated with them. For these rumor cascades, emotion scores were calculated following the procedure from \citep{Vosoughi.2018}. Specifically, for all reply tweets, positive and negative emotion scores were determined based on the NRC emotion lexicon. The emotion scores were then aggregated and averaged to obtain mean emotion scores for each rumor cascade. Note that we measure emotions at cascade level (not at rumor level) and, thus, replies to different cascades of the same rumor can convey different emotions. We use the emotion scores to separate rumors embedding positive emotions from rumors embedding negative emotions. Then we again estimate two separate regression models for each dependent variable.

\Cref{tbl:regression_subsets_emotions_ame} tabulates the average marginal effects. Our results yield the following findings. Overall, we obtain results that are largely consistent with those in the main paper. Nevertheless, we also reveal some variability across positive vs. negative rumors. 

For the negative binomial regression with retweet count as the dependent variable, the difference in the AME of \emph{Lifetime} between true vs. false rumors is more negative for rumor cascades embedding negative emotions (difference of $-0.073$, $p<0.001$) than for those embedding positive emotions (difference of $-0.020$, $p<0.001$). In contrast, we find that the (positive) offset in the AME of \emph{RetweetDepth} between true vs. false rumors is higher for cascades embedding positive emotions (difference of $0.445$, $p<0.001$) vs. negative emotions (difference of $0.374$, $p<0.001$). We observe a similar pattern for the model with response time as the dependent variable. Here the difference in the AME of \emph{Lifetime} is more positive for rumor cascades embedding negative emotions (difference of $0.408$, $p<0.001$) than for those embedding positive emotions (difference of $0.083$, $p<0.001$). Furthermore, the difference in the AME of \emph{RetweetDepth} is more negative for true vs. false rumors embedding positive emotions (difference of $-0.168$, $p<0.001$) and not significantly different for true vs. false rumors embedding negative emotions.

\begin{table}[H]
%\TABLE
\caption{{Average Marginal Effects for Rumor Cascades With Positive vs. Negative Emotions.}\label{tbl:regression_subsets_emotions_ame}}
{
%\OneAndAHalfSpacedXI
\scriptsize
\begin{tabularx}{\textwidth}{@{\hspace{\tabcolsep}\extracolsep{\fill}}l *{4}{S} } 
\toprule
		& \multicolumn{2}{c}{{\begin{tabular}{@{}c@{}}Dependent Variable:\\ \emph{Retweet Count}\\\end{tabular}}}
		& \multicolumn{2}{c}{{\begin{tabular}{@{}c@{}}Dependent Variable:\\\emph{Response Time (log)}\\\end{tabular}}} \\	
	\cmidrule(lr){2-3} \cmidrule(lr){4-5} %\cmidrule(lr){6-7} 
 & {\textbf{Positive Emotions}} & {\textbf{Negative Emotions}} & {\textbf{Positive Emotions}} & {\textbf{Negative Emotions}}\\
 & {\textbf{(1)}} & {\textbf{(2)}} & {\textbf{(3)}} & {\textbf{(4)}} \\ 
\midrule
  Lifetime, \emph{False Rumors}  & -0.095^{***} & -0.143^{***} & 0.651^{***} & 1.286^{***} \\ 
  & (0.002) & (0.004) & (0.002) & (0.003) \\ 
	Lifetime, \emph{True Rumors}  & -0.075^{***} & -0.070^{***} & 0.569^{***} & 0.878^{***} \\ 
  & (0.010) & (0.018) & (0.010) & (0.018) \\ 
	Retweet Depth, \emph{False Rumors} & -0.251^{***} & -0.203^{***} & -0.562^{***} & -0.407^{***} \\ 
  & (0.004) & (0.003) & (0.003) & (0.002) \\ 
	Retweet Depth, \emph{True Rumors} & -0.696^{***} & -0.577^{***} & -0.394^{***} & -0.427^{***} \\ 
  & (0.017) & (0.017) & (0.015) & (0.017) \\ 
\addlinespace
\multicolumn{5}{l}{\underline{$\Delta$ between true and false rumors}} \\
%\addlinespace
\quad {Lifetime} & -0.020^{***} & -0.073^{***} & 0.083^{***} & 0.408^{***} \\ 
\quad {Retweet Depth} & 0.445^{***} & 0.374^{***} & -0.168^{***} & 0.020  \\ 
\bottomrule
\multicolumn{5}{r}{Significance levels: $^*p< 0.05$, $^{**}p< 0.01$, $^{***}p< 0.001$; standard errors in parentheses} \\
\end{tabularx}
}
\subcaption*{\emph{Note:} Average marginal effects for rumor cascades with positive vs. negative emotions in replies. Columns (1) and (2) report the AME for negative binomial regressions explaining the number of retweets. 
Columns (3) and (4) report the AME for OLS regressions explaining the time difference between a user's tweet and the parent tweet. Control variables and random effects (cascade level) are included. Rumor cascades that did not receive replies are excluded. The significance levels for differences in marginal effects are calculated using $t$-tests. 
}
\end{table}

Across both positive and negative emotions, findings consistent with those in the main paper were obtained. Hence, both lifetime and crowd effects are important determinants for explaining the underlying mechanism of sharing behavior. We find that the extent to which false and true rumors spread differently depending on the lifetime is stronger for rumor cascades embedding negative emotions. In contrast, we observe a higher difference in the effect of retweet depth for false vs. true rumor cascades embedding positive emotions. Here the combination of false veracity and replies conveying positive emotions is relatively more viral at higher depths. 

\newpage
\section{Analysis of Social Influence of Root}
\label{appendix:root_social_influence}

We perform an additional analysis where we study the contribution of the user from the initial broadcast (\ie, the root in the cascade) to a rumor's viral effects. For this, we repeat the above analysis from the main paper but additionally include user-level characteristics from the root to control for her/his social influence. As before, we control for unobserved heterogeneity at the cascade level through a random effect specification. The results are reported in \Cref{tbl:regression_siroot}.

Reassuringly, the above findings regarding lifetime and crowd effects remain consistent. User-level characteristics (\ie, social influence of the root) are important determinants for retweet count and response time. Users with young accounts are associated with a more viral diffusion of rumors. We find no indication that a root's engagement is linked to the retweet count. However, users share content faster from content creators who are verified. Surprisingly, more followers for the root user are associated negatively with retweet count and positively with response time. This implies that, after controlling for between-cascade heterogeneity, users share true rumors cautiously. This is line with the fact that rumors refer to content that is unverified. 

Interesting findings are observed when we compare the contribution of a creator's social influence to diffusion dynamics across true vs. false rumors. To test this, we include interactions in our regression model; \eg, $\mathit{Falsehood} \times \mathit{Followers}$ (Root), etc. We find that false rumors from verified accounts are shared less frequently than true rumors. This is captured by the interaction $\mathit{Falsehood} \times \mathit{Verified Account}$ (Root). Hence, a \textquote{verified} badge is linked to a lower retweet volume in the case of false (as opposed to true) rumors. However, for false rumors, the \textquote{verified} badge is linked to a faster spread, implying that users spend less time before making a retweet decision.

\begin{table}[H]
%\TABLE
\caption{Analysis of Social Influence of Root Tweet. \label{tbl:regression_siroot}}
{
%\OneAndAHalfSpacedXI
\scriptsize
\begin{tabularx}{\textwidth}{@{\hspace{\tabcolsep}\extracolsep{\fill}}l *{2}{S} } 
\toprule
		& \multicolumn{1}{c}{{\begin{tabular}{@{}c@{}}Dependent Variable:\\ \emph{Retweet Count}\\\end{tabular}}} 
		& \multicolumn{1}{c}{{\begin{tabular}{@{}c@{}}Dependent Variable:\\\emph{Response Time (log)}\\\end{tabular}}} \\
\midrule
 & {\textbf{Model (1)}} & {\textbf{Model (2)}} \\ 
\midrule
Falsehood & 0.212^{***}  & 0.227^{***}  \\
                                           & (0.012)      & (0.035)      \\
Falsehood $\times$ Lifetime & -0.040^{***} & 0.218^{***}  \\
                                           & (0.004)      & (0.006)      \\
Falsehood $\times$ Retweet Depth & 0.175^{***}  & -0.079^{***} \\
                                           & (0.005)      & (0.007)      \\
Falsehood $\times$ Followers (Root)        & -0.003      &  0.065^{**}   \\
                                           & (0.009)     &   (0.024)      \\
Falsehood $\times$ Followees (Root)        & 0.045^{***}  & 0.103^{***}  \\
                                           & (0.007)      & (0.019)      \\
Falsehood $\times$ Account Age (Root)      & -0.007       & -0.041^{***} \\
                                           & (0.004)      & (0.010)      \\
Falsehood $\times$ User Engagement (Root)  & -0.004       & -0.057^{***} \\
                                           & (0.006)      & (0.015)      \\
Falsehood $\times$ Verified Account (Root) & -0.109^{***} & -0.179^{***} \\
                                           & (0.019)      & (0.053)      \\
Followers (Root)                           & -0.026^{**}  & 0.711^{***}  \\
                                           & (0.008)      & (0.023)      \\
Followees (Root)                           & -0.096^{***} & -0.178^{***} \\
                                           & (0.007)      & (0.017)      \\
Account Age (Root)                         & -0.014^{***} & -0.022^{*}   \\
                                           & (0.004)      & (0.009)      \\
User Engagement (Root)                     & -0.012       & -0.033^{*}   \\
                                           & (0.006)      & (0.015)      \\
Verified Account (Root)                    & 0.015        & -0.177^{***} \\
                                           & (0.018)      & (0.048)      \\
Lifetime                                   & -0.041^{***} & 0.659^{***}  \\
                                           & (0.004)      & (0.006)      \\
Retweet Depth                              & -0.299^{***} & -0.271^{***} \\
                                           & (0.005)      & (0.007)      \\
Followers                                  & 0.527^{***}  & -0.051^{***} \\
                                           & (0.001)      & (0.002)      \\
Followees                                  & -0.116^{***} & 0.154^{***}  \\
                                           & (0.001)      & (0.001)      \\
Account Age                                & -0.044^{***} & -0.061^{***} \\
                                           & (0.001)      & (0.001)      \\
User Engagement                            & -0.073^{***} & -0.025^{***} \\
                                           & (0.001)      & (0.001)      \\
Verified Account                           & 0.061^{***}  & -0.367^{***} \\
                                           & (0.007)      & (0.020)      \\
Intercept                                  & -0.147^{***} & 1.289^{***}  \\
                                           & (0.012)      & (0.033)      \\
Random effects (cascade level) & {Yes} & {Yes} \\								
\bottomrule
\multicolumn{3}{r}{Significance levels: $^*p< 0.05$, $^{**}p< 0.01$, $^{***}p< 0.001$; standard errors in parentheses} \\
\end{tabularx}
}
\subcaption*{\emph{Note:} The negative binomial regression in Model (1) explains the number of retweets. The OLS regression results in Model (2) explain the time difference between a user's tweet and the parent tweet. Unit of analysis is the retweet level ($N=\,$\num[group-separator={,},group-minimum-digits=1]{3724197}). Cascade-specific random effects are included.
}
\end{table}

\newpage
\section{Analysis With Cascade-Size Fixed Effects}
\label{appendix:size_fe}

A recent study by Juul \& Ugander (2021) \cite{Juul.2021} observed that structural differences at the cascade level (\ie, for folded retweet cascades) between real and false rumors can be largely explained by differences in the cascade size. However, different from this study, we analyze users' sharing behavior at the retweet level. In our regression analyses, we account for heterogeneity in the structural properties of cascades through our random-effects specification, which implicitly controls for the overall size of the cascade. As an additional check, we repeated our analysis with cascade-size fixed effects to control for the total number of retweets of the rumor cascades explicitly. The regression results are reported in \Cref{tbl:regression_cascade_size_fe}. We find that all findings regarding lifetime and crowd effects remain consistent.

\begin{table}[H]
	%\TABLE
	\caption{Analysis With Cascade-Size Fixed Effects. \label{tbl:regression_cascade_size_fe}}
	{
		%\OneAndAHalfSpacedXI
		\scriptsize
		\begin{tabularx}{\textwidth}{@{\hspace{\tabcolsep}\extracolsep{\fill}}l *{2}{S} } 
			\toprule
			& \multicolumn{1}{c}{{\begin{tabular}{@{}c@{}}Dependent Variable:\\ \emph{Retweet Count}\\\end{tabular}}} 
			& \multicolumn{1}{c}{{\begin{tabular}{@{}c@{}}Dependent Variable:\\\emph{Response Time (log)}\\\end{tabular}}} \\
			\midrule
			& {\textbf{Model (1)}} & {\textbf{Model (2)}} \\ 
			\midrule
			Falsehood                        & 0.186^{***}  & 0.093^{***}  \\
			& (0.006)      & (0.012)      \\
			Falsehood $\times$ Lifetime      & -0.041^{***} & 0.217^{***}  \\
			& (0.004)      & (0.006)      \\
			Falsehood $\times$ Retweet Depth & 0.190^{***}  & -0.113^{***} \\
			& (0.005)      & (0.007)      \\			
			Lifetime                         & -0.041^{***} & 0.662^{***}  \\
			& (0.004)      & (0.006)      \\
			Retweet Depth                    & -0.315^{***} & -0.235^{***} \\
			& (0.005)      & (0.007)      \\
			Followers                        & 0.528^{***}  & -0.044^{***} \\
			& (0.001)      & (0.002)      \\
			Followees                        & -0.119^{***} & 0.153^{***}  \\
			& (0.001)      & (0.001)      \\
			Account Age                      & -0.046^{***} & -0.064^{***} \\
			& (0.001)      & (0.001)      \\
			User Engagement                  & -0.078^{***} & -0.029^{***} \\
			& (0.001)      & (0.001)      \\
			Verified Account                 & 0.052^{***}  & -0.377^{***} \\
			& (0.007)      & (0.020)      \\
			%cascade\_max\_size               & 0.318^{***}  & 2.344^{***}  \\
%			& (0.012)      & (0.042)      \\			
			Intercept                      & 0.109^{***}  & 2.230^{***}  \\
			& (0.011)      & (0.036)      \\
			\midrule
			Random effects (cascade level) & {Yes} & {Yes} \\																					
			Fixed effects (cascade size)  & {Yes} & {Yes} \\								
			\bottomrule
			\multicolumn{3}{r}{Significance levels: $^*p< 0.05$, $^{**}p< 0.01$, $^{***}p< 0.001$; standard errors in parentheses} \\
		\end{tabularx}
		}
	\subcaption*{\emph{Note:} The negative binomial regression in Model (1) explains the number of retweets. The OLS regression results in Model (2) explain the time difference between a user's tweet and the parent tweet. Unit of analysis is the retweet level ($N=\,$\num[group-separator={,},group-minimum-digits=1]{3724197}). Cascade-size fixed effects control for the overall size of the retweet cascades. Cascade-specific random effects are included.
}
\end{table}

%\clearpage
%\appendix
%
%\onecolumn
%

\end{document}